\documentclass[11pt]{article}
\usepackage[letterpaper, left=1in, top=1in, right=1in, bottom=1in,nohead,includefoot]{geometry}
\usepackage{color,charter,graphicx,natbib,here,setspace,multirow,amsmath,amssymb,amsfonts,enumitem,placeins}

\usepackage[utf8]{inputenc}

\usepackage[dvipsnames]{xcolor}
	\usepackage[pdftex]{hyperref}
	\hypersetup{colorlinks,
	citecolor=NavyBlue  
	}

\bibpunct{(}{)}{;}{a}{}{,} 

\def\figfile{jpg}
\def\figdir{figurescolorjpg} 
\def\suppsimfigdir{suppsimfigurescolorjpg} 

\def\a{\mbox{\boldmath$a$}}
\def\h{\mbox{\boldmath$h$}}
\def\H{\mbox{\boldmath$H$}}
\def\A{\mbox{\boldmath$A$}}

\def\C{\mbox{\boldmath$C$}}

\def\F{\mbox{\boldmath$F$}}

\def\R{\mbox{\boldmath$R$}}
\def\I{\mbox{\boldmath$I$}}
\def\zero{\mbox{\boldmath$0$}}
\def\one{\mbox{\boldmath$1$}}
\def\x{\mbox{\boldmath$x$}}

\def\b{\mbox{\boldmath$b$}}

\def\r{\mbox{\boldmath$r$}}

\def\bPhi{\mbox{\boldmath$\Phi$}}

\def\btheta{\mbox{\boldmath$\theta$}}

\def\W{\mbox{\boldmath$W$}}

\def\bomega{\mbox{\boldmath$\omega$}}

\def\m{\mbox{\boldmath$m$}}
\def\M{\mbox{\boldmath$M$}}
\def\h{\mbox{\boldmath$h$}}

\def\mA{\mathcal{A}}
\def\mD{\mathcal{D}}
\def\mH{\mathcal{H}}
\def\seq#1#2{#1{:}#2}
\def\eqno#1{eqn.~(\ref{eq:#1})}

\begin{document}

\title{Dynamic Bayesian Predictive Synthesis\\ in Time Series Forecasting}  
\author{Kenichiro McAlinn\thanks{Department of Statistical Science, Fox School of Business, Temple University,  Philadelphia, PA 19004. {\scriptsize  Email: kenichiro.mcalinn@temple.edu} }
\,  \& Mike West\thanks{Department of Statistical Science, Duke University, Durham,  NC 27708-0251. \newline \indent\,\,\,\,
 {\scriptsize  Email: Mike.West@duke.edu. Tel: +1 919 684 4210. Fax: +1 919 684 8594}   
}}
 
\maketitle\thispagestyle{empty}\setcounter{page}0

\begin{abstract}

We discuss model and forecast combination in time series forecasting.
A foundational Bayesian perspective based on agent opinion analysis theory defines a new framework for density forecast combination, and encompasses  several existing forecast pooling methods.
We develop a novel class of dynamic latent factor models for time series forecast synthesis; simulation-based computation enables implementation. 
These models can dynamically adapt to time-varying biases, miscalibration and inter-dependencies among multiple models or forecasters. 
A macroeconomic forecasting study highlights the dynamic relationships among synthesized forecast densities, as well as the potential for improved forecast accuracy at multiple horizons.

\end{abstract}

\newpage

\section{Introduction \label{sec:intro} }

Recent research at the interfaces of applied/empirical macroeconomics and Bayesian methodology development reflects renewed interest in questions of model and forecast comparison, calibration, and combination.    A number of creative 
ideas for forecast density pooling define new empirical models fitted by Bayesian
methods~\citep[e.g.][]{Terui2002,HallMitchell2007,Amisano2007,Hooger2010,Kascha2010,Geweke2011,Geweke2012,Billio2012,Billio2013,Aastveit2014,
Fawcett2014,Aastveit2015,Pettenuzzo2015,Negro2016}, with examples of 
improved forecast performance in studies in macroeconomics and finance. 
This growing body of research responds in part to the need to 
improve information flows to policy and decision makers at national and international levels.  
From a methodological viewpoint,  the field is also stimulated by the increased availability of formal 
forecasting models that yield full density forecasts; whether these arise from multiple 
professional forecasters and models, and sets of  competing econometric models, the result is a need to 
define formal,  reliable methodology for integrating such predictive information.  Importantly, such methods should
allow for learning  and integrating information about the forecasters or models, characterizing 
their anticipated biases, relationships, and dependencies among them, and how these characteristics change with time.


We contribute to this field with a class of models that implement {\em Bayesian predictive synthesis (BPS)} in a 
sequential, dynamic setting for time series forecasting.   Our work is partly 
motivated by a concern for foundational/theoretical underpinnings of methodology. 
A forecast combination rule/algorithm may demonstrate success in a specific case study, 
but understanding potential conceptual and theoretical foundations is of interest in order to advance
broader understanding-- through transparency of implicit underlying assumptions-- and hence open paths to 
possible   generalizations of practical importance.   Our study of this links to past literature on subjective Bayesian 
\lq\lq agent/expert opinion analysis"~\citep[e.g.][]{DVL1979,West1984,GenestSchervish1985,West1988,West1992c,West1992d,APDetal1995,French2011}.  In that context, 
a decision maker regards multiple models or forecasters as providers of \lq\lq forecast data" to be used in her 
prior-posterior updating~\cite[see also ][Sect 16.3.2]{WestHarrison1997book2}. It turns out that this perspective-- referred to as Bayesian predictive synthesis (BPS)-- theoretically identifies the existence of a  subclass of Bayesian updating rules in the form
of latent factor models; explicitly,  {\em the set of forecast densities are those of inherent latent factors linked to the 
outcome of interest}. This opens a path to exploring 
BPS approaches based on various assumptions about latent factor regression model forms.  We explore this in 
some illuminating examples, showing special cases   linked to various existing density pooling methods.   

The direct time series extension involves 
{\em dynamic latent factor models} in which sequences of forecast densities define time-varying priors for
inherent latent factor processes linked to the time series of interest.  We discuss   and  
develop this new class of   models 
for dynamic BPS.   A study of forecasting U.S. inflation demonstrates application,  showing how dynamic BPS is able
to adaptively learn over time about  time-varying biases and miscalibration of multiple models or forecasters, 
and generate useful insights into time-varying patterns of  dependencies among them while also improving forecast accuracy.  
The study includes illuminating comparison with other standard and recent methods of forecast density pooling.
Importantly, we argue  for defining BPS models specific to forecasting goals, and demonstrate this in multi-step ahead 
forecasting in the case study.  
 
Section~\ref{sec:foundation}    summarizes the 
foundations of BPS in agent opinion analysis theory free from the time series context, with key 
examples. Full details of the emerging class of dynamic 
latent factor models are then developed  for time series in Section~\ref{sec:dynamicBPS}.  
In Section~\ref{sec:Inf}, we develop a detailed case study in forecasting quarterly U.S. inflation, comparisons to 
other approaches, and aspects of analysis of synthetic data.  Additional comments 
in Section~\ref{sec:summary} conclude the paper.  Three appendices provide full details of the MCMC methodology for 
Bayesian analysis of dynamic latent factor BPS models underlying the case study, as well as additional graphical summaries 
from that analysis and of a separate synthetic data set created to validate the results in the case study. 

\paragraph{Some notation:} We use lower case bold font for vectors and upper case bold font for matrices. Vectors are columns by default. Distributional notation 
$y\sim N(f,v),$ $\x\sim N(\a,\A)$ and $k\sim G(a,b)$ are   for the 
univariate normal, multivariate normal and gamma distributions, respectively.   The delta Dirac function is $\delta_x(y),$ the probability mass function of 
$y$ degenerate at a point $x.$ 
We use, for example, $N(y|f,v)$ to denote the actual density function of $y$ when $y\sim N(f,v).$ 
Index sets $s{:}t$ stand for $s,s+1,\ldots,t$ when $s<t$, such as in   $y_{\seq1t} = \{ y_1,\ldots,y_t\}.$
  
\section{Background and Foundations \label{sec:foundation} }

\subsection{Subjective Bayesian Forecasting and Information Processing}

We begin with background and examples free from the time series context, for clarity.    Section 3
then moves to the dynamic setting with direct translation of the theory summarized here.
 
In agent opinion analysis~\citep{West1992d} 
a Bayesian decision maker  $\mD$ aims to predict an outcome $y$ and to use information from $J$ individual agents (models, forecasters, or forecasting agencies, etc.) labelled $\mA_j$, $(j=\seq1J).$ To begin, 
$\mD$  has prior $p(y)$; the agents provide her with forecast information in terms of their pdfs $h_j(y).$   
These forecast densities represent the individual inferences from the  agents; for $\mD,$ they are data and
define the information set 
$\mH = \{ h_1(\cdot), \ldots, h_J(\cdot) \}.$  Thus 
$\mD$ will predict $y$ using her implied posterior $p(y|\mH)$.   Given the complex nature of $\mH$-- a set of
$J$ density functions, in a setting where there will  be varying dependencies among
agents as well as individual biases-- a fully specified Bayesian model $p(y,\mH)=p(y)p(\mH|y)$ is not easily conceptualized. 

The relevant agent opinion analysis theory~\citep{GenestSchervish1985,West1992c,West1992d} 
is summarized as follows.  Before observing $\mH,$ suppose  that 
$\mD$ specifies her prior $p(y)$ and her {\em prior expectation} 
of the product of agent densities, denoted by $m(\x) = E[ \prod_{j=\seq1J}h_j(x_j) ].$  Here $\x=x_{\seq1J} = (x_1,\ldots,x_J)'$ is a vector of
$J$ dummy variables and, critically, the expectation is over the {\em a priori} uncertain functions $h_j(\cdot).$ 
 That is, $\mD$  specifies only these two summary aspects of her full, but otherwise incompletely specified 
joint prior $p(y,\mH).$  Then, it follows that there 
exists a subset of Bayesian models $p(y,\mH)$ under which the required posterior has the form
\begin{equation}\label{eq:theorem1}
	p(y|\mH)=\int_{\x} \alpha(y|\x)\prod_{j=\seq1J}h_j(x_j)d\x
\end{equation}
where $\x=x_{\seq1J} = (x_1,\ldots,x_J)'$ is a $J-$dimensional latent vector and $\alpha(y|\x)$ 
a conditional pdf for $y$ given $\x.$     Critically, this shows that there exist latent variables $\x$ with the interpretations that:
{\em (i)} the forecast distribution of agent $j$ is that of latent variable $x_j,$ {\em (ii)} the latent variables are generated
independently from the $h_j(\cdot),$ and (iii) the latent variables relate as a group to the outcome $y$ through a conditional
(regression) distribution $\alpha(y|\x).$ 

Interpret  the latent factor vector $\x$ as follows. 
Suppose the (purely hypothetical) agent information $h_j(y) = \delta_{x_j}(y)$-- delta Dirac functions-- for $j=\seq1J.$
That is, $\mA_j$ makes a perfect prediction  $y=x_j$ for some specified value $x_j.$ 
$\mD$'s posterior is then $\alpha(y|\x)$ which   reflects 
her views of $y$ based on supposedly exact predicted values (or \lq\lq oracle" values)  from the agents. 
We refer to the $x_j$ as the {\em latent agent states} and to $\alpha(y|\x)$ as $\mD's$ {\em calibration function}. 

Critically, the representation of \eqno{theorem1} does not require a full specification of $p(y,\mH)$,
and does  not indicate what the functional form of $\alpha(y|\x)$ is; this opens the 
path to developing models based on different specifications of $\alpha(y|\x).$   
This posterior $p(y|\mH)$ must be consistent with the specified $p(y)$
and $m(\x)$, seen by integrating eqn.~(\ref{eq:theorem1}) with respect to $\mH$ to yield  
$p(y)=\int \alpha(y|\x) m(\x) d\x.$   Note the implied role and interpretation of $m(\x)$ here: prior to 
observing $\mH,$ the pdf $m(\x)$ is $\mD's$  prior for the latent agent states; on observing $\mH,$ her posterior 
for $\x$ is the product of the $h_j(x_j).$ 
This shows how $\mD$ can
incorporate views, and historical information, about agent-specific biases, patterns of 
miscalibration, inter-dependencies among agents and their relative expertise and expected
forecast accuracy   
implicitly through $m(\x)$ and/or more directly through specification of $\alpha(y|\x).$
This, and the nature of the overall framework,  is best seen through examples. 

\subsection{Examples: Mixture BPS and Connections to Other Methods\label{ex:BPSmix}}

A first general point is that BPS defines a theoretical foundation for {\em any} pooling method 
that implicitly relies on the idea of latent factors generated from the $h_j(\cdot).$   \cite{Aastveit2015}
is a key example, as their pooling method/model is explicitly constructed based on simulation from the
$h_j(\cdot)$ (using sequential Monte Carlo methods).  These authors use a specific class of models that
can now be seen as implicitly defining-- in a time series extension-- BPS models with particular (rather 
complicated and time-varying) calibration functions.  The BPS theory indicates that this is now
justified from a foundational Bayesian perspective. 

An example class takes   
$\alpha(y|\x) = a_0(\x) \pi_0(y) + \sum_{\seq1J} a_j(\x)\delta_{x_j}(y)$, a 
mixture of point masses with a base prediction $\pi_0(y)$ using 
{\em agent state-dependent probabilities} $a_{\seq0J}(\x).$   
With this specific form of calibration function, it follows from 
eqn.~(\ref{eq:theorem1}) that 
\begin{equation} \label{eq:BPSmix} 
p(y|\mH)= q_0(y|\mH) \pi_0(y) + \sum_{\seq1J} q_j(y|\mH)h_j(y)
\end{equation}
where the probability weights
$q_j(\mH) = \int a_j(\x)\delta_{x_j}(y)\prod_j h_j(x_j)d\x$  (for $j=\seq1J)$ depend on  $\mH$ and naturally act to 
recalibrate the contributions from each $h_j(\cdot)$ modulo the specified forms of the calibration weights $a_j(\x).$ 
Some specific examples connect with existing methods as well as to new and practically relevant features. 

{\bf (A) Constant weights:} If the calibration function mixture weights are constant, $a_j(\x)=a_j$ for $j=\seq 1J,$ 
 the implied posterior in eqn.~(\ref{eq:BPSmix}) has  agent density $j$ weights $q_j(y|\mH) = a_j.$ 
Thus, the BPS framework has several special cases: {\em (i)}  the traditional Bayesian model averaging (BMA) approach, in 
which the $a_j$ are assigned as model probabilities based on past data and experience; 
{\em (ii)}  other mixture approaches in which $\mD$ is free to choose the weights $\a_j$ to optimize any 
objective function as in stylized optimal prediction pools~\citep[e.g.][]{Geweke2011}, or to specify them based on past performance in forecasting or decision problems~\citep[e.g.][]{Pettenuzzo2015}.

{\bf (B) Agent-specific outcome dependent weights: } Suppose the calibration function mixture weight for agent $j$ depends only on latent agent factor $j,$ 
 i.e., $a_j(\x)=a_j(x_j)$ for $j=\seq 1J.$ Then the implied posterior in eqn.~(\ref{eq:BPSmix}) has  agent density $j$ weights $q_j(y|\mH) = a_j(y)$ which acts to modify and \lq\lq recalibrate'' the contribution of forecast $h_j(y).$  This defines an
 \lq\lq outcome dependent'' mixture posterior, the weights on agent forecast densities depending on $y,$  
 i.e., defines \lq\lq generalized density combination'' methods~\citep{Fawcett2014} as special cases. 
  
{\bf (C) General dependence weights:} Generally, 
 $a_j(\x)$ can be chosen to incorporate $\mD's$ views of agent-specific biases, miscalibration and--
critically and not represented in other   approaches-- {\em dependencies among agents}.   
$\mD$ might choose 
weight functions $a_j(\x)$ as \lq\lq kernels'' to correct for expected location biases, to 
decay at values of $\x$ viewed as extreme under $m(\x)$, to take smaller values in cases of expected 
high positive dependency among agents when the $x_j$ values are clustered together,  to reflect increased 
uncertainty in cases of high dispersion of the $x_j,$  and so forth. In the   posterior of
eqn.~(\ref{eq:BPSmix}), the weight function $q_j(y|\mH)$  on $h_j(y)$  is outcome dependent as in example (B) and \cite{Fawcett2014},  but   now-- depending on choices of $a_j(\x)$-- can capture aspects of cross-agent dependencies as well as 
  agent-specific calibration features. 
  
These examples serve  as relevant background and connect to some prior approaches.  Parallel research reported in
\cite{JohnsonWest2016} develops Bayesian methodology for general classes of such mixture models, based on various 
assumed forms of calibration weights $a_j(\x)$.   For the current paper,  we move to quite different and 
core examples based on normal theory
(rather than mixtures) and that underlie our main methodological extensions to dynamic latent factor models.

\subsection{Examples: Normal Latent Factor BPS  \label{ex:BPSnormal}}
Directly relevant to the  dynamic BPS models below, a  class of examples arises when the implied 
joint prior $\alpha(y|\x)m(\x)$
is multivariate normal or T, which easily and intuitively allow $\mD$ to incorporate views 
of  agent biases, miscalibration, and dependencies. 
Suppose a case in which $(y,\x)$ are jointly normal 
with margins  $p(y) = N(y|f,q)$ and $m(\x) = N(\x| \m, \M)$
where: {\em (i)}  $\m = f\one+\b$ for some $J-$vector $\b=(b_1,\ldots,b_J)'$ 
and where $\one$ is the $J-$vector of ones; {\em (ii)} $\M$ is a
$J\times J$ variance  matrix $\M$ with diagonal elements $q s_j$ for some scales $s_j>0.$ 
Here $\mD$  expects agent forecast distributions to be normal (although of course her full, but incompletely specific prior
recognizes that is just an expectation),  and includes explicit, agent-specific location bias $b_j$ and scale bias $s_j$ 
relative to $\mD's$ prior.      The covariances in $\M$ explicitly recognize cross-agent dependencies; if $\mD$ regards agents $i,j$ 
as very highly dependent, then the $i,j$ correlation should be large and positive.  
The full joint normal for $(y,\x)$ is completed with a covariance vector $\r=C(\x,y) = (r_1,\ldots,r_J)'.$ The implied conditional 
$\alpha(y|\x)$ is then normal with 
$E(y|\x) = f + \sum_{j=\seq1J} \theta_j (x_j-m_j)$
and $V(y|\x) = v$ given by $v = q - \sum_{j=\seq1J} r_j\theta_j $  where the $\theta_j$ are the elements of the 
$J-$vector $\M^{-1}\r.$   

Suppose now the agents provide normal forecasts, $h_j(y) = N(y| h_j, H_j)$ for $j=\seq1J.$  Then
$\mD's$ posterior $p(y|\mH)$ is, from eqn.~(\ref{eq:BPSmix}), normal with 
$E(y|\mH) =  f + \sum_{j=\seq1J} \theta_j (h_j-m_j)$ and variance 
$V(y|\mH) = v + \sum_{j=\seq1J} \theta_j^2H_j.$ 
Thus, the posterior effectively corrects for quantified biases in location and 
scale for each agent, while also now incorporating the quantified views of cross-agent dependencies implicitly through the
implied regression coefficients $\theta_j.$     In special cases, this justifies approaches using 
simple linear pooling of point forecasts, but now with explicit bias corrections and uncertainty quantification via the full posterior. 
Also, the posterior mean and variance of $p(y|\mH)$ above   arise with non-normal agent densities when $(h_j,H_j)$ 
are agent $j$'s mean and variance.

From a methodological viewpoint,  it is the role of the bias and dependence parameters in defining the regression model
$\alpha(y|\x)$ itself that is key.   
This calibration density can be rewritten as 
\begin{equation} \label{eq:BPSnormalalpha}
\alpha(y|\x) = N(y|\F'\btheta,v) \quad\textrm{with}\quad \F=(1, \x')' \quad\textrm{and}\quad
\btheta=(\theta_0,\theta_1,...,\theta_J)'
\end{equation} 
where  $\theta_0 = f-\sum_{j=\seq1J}\theta_jm_j.$   Thus, the practically relevant {\em effective} calibration 
parameters are $(\btheta,v)$, and historical data will inform $\mD$ on these.
This defines the starting point for analysis and extension to
time series contexts, as now follows. 

%

\section{Dynamic BPS using Latent Factor Regression Models \label{sec:dynamicBPS}} 

The new methodological developments forming the core of this paper adapt and 
extend the basic BPS framework   to sequential forecasting in 
time series. In particular, we develop dynamic extensions  of Section~\ref{ex:BPSnormal}
involving  time-varying parameters to characterize and 
formally allow for agent-specific biases, patterns of 
miscalibration, inter-dependencies, and relative expertise/forecast accuracy as
time evolves and data is processed.  We do this in the context of a scalar time series. 

Among the connections in recent literature we have already noted that 
\cite{Hooger2010} and \cite{Aastveit2015} relate directly in key aspects of technical structure. 
In addition to opportunities for time-varying parameter models-- a special
case of the broader DLM setting developed in the following sections-- these authors develop 
empirical methods using forecasts {\em simulated} from the sets of agent models. Based on the BPS theory, 
this is now understood to be foundationally justified since 
agent forecast distributions are those of implied latent factors that relate to 
the outcome of interest.  As we see below, practical Bayesian analysis of dynamic BPS models naturally 
involves simulation of these latent states from the agent distributions in forecasting computations;
however, they must be simulated from different distributions-- the appropriate conditional posteriors-- 
for model fitting and analysis.

\subsection{Dynamic Sequential Setting} 

The decision maker $\mD$ is sequentially predicting a time series $y_t, t=1,2, \ldots,$ and
at each time point receives forecast densities from each agent.  At each time $t-1,$ $\mD$ aims to 
forecast $y_t$ and receives current forecast densities 
$\mH_t = \{ h_{t1}(y_t),\ldots, h_{tJ}(y_t) \}$ from the set of agents. The full 
information set used by $\mD$ is thus 
$\{ y_{\seq1{t-1}}, \ \mH_{\seq1t} \}.$ 
 As data accrues, 
$\mD$ learns about relationships among agents, their forecast and dependency 
characteristics, so that a Bayesian model will involve parameters that define the
BPS framework and for which $\mD$ updates information over time.   The implication for 
the temporal/dynamic extension of the BPS model of Section~\ref{sec:foundation}
is that $\mD$ has a time $t-1$ distribution for $y_t$ of the form
\begin{equation}\label{eq:theorem}
	p(y_t|\bPhi_t,y_{\seq1{t-1}},\mH_{\seq1t}) 
	\equiv p(y_t|\bPhi_t,\mH_t)=\int \alpha_t(y_t|\x_t,\bPhi_t)
		\prod_{j=\seq1J}h_{tj}(x_{tj})dx_{tj}
\end{equation}
where $\x_t=x_{t,\seq1J}$ is a $J-$dimensional latent agent state vector at time $t$, 
$\alpha_t(y_t|\x_t,\bPhi_t)$ is $\mD$'s conditional calibration  pdf for $y_t$ given $\x_t,$ 
and $\bPhi_t$ represents time-varying parameters defining the calibration pdf-- parameters 
for which $\mD$ has current beliefs represented in terms of a current
(time $t-1$) posterior $p(\bPhi_t|y_{\seq1{t-1}},\mH_{\seq1{t-1}}).$    The methodological focus
can now rest on evaluation of models based on various assumptions about the form 
of $\alpha_t(y_t|\x_t,\bPhi_t)$ and its defining dynamic state parameters $\bPhi_t.$ 
Naturally, we look to tractable dynamic linear 
regression models, a subset of the broader class of dynamic linear models, or DLMs~\citep{WestHarrison1997book2,Prado2010}, as a first approach to defining a  computationally accessible yet flexible 
framework for dynamic BPS. 

\subsection{Latent Factor Dynamic Linear Models \label{sec:dynamic}}

Consider a dynamic regression for BPS calibration that 
extends the basic example of \eqno{BPSnormalalpha} to the time series setting. 
That is, \eqno{BPSnormalalpha} becomes the dynamic version
\begin{equation} \label{eq:BPSnormalalphadynamic}
\alpha_t(y_t|\x_t,\bPhi_t) = N(y_t|\F_t'\btheta_t,v_t) \quad\textrm{with}\quad \F_t=(1, \x_t')' \quad\textrm{and}\quad
\btheta_t=(\theta_{t0},\theta_{t1},...,\theta_{tJ})',
\end{equation} 
the latter being the $(1+J)-$vector of time-varying bias/calibration coefficients and the conditional variance $v_t$ defining residual 
variation beyond that explained by the regression on latent agent factors.   Explicitly, the functional model parameters are now
  $\bPhi_t = (\btheta_t,v_t)$ at each time $t.$ 
This BPS specification defines the first component of the standard conjugate form dynamic linear model (DLM)~\citep[Section 4]{WestHarrison1997book2}
\begin{subequations}
\label{DLM}
\begin{align}
	y_t&=\F_t'\btheta_t+\nu_t, \quad \nu_t\sim N(0,v_t), \label{eq:DLMa} \\
	\btheta_t&=\btheta_{t-1}+\bomega_t, \quad \bomega_t\sim N(0, v_t\W_t)\label{eq:DLMb}
\end{align}
\end{subequations}
where  $\btheta_t$ evolves  in time according to a linear/normal random walk with
innovations variance matrix $v_t\W_t$ at time $t$,  and $v_t$ is the residual variance in predicting 
$y_t$ based on past information and the set of agent forecast distributions.
The residuals $\nu_t$ and evolution {\em innovations} $\bomega_s$ are independent over time and mutually independent
for all $t,s.$

The DLM specification is completed using standard discount factor based methods, long used in the core Bayesian 
forecasting literature~\cite[e.g.][]{WestHarrison1997book2,Prado2010} and of increasing use in econometric and financial 
applications in more recent times, with discount factors sometimes now referred to as \lq\lq forgetting factors''~\citep[e.g.][]{dangl2012,koop2013,GruberWest2016BA,GruberWest2017ECOSTA,ZhaoXieWest2016ASMBI}. First,  
the time-varying intercept and agent coefficients $\btheta_t$ 
follow the random walk evolution of \eqno{DLMb} where $\W_t$ is defined via a standard, 
single discount factor specification~(\citealt[][Sect 6.3]{WestHarrison1997book2}; \citealt[][Sect 4.3]{Prado2010}),
using a state evolution discount factor $\beta\in (0,1].$  Second,  
the residual variance $v_t$ follows a standard beta-gamma random walk volatility model 
~(\citealt[][Sect 10.8]{WestHarrison1997book2}; \citealt[][Sect 4.3]{Prado2010}), with 
$v_t = v_{t-1}\delta/\gamma_t$ for some discount factor $\delta\in (0,1]$ and 
where $\gamma_t$ are beta distributed innovations, independent over time and independent 
of $\nu_s,\bomega_r$ for all $t,s,r$.  
Given choices of discount factors underlying these two components, and a (conjugate normal/inverse-gamma) 
initial prior for $(\btheta_0, v_0)$ at $t=0,$ the model is fully specified. 

Eqns.~(\ref{DLM}) define a dynamic latent factor model: the $\x_t$ vectors in each $\F_t$ 
are latent variables. At time $t-1,$ the set of agent densities becomes available for
forecasting $y_t;$  then, from the BPS foundation,  each $x_{tj}$ is a latent draw from $h_{tj}(\cdot).$
Note that the latent factor generating process has the 
$x_{tj}$ drawn {\em independently} from their $h_{tj}(\cdot)$  and externally to the 
BPS model. That is, coupled with eqns.~(\ref{eq:DLMa},\ref{eq:DLMb}),  we have 
\begin{equation}\label{eq:dfmh}
 p(\x_t| \bPhi_t,y_{\seq1{t-1}},\mH_{\seq1t}) \equiv p(\x_t|\mH_t) = \prod_{j=\seq1J} h_{tj}(x_{tj}) 
\end{equation}
for all time $t$ and with $\x_t,\x_s$ conditionally independent for all $t\ne s.$ 
Critically, the independence of the $x_{tj}$ {\em conditional on the $h_{tj}(\cdot)$} must not be confused with the question
of $\mD$'s modeling and estimation of the 
{\em dependencies among agents};  this is simply central and integral, and reflected through the 
effective DLM parameters $\bPhi_t = (\btheta_t,v_t).$

We use standard DLMs to define synthesis functions $\alpha_t(y_t|\x_t,\bPhi_t)$  in view of the central role
of such models in Bayesian forecasting, and the flexibility they have in 
adapting to time-varying structure and relationships.  That said, the BPS theory does not imply any specific structure for the synthesis functions; the decision maker $\mD$ is free to make alternative  model specifications.

Some methods of point forecast combination restrict weights on forecasts to the unit simplex and to sum to one, i.e., as 
probabilities.  
In the dynamic BPS setting,  restricting the dynamic regression coefficients would be 
technically challenging, but more importantly we do not see it as relevant. We can expect such constraints to
 lead to BPS models that underperform compared to the unrestricted case.  
For example, consider the case where all agents overestimate the quantity of interest by some positive value.
Under the restrictive case, there is no combination of weights that can achieve that quantity, while the unrestricted analysis
will appropriately adapt. If of interest, the posteriors for 
$\btheta_t$ over time can of course be explored to investigate whether or not the model:data combination provides support for any kind of such restriction.

\subsection{Bayesian Analysis and Computation \label{sec:comp}}

At any current time $t,$    $\mD$ has available the history of the BPS analysis to that point, 
including the now historical information $\{ y_{\seq1t}, \mH_{\seq1t}\}.$     Over times $\seq1t,$ 
the BPS analysis will have involved inferences on both the latent agent states $\x_{\seq1T}$ as well as
the dynamic BPS model parameters $\bPhi_{\seq1T}.$  Importantly,  inferences on the former 
provide insights into the nature of dependencies among the agents, as well as individual agent 
forecast characteristics. The former addresses key and topical issues of overlap and redundancies
among groups of forecasting models or individuals, as well as information sharing and potential  
herding behavior within groups of forecasters.  The \lq\lq output" of full posterior 
summaries for the $\x_t$ series is thus a key and important feature of BPS.

For posterior analysis, the holistic view is that $\mD$ is interested in computing the posterior for 
the full set of past latent agent states (latent factors) and dynamic parameters   $\{ \x_{\seq1t}, \bPhi_{\seq1t}\}$,
rather than restricting attention to forward filtering to update posteriors for current values
  $\{ \x_t, \bPhi_t\}$; the latter is of course implied by the former.    This  analysis is enabled by
Markov chain Monte Carlo (MCMC) methods, and then forecasting from time $t$ onward follows by 
theoretical and simulation-based extrapolation of the model; both aspects involve novelties in the BPS framework 
but are otherwise straightforward extensions of traditional   methods 
in Bayesian time series~(\citealt[][Chap 15]{WestHarrison1997book2}; \citealt{Prado2010}).

\paragraph{Posterior Computations via MCMC.}     The dynamic latent factor model   of 
eqns.~(\ref{eq:DLMa},\ref{eq:DLMb},\ref{eq:dfmh})
leads   to a two-component block Gibbs sampler for sets of the 
latent agent states $\x_t$ and DLM dynamic parameters $\bPhi_t$. These are iteratively 
resimulated from   two conditional posteriors noted below, with obvious 
initialization based on agent states drawn independently from priors $h_\ast(\ast).$  

First, conditional on values of agent states, the next MCMC step draws  new parameters from 
  $  p( \bPhi_{\seq1t} |  \x_{\seq1t}, y_{\seq1t} ).$
By design, this is a discount-based dynamic linear regression model, and sampling 
uses the standard forward filtering, backward sampling (FFBS) algorithm 
(e.g.~\citealt{Schnatter1994}; \citealt[][Sect 15.2]{WestHarrison1997book2}; \citealt[][Sect 4.5]{Prado2010}). 

Second,  conditional on values of dynamic parameters, the  MCMC  draws new agent states from 
$ p( \x_{\seq1t} |  \bPhi_{\seq1t}, y_{\seq1t}, \mH_{\seq1t} ).$    It is immediate that the $\x_t$ are
conditionally independent over time $t$ in this conditional distribution, with time $t$ 
conditionals 
$$ p( \x_t|  \bPhi_t, y_t, \mH_t) \propto N(y_t|\F_t'\btheta_t, v_t) \prod_{j=\seq1J} h_{tj}(x_{tj}) 
	\quad\textrm{where}\quad  \F_t=(1, x_{t1},x_{t2},...,x_{tJ})'.$$
In cases when each of the agent forecast densities is normal,  this yields a multivariate normal 
for $\x_t$ that is trivially sampled.  In other cases,  this will involve either a Metropolis-Hastings simulator 
or an augmentation method. A central, practically relevant case is when  
agent forecasts are T distributions;  each $h_{tj}(\cdot)$ can then be 
represented as a scale mixture of normals, and augmenting the posterior MCMC to 
include the implicit underlying latent scale factors generates conditional normals for each $\x_t$ coupled
with conditional inverse gammas for those scales. This is again a standard MCMC approach
and much used in Bayesian time series, in particular (e.g.~\citealt{Schnatter1994}; \citealt[][Chap 15]{WestHarrison1997book2}).

Full technical details of the MCMC computations, and additional discussion,  
is given in the supplementary Appendix~\ref{supp:comp}.
 
\paragraph{Forecasting 1-Step Ahead.}   
At time $t$ we forecast 1-step ahead by generating \lq\lq synthetic futures" from the BPS 
model, as follows: (i) For each sampled $\bPhi_t$ from the posterior MCMC above, 
draw $v_{t+1}$ from its discount volatility evolution
model, and then  $\btheta_{t+1}$ conditional on $\btheta_t,v_{t+1}$
from the evolution model~\eqno{DLMb}-- this  gives a draw 
$\bPhi_{t+1} = \{ \btheta_{t+1}, v_{t+1} \}$ 
from $p(\bPhi_{t+1} |y_{\seq1t}, \mH_{\seq1t} )$; 
(ii) Draw $\x_{t+1}$ via independent sampling of the $h_{t+1,j}(x_{t+1,j}),$  $(j=\seq1J);$   
(iii) Draw $y_{t+1}$ from the conditional normal of~ \eqno{DLMa} given these sampled
parameters and agent states. Repeating this generates a random sample from the 1-step ahead
forecast distribution for time $t+1$.

\subsection{Multi-Step Ahead Forecasting \label{sec:k-step}}

Forecasting over multiple horizons is often of   greater importance than 1-step ahead forecasting.   
Economic policy makers, for example,  forecast/assess macroeconomic variables over a year or multiple years, 
drawing from their own forecast models, judgemental inputs, other economists and forecasters, in order to advise
policy decisions.
However, forecasting over longer horizons is typically more difficult than over 
shorter horizons, and models calibrated on the short-term basis can often 
be quite poor in the longer-term. As noted in Section~\ref{sec:intro}, fitting of time series models
is inherently based on 1-step ahead, as DLM (and other) model equations make explicit. So, when
multi-step ahead forecasting is primary,  new ideas for forecast calibration and combination are needed.
BPS provides a natural and flexible  framework to synthesize forecasts over multiple horizons, 
with potential to improve forecasting at multiple horizons simultaneously, as we now discuss.

\paragraph{Direct projection for multi-step forecasting.} 
At time $t,$  the agents provide $k$-step ahead forecast densities $h_{t,\seq1J}(x_{t+k})$. 
The direct approach follows traditional DLM updating and forecasting via 
simulation as for 1-step ahead. That is: (i) project the 
BPS model forward from time $t$ to $t+k$ by simulating the dynamic model parameters 
$\bPhi_{t+1}, \bPhi_{t+2}, \ldots, \bPhi_{t+k}$ using sequential, step ahead extension of the
1-step case; (ii) draw independently from each of the 
$h_{t,\seq1J}(x_{t+k})$ to give a sampled vector $\x_{t+k};$ then (iii) draw $y_{t+k}$   
from the conditional normal given these sampled parameters and states.   While this is theoretically
correct, it fails to update/calibrate based on the horizon of interest, relying wholly on the
model as fitted-- with its essential basis in 1-step forecasting accuracy-- even though $\mD$ 
may be mainly interested in forecasting several steps ahead. 

\paragraph{BPS$(k)$ for customized multi-step forecasting.} 
BPS can be customized to forecasting goals, allowing $\mD$  
to focus  on the horizon of interest.   This responds in part  to the reality that
a model that forecasts well in the short-term may be useless for multiple steps ahead, while 
another model may be ideal for several steps ahead but poorer in the short-term. As noted by others~\citep[e.g.][and references therein]{Aastveit2014} this calls for  combination methods specific to the forecast horizon.  We therefore consider 
multiple BPS models in parallel for different forecast horizons. 

This involves a   modification of  Section~\ref{sec:dynamicBPS} in which the model at time $t-1$ 
for predicting $y_t$ changes as follows.  With a {\em specific forecast horizon} $k>1$, 
modify the BPS so that the agents' $k$-step ahead forecast densities 
made at time $t-k,$  i.e., $h_{t-k,j}(x_{tj})$ replace $h_{tj}(x_{tj})$ in  the resulting 
model analysis. This changes the interpretation of the dynamic model parameters $\{ \btheta_t, v_t\}$ 
to be explicitly geared to the $k$-step horizon. Bayesian model fitting then naturally 
\lq\lq tunes" the model to  the horizon $k$ of interest.   Forecasting the chosen $k$-steps ahead
now simply involves extrapolating the model via simulation, as above, but now in this modified and
horizon-specific BPS model. 

We denote this customized model strategy by BPS$(k)$ to distinguish it from the
direct extrapolation of BPS. 
Note that this is fundamentally different from the traditional method of model extrapolation 
as it directly updates, calibrates, and learns using the horizon of interest.  The applied study in 
Section~\ref{sec:Inf} below bears out the view that this can be expected to improve 
forecasting accuracy over multiple horizons.  One cost, of course, is that a bank of BPS models
is now required for any set of horizons of interest; that is, different models will be built for 
different horizons $k,$ so increasing the computational effort required.
We do not link analyses across horizons, i.e., do not relate dynamic BPS model 
 parameters in BPS(4) to those of BPS(1), for example.    A model that is good at short-term forecasting may be 
 hopeless at longer horizons, and vice-versa,  so there is no notion of a formal relationship between BPS models, 
 and their parameters, at different horizons.

We further note  contextual relevance of this perspective in macroeconomics
when $\mD$ is a consumer of forecasts from groups, agencies or model developers. 
Such  agents may use different models, data, advisors, and approaches for different horizons.  
When the forecast generating models/methods are known,  $\mD$ may redefine  the BPS model accordingly;
however, generally in practice these underlying models, strategies, data, and advisors 
will not be wholly known and understood.  

\section{US Macroeconomic Time Series \label{sec:Inf}}

\subsection{Data, Forecasting Models and Implementation \label{sec:data}}

\paragraph{Time Series Data.} 
We analyze quarterly U.S. macroeconomic data, focusing on forecasting inflation rates with
both 1-quarter and 4-quarter ahead interests. 
The study involves three quarterly macro series:  annual inflation rate $(p)$, short-term nominal interest rate $(r)$, 
and unemployment rate $(u)$ in the U.S. economy from 
1961/Q1 to 2014/Q4, a context of topical interest~\citep{Cogley2005,Primiceri2005,Koop2009,Nakajima2010}.
The inflation rate is the annual percentage change in a chain-weighted GDP price index, the interest rate is the yield on three-month Treasury bills, and the unemployment rate is seasonally adjusted and includes all workers over 16 years of age.
Data are recorded as the latest  last vintage when the data was collected (2015/1Q).
Prior studies~\citep[e.g.][]{Nakajima2010} use data over the period of 1963/Q1-2011/Q4; we extend 
this to more recent times, 1961/Q1 to 2014/Q4 (see data display  
in supplementary Appendix B). We focus on forecasting 
inflation using  past values of the three indices as candidate predictors underlying a set of four time series models-- the $J=4$ agents--  to be evaluated, calibrated, and synthesized.   
The time frame includes key periods that warrant special attention: the early 1990s recession, the Asian and Russian financial crises in the late 1990s, the dot-com bubble in the early 2000s, and the sub-prime mortgage crisis and great recession 
of the late 2000s.
These periods exhibit sharp shocks to the U.S. economy generally, and test the predictive ability of any models and strategies under stress. For any forecast calibration and aggregation strategy to be effective and useful, its predictive performance 
must be robust under these conditions; most traditional macroeconomic models suffer significant deficiencies in such 
times. 

\paragraph{Agent Models and BPS Specification.} 
The $J=4$ agents represent the two major structures of time series forecast models: factor and lag. 
Labeling them M*, the agent models for inflation $y_t\equiv p_t$ use predictors:   
M1-  $p_{t-1}$;  
M2- $p_{t-1:3}, r_{t-1:3}, u_{t-1:3}$; 
M3- $p_{t-1:3}$; 
M4- $p_{t-1}, r_{t-1}, u_{t-1}$. 
Thus, each has a time-varying autoregressive term in inflation rate $p$, while two also have dynamic regressions on lagged interest rate $r$ and unemployment rate $u$, the differences being in lags chosen and model
complexity. In each, residual volatility follows a standard beta-gamma random walk. 
Each M* is a standard DLM so that 
model fitting and generation of forecasts is routine.  
While these models are simple compared to more sophisticated models used to forecast inflation (such as \citealp{SW99,SW07,Stella12,Belmonte14} as well as Bayesian TVP-VARs seen in \citealp{Nakajima2010}), part of the utility and appeal of predictive synthesis, and in forecast combinations in general, is in gaining improvements using relatively simple models and not resorting to complicated models that have technical/computational difficulties.
Additionally, we can expect that adding these sophisticated models to the set of agents will only improve the synthesis.
That said, the chosen models are, in fact,  not so trivial in terms of forecasting performance 
{\em per se}, and most more elaborate inflation models 
are   limited in forecasting ability, especially at longer horizons. So, while the selected simpler univariate models  
are not   by any means  \lq\lq proposed'' as state-of -the-art, they are   of non-trivial interest as well as providing inputs to the
main subject of the paper and, as we detail below,  serve the main purposes in demonstrating application of 
dynamic latent factor BPS models in this now foundationally defined Bayesian approach.   
Using these easily implemented models, we see 
the BPS analysis exhibiting the ability to adapt to   
time-varying biases in individual models, automatically adapt/reweight models differently at differing time periods based on \lq\lq local'' performance, identify dependencies among forecast models (and adapt the synthesized
forecasts to them), and define improved forecasting at multiple horizons with BPS models customized to horizon.

Prior specifications for the DLM state vector and 
discount volatility model in each of the $J$ agent models is based on $\btheta_0|v_0\sim N(\zero, (v_0/s_0)\I)$
and $1/v_0\sim G(n_0/2,n_0s_0/2)$ with $n_0=2, s_0=0.01,$ 
using the usual $(\btheta,v)$ DLM notation~\citep[][Chap 4]{WestHarrison1997book2}. 
Each agent model uses standard discount factor $(\beta)$ specification for
state evolution variances and discount factor $(\delta)$ for 
residual volatility; we use 
$(\beta,\delta)=(0.99,0.95)$ in each of these agent models.  
All  DLM-based agent forecast densities $h_\ast(\cdot)$  are then those of predictive T distributions.   

In the dynamic BPS models for forecast horizons $k=1$ and $k=4,$ we take initial priors as 
$\btheta_0|v_0\sim N(\m_0, (v_0/s_0)\I)$ with $\m_0=(0,\one'/J)'$  and $1/v_0\sim G(n_0/2,n_0s_0/2)$ 
with $n_0=10,s_0=0.002$. 
 BPS for 1-step ahead forecasting is based on 
$(\beta,\delta)=(0.95,0.99)$, while
BPS$(4)$, customized to 4-quarter ahead forecasting as discussed in Section~\ref{sec:k-step},
uses $\btheta_0|v_0\sim N(\m_0, 10^{-4}(v_0/s_0)\I)$ and $(\beta,\delta)=(0.99,0.99).$     
Differences by forecast horizon echo
earlier discussion about different model choices being relevant to different forecast goals. 

In general, discount factors should be set between $0.9-0.99$, expressing views on the variability of $\btheta_t$, through $\upsilon_tW_t$ (via $\beta$), and   $\upsilon_t$ (via $\delta$).
If the decision maker $\mD$ believes that the synthesis will benefit from weights that are extremely flexible (e.g. where good agents change over multiple periods of time), then she can specify a lower discount factor  $\beta$.
In responding to dynamics, taking discount factors too small will lead to overly adaptive models that may 
show some small improvements in short-term forecasting accuracy but be quite poor at longer horizons, whereas
discount values too close to 1 lead to under-adaptive and \lq\lq over-smoothing'' models.  Relevant choices are, of course, 
always context dependent.   In this example, we can assume that the simple models used will significantly underperform in the long term forecasts, as they overfit to short term dynamics.
For this reason, we set a higher discount factor $\beta$ to smooth out the parameters because we can expect that there is very little signal to extract from the agents.    


The prior mean $\m_0$ of $\btheta_0$ equally weights the latent agent states as a default neutral position, while the prior
conditional variances are large, and the degrees of freedom low, so that initial data will have a substantial impact in modifying the
implied, relatively vague priors.   
We have explored analyses across ranges of priors and discount factors, and chosen these values as they lead to good agent-specific and BPS forecasting accuracy; our central conclusions 
with respect to BPS do not change materially with different values close to those chosen for the
summary examples. 
It is always relevant to use   rather diffuse initial priors and to experiment with those and
discount factors on an initial period of training data, and then monitor forecasting performance over time based on a chosen 
model and its defining discount factors. The sequential framework allows for intervention to change model posteriors, and
fixed parameters including the key discount factors, and practitioners should always be open to intervention opportunities. 
Looking at the behavior of the parameters during an initial learning period with training data gives   insight into-- among other things-- 
choice of discount factors and prior specifications.
For example, if the coefficients look jumpy/static during the learning period, it might be wise to increase/decrease the discount factor $\beta$ to achieve a good level of adaptability.

\paragraph{Data Analysis and Forecasting.} 
The 4 agent models are analyzed and synthesized as follows. 
First, the agent models are analyzed in parallel over 1961/Q1-1977/Q1 as a training period, simply 
running the DLM forward filtering to the end of that period to calibrate the agent forecasts.  This continues over 1977/Q2-1989/Q4 now accompanied by the calibration of the other forecast combination methods. Also, at each quarter $t$  during this period, 
the MCMC-based BPS analysis is run using data from 1977/Q2 up to time $t$; that is, we repeat the 
analysis with an increasing \lq\lq expanding window" of past data as we move forward in time. 
We do this for the traditional 1-step ahead focused BPS model, and-- separately and in parallel-- for a 4-step
ahead focused BPS(4) model, as discussed in  Section~\ref{sec:k-step}.
This continues over the third period to the end of the series,  1990/Q1-2014/Q4; now we also 
record and compare forecasts as they are sequentially generated. This testing period spans over a quarter century, and we are able to explore predictive performance over periods of drastically varying economic circumstances, check robustness, and compare benefits and characteristics of each strategy.
Out-of-sample forecasting is thus conducted and evaluated in a way that mirrors the realities facing 
decision and policy makers.

\paragraph{Forecast Accuracy and Comparisons.}
We compare forecasts from BPS with standard Bayesian model uncertainty analysis (or Bayesian model averaging-- BMA) in which the agent densities are mixed with respect to sequentially updated model 
probabilities~\citep[e.g.][Sect 12.2]{PJHandCF1976,WestHarrison1997book2}, and with the state-of-the-art density combination method of \cite{Billio2013} using the DeCo package \citep{Casarin2015}.
In addition, we compare with simpler, equally-weighted averages of agent forecast densities 
using both linear pools (equally-weighted arithmetic means of forecast densities) and  logarithmic pools
(equally-weighted harmonic means of forecast densities), with some theoretical underpinnings~(e.g.~\citealt{West1984}). 
While these strategies might seem overly simplistic, they have been shown 
to dominate some more complex aggregation strategies in some contexts, at least in terms of direct point forecasts in empirical studies~\citep[e.g.][]{Genre2013}.
For point forecasts from all methods, we compute and compare mean squared forecast error 
(MSFE) over the forecast horizons of interest.   In comparing density forecasts with BPS,
we also 
evaluate log predictive density ratios (LPDR); 
at horizon $k$ and across time indices $t$, this is 
\begin{align*}
	\mathrm{LPDR}_{\seq1t}(k)=\sum_{i=\seq1t}\mathrm{log}\{p_{s}(y_{t+k}|y_{1{:}t})/p_{\mathrm{BPS}}(y_{t+k}|y_{1{:}t})\}
\end{align*}
where $p_s(y_{t+k}|y_{1{:}t})$ is the predictive density under model or model combination 
aggregation strategy indexed by $s$, compared against the corresponding BPS forecasts at this horizon.
As used by several authors recently~\cite[e.g.][]{Nakajima2010,Aastveit2015},  LPDR measures provide a direct
statistical assessment of relative accuracy at multiple horizons that extend traditional 
1-step focused Bayes' factors. They weigh and compare dispersion of forecast densities
along with location, to elaborate on raw MSFE measures; 
comparing both measurements, i.e., point and density forecasts,  gives a broader understanding of the predictive abilities of the different strategies.

\subsection{Dynamic BPS and Forecasting \label{sec:outsample}}
 
Comparing predictive summaries over the out-of-sample period, BPS improves 
forecasting accuracy relative to the 4 agent models, and dominates BMA, DeCo, and the 
pooling strategies; see numerical summaries in Table~\ref{table:inf}.
Looking at point forecast accuracy, BPS exhibits improvements of no less than 10$\%$ over all models and strategies for 1- and 4-step ahead forecasts (BPS($k$) at $k=4$ for the latter).
As might be expected, BPS substantially improves characterization of forecast uncertainties
as well as adaptation in forecast locations, reflected in the LPDR measures.  
Further, our expectations of improved multi-step forecasting 
using horizon-specific BPS are borne out: direct projection of the standard BPS model 
to 4-step ahead forecasts perform poorly, mainly as a result of 
under-dispersed forecast densities from each agent. In contrast,   
BPS(4) model performs substantially better, being customized to the 4-quarter horizon.

We further our analysis by reviewing summary graphs showing aspects of analyses evolving over time 
during the testing period, a period that includes challenging economic times that impede good predictive performance. 
We take the 1-step and 4-step contexts in sequence.   Additional summaries and  figures appear in supplementary Appendix B.

\begin{figure}[htbp!]
\centering
\includegraphics[width=0.75\textwidth]{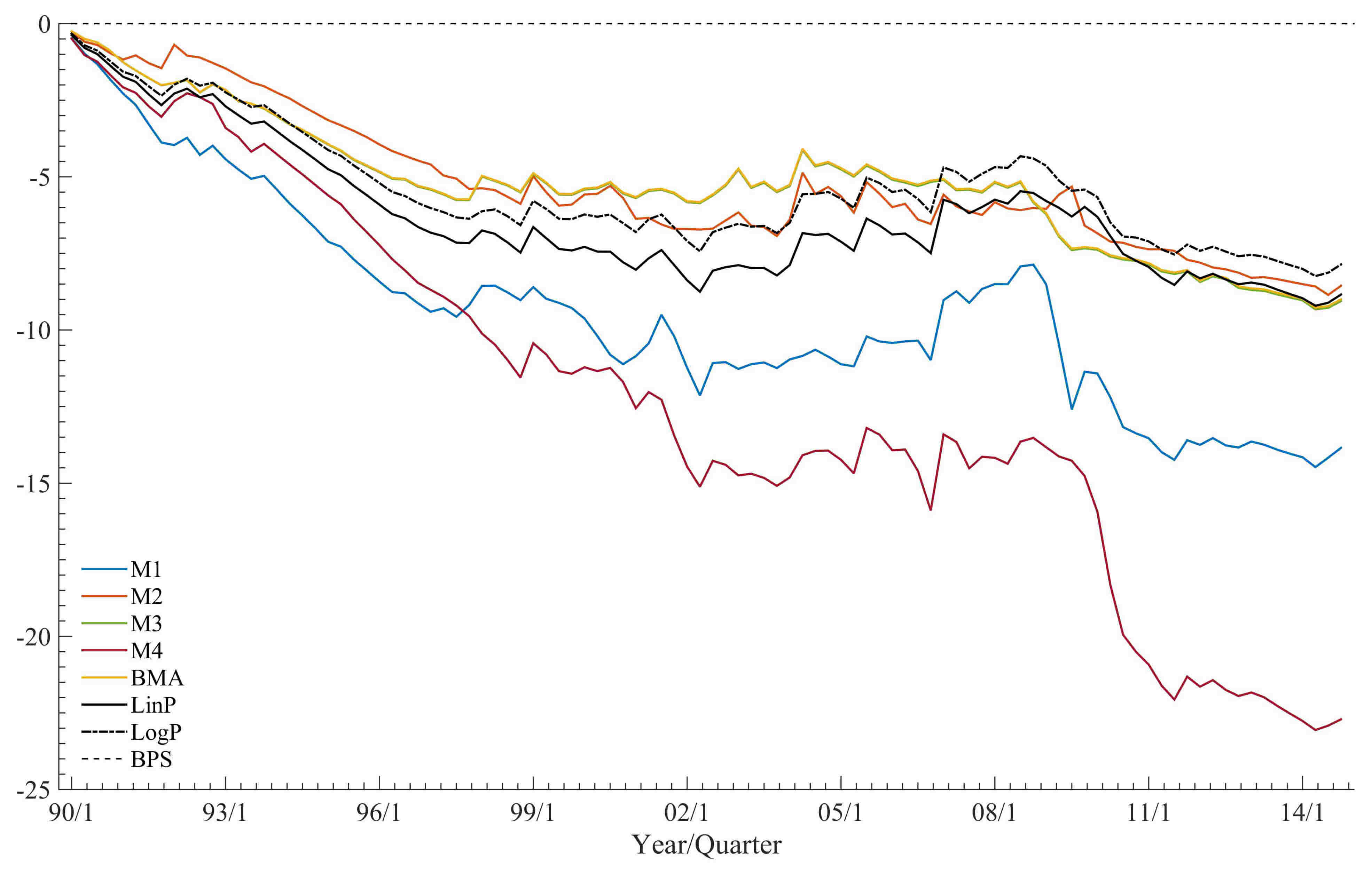} 
\caption{US inflation rate forecasting 1990/Q1-2014/Q4: 1-step ahead log predictive density ratios LPDR$_{1{:}t}(1)$  sequentially revised at each of the $t=\seq1{100}$ quarters. The baseline at 0 over all $t$ corresponds to the standard BPS model. 
\label{1lpdr}}
\end{figure}

\paragraph{1-Step Ahead Forecasting.} 
Several figures summarize sequential analysis for 1-step forecasting.  
 Fig.~\ref{1lpdr} confirms that BPS performs uniformly better than, or on par with, the agent models and other
 methods 
based on LPDR measures that reflect relative  dispersion of forecast densities 
as well as location.  Major shocks and times of increased volatility have substantial impact on the relative performance.
Four notable \lq\lq shock" periods are: 1992/Q3-Q4 (early 90s recession), 1997/Q4-1998/Q1 (Asian and Russian financial crisis), 2001/Q2-2003/Q1 (dot-com bubble), and 2009/Q2-2010/Q1 (sub-prime mortgage crisis).
Even under the influence of these shocks, BPS is able to perform well with most of its improvements over other models and strategies coming from swift adaptation.

\begin{figure}[htbp!]
\centering
\includegraphics[width=0.75\textwidth]{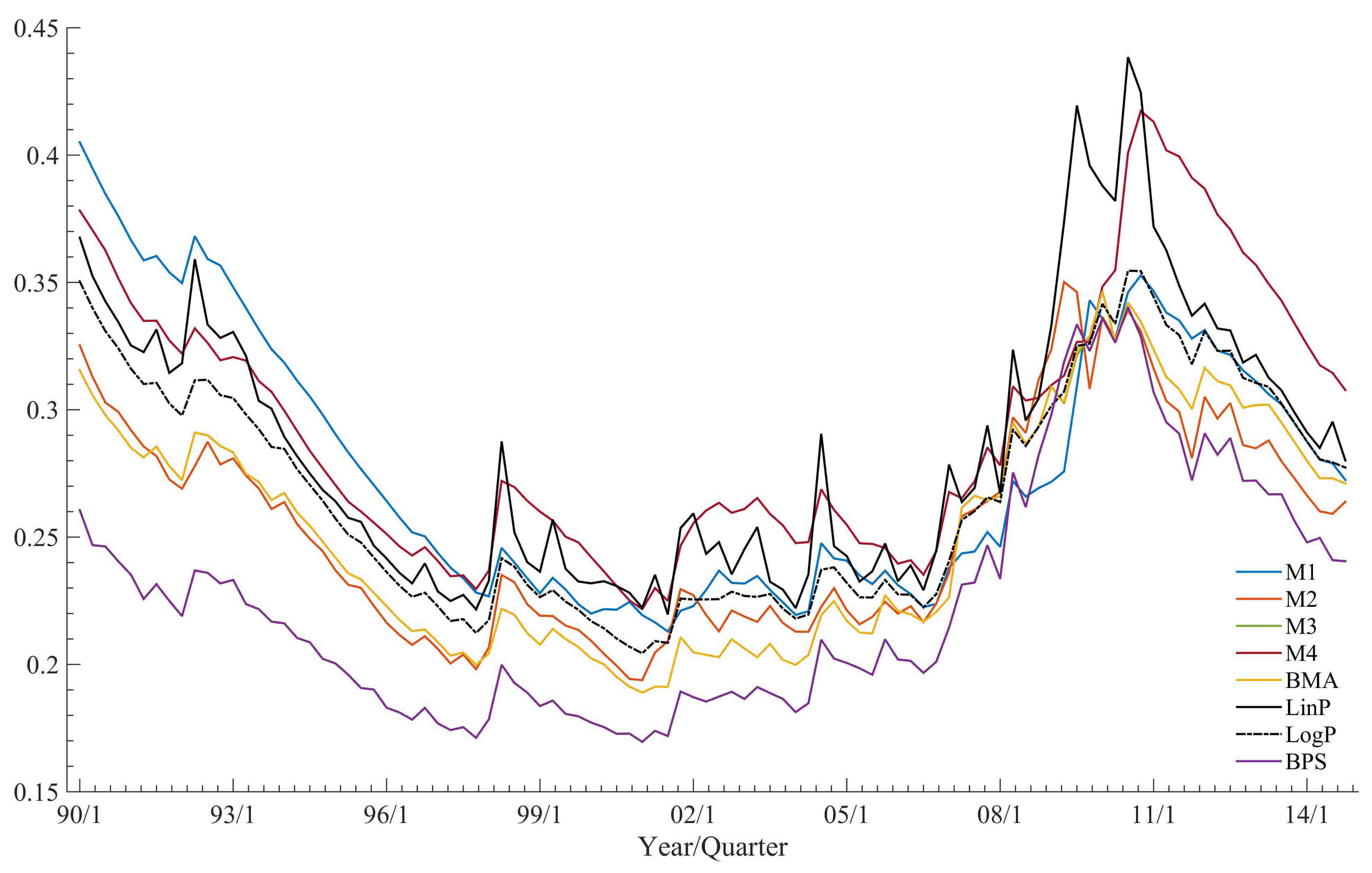} 
\caption{US inflation rate forecasting 1990/Q1-2014/Q4: 1-step ahead forecast standard deviations
 sequentially computed at each of the $t=\seq1{100}$ quarters. 
\label{1sd}}
\end{figure}

Fig.~\ref{1sd} compares on-line 1-step ahead forecast standard deviations. 
Economic (and other) decision makers are often faced with forecasts that have large forecast uncertainties;
while honest in reflecting uncertainties, resulting optimal decisions may then be so unreliable as to be useless.
Large economic models that require complex estimation methods, but have useful properties for policy makers, often produce large forecast standard deviations that might come from the complexity of the model, data, estimation method, or all of the above without necessarily knowing the source of uncertainty.
BPS, on the other hand, synthesizes the forecasts and by doing so, has the ability to 
{\em decrease} forecast uncertainties relative to the agents, without overly underestimating 
real risks; this is evident in the example here, where BPS leads the agents (and other strategies) in
terms of LPDR performance. Fig.~\ref{1sd} shows that some part of this comes from generally 
reduced forecast uncertainties-- coupled with more accurate point forecasts-- at this 1-step horizon. We caution that reduced uncertainties 
are not always expected or achieved, as exemplified below. 

Given the pictures for LPDR scores  and 1-step forecast uncertainties above,  it is no surprise that 
BPS almost uniformly dominates in terms of raw point forecast accuracy as well. Apart from an initially unstable period 
at  beginning of the time frame,  MSFE measures bear this out (see supplementary Appendix B). 
Point forecast accuracy is particularly improved at crisis periods, with 
MSFE staying relatively level under BPS while  significantly increasing for other models and methods. 
In summary, 
BPS is able to adapt to maintain improved forecasting performance both in terms of point forecasts 
and risk characterization, a key positive feature for decision makers who are tasked with forecasting risk and quantiles, especially under critical situations such as economic crises.

\begin{figure}[htbp!]
\centering
\includegraphics[width=0.75\textwidth]{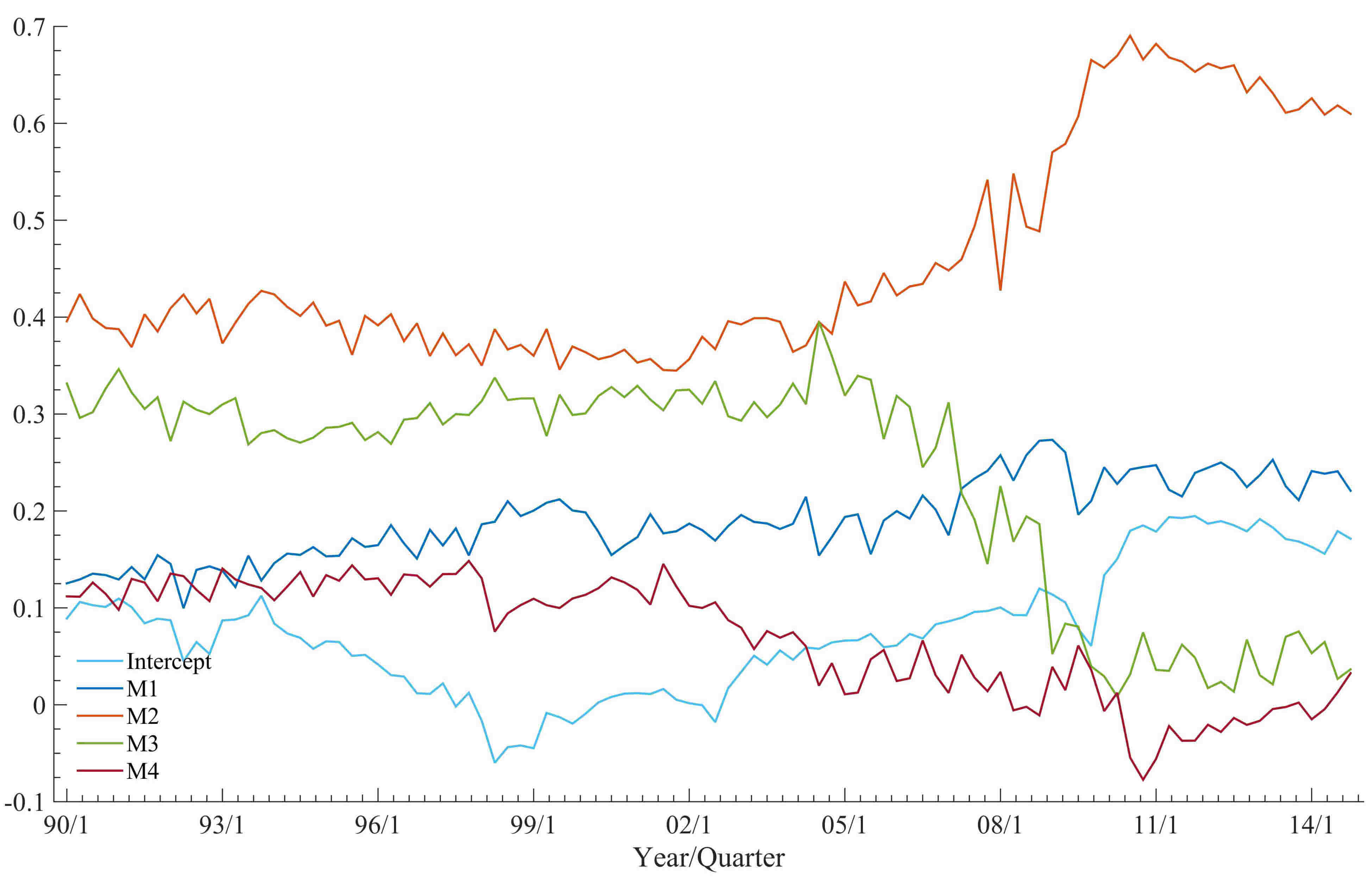} 
\caption{US inflation rate forecasting 1990/Q1-2014/Q4:  On-line posterior means of BPS model
 coefficients sequentially computed at each of the $t=\seq1{100}$ quarters. 
\label{1coeff}}
\end{figure}

We also note that, over the prior period 1977/Q2-1989/Q4,  BMA-- characteristically-- 
effectively degenerated, with posterior probabilities increasingly favoring agent M3; thus, at the
start of the test period, BMA-based forecast densities are very close to those from M3 alone. 
BPS, on the other hand, allows for continual adaptation as agent models change in their relative 
forecasting abilities; over the test period, agent M2 tends to play a dominant role in BPS, notable in terms of the on-line estimates of BPS agent coefficients in $\btheta_{tj}$; see Fig.~\ref{1coeff}.  An interesting point to note is how BPS successfully adapts its coefficients during the sub-prime mortgage crisis by significantly down-weighting M3. 
As a dynamic model, BPS
will not degenerate, continually allowing for \lq\lq surprises" in changes in relative forecast performance across
the agents.

\begin{figure}[htbp!]
\centering
\includegraphics[width=0.75\textwidth]{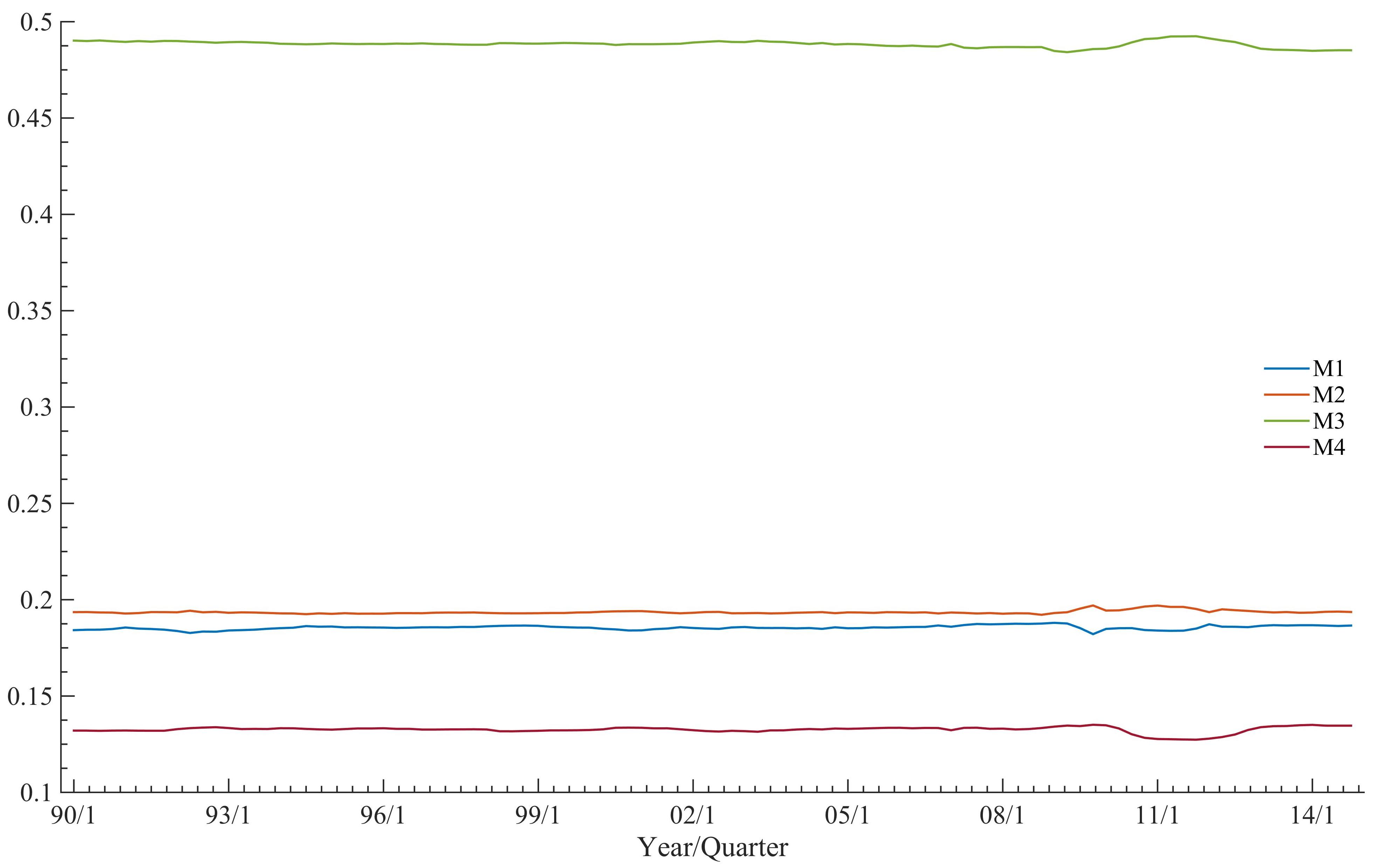} 
\caption{US inflation rate forecasting 1990/Q1-2014/Q4:  On-line means of DeCo model
 weights sequentially computed at each of the $t=\seq1{100}$ quarters. 
\label{1decocoeff}}
\end{figure}

For comparison, the DeCo analysis generates weights that heavily favor M3 from the training period;
see Fig.~\ref{1decocoeff}. While dynamic in theory, the inferred weights end up being quite stable over time. 
Contrasting this to BPS, it is clear that the benefits of BPS stem largely from   dynamic adaptability. 
 As the posterior mean trajectories of Fig.~\ref{1coeff} indicate, BPS is constantly updating and calibrating without over-learning, partly due to discount learning (forgetting) and the latent biases/dependencies being effectively transferred to the coefficients.
 Through understanding what aspects of BPS (intercept, discounting, latent biases/dependencies, etc...) contribute to the gains seen in this example is of interest, it is clear that they allow the coefficients to quickly adapt over time and improve forecasts.

\begin{figure}[htbp!]
\centering
\includegraphics[width=0.75\textwidth]{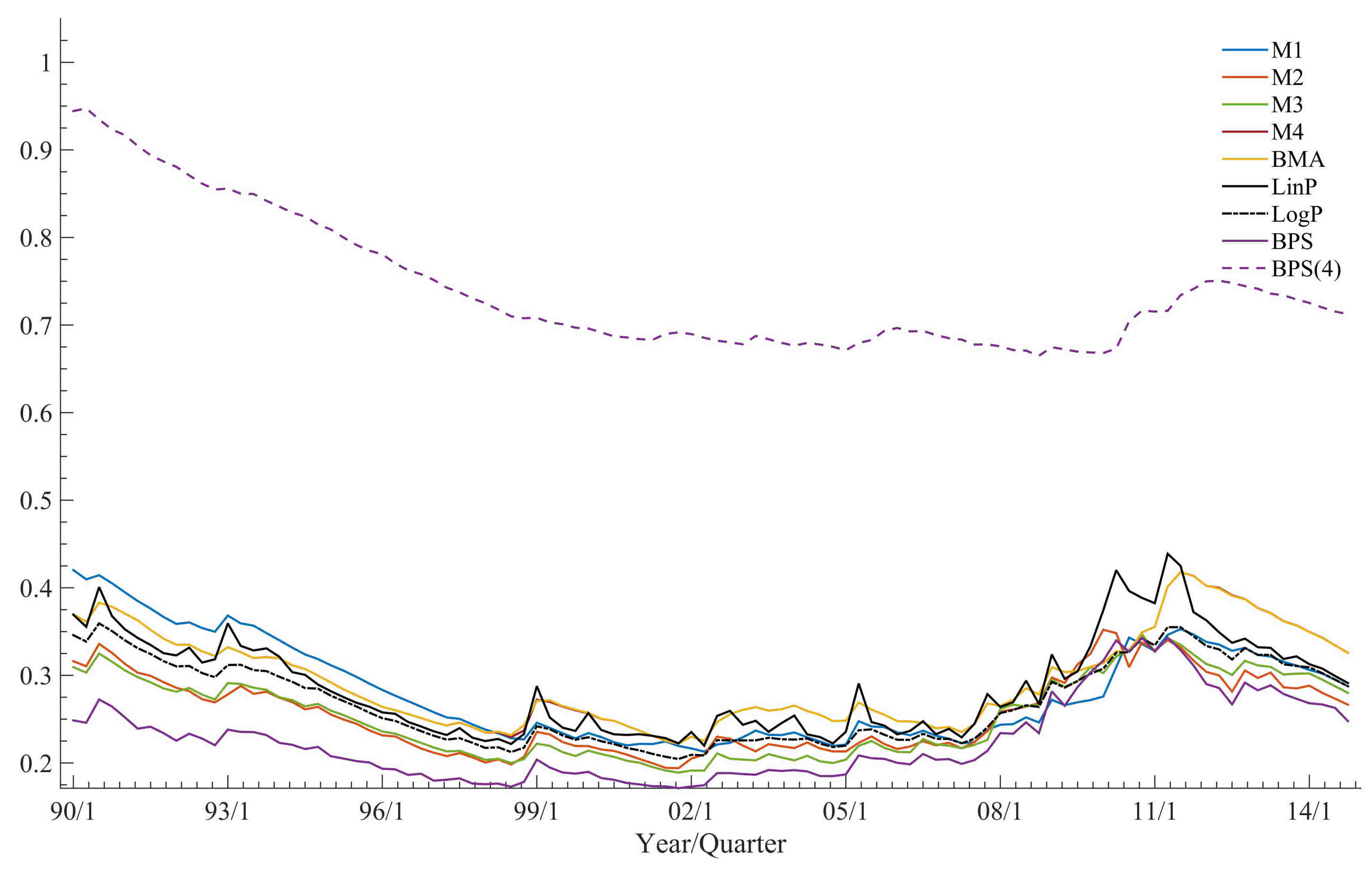} 
\caption{US inflation rate forecasting 1990/Q1-2014/Q4: 4-step ahead forecast standard deviations
sequentially computed at each of the $t=\seq1{100}$ quarters. 
\label{14sd}}
\end{figure}

\paragraph{4-Step Ahead Forecasting.} 
Several figures summarize sequential analysis for 4-step forecasting, using 
both the direct extrapolation to 4-quarters ahead under the BPS model and the 
customized BPS(4) model. Each BPS strategy performs consistently better than agents and other
strategies in point forecasting, while BPS(4) makes significant improvements in terms of both point and 
distribution forecasts compared to direct BPS extrapolation.  This is clearly seen in the time trajectories of 
LPDR under BPS(4) in Fig.~\ref{4lpdr}, and in point forecast accuracy measured by mean square forecast errors (see
supplementary Appendix B).   
Direct BPS extrapolation performs relatively poorly in terms of both point forecast accuracy and LPDR (see figures in 
supplementary Appendix B) as it is 
inherently calibrated to 1-step model fit. In particular, it fails to adequately represent the increased uncertainty associated 
with longer term forecasts.
Looking at the forecast standard deviations in Fig.~\ref{14sd}, it is clear that BPS(4) is able to improve 
by adjusting to the increased forecast uncertainties.  Then, even though forecast uncertainties increase substantially, they
are clearly more than balanced by improved location forecasts.
This again bears out the recommendation to directly synthesize forecasts on the horizon of interest. 

On-line trajectories of estimates of the BPS(4) coefficients $\btheta_t$ as they
are sequentially updated and adapt (see figure in supplementary Appendix B) show 
notable reduction in adaptability over time relative to the 1-step BPS coefficients of Fig.~\ref{1coeff}). 
This is to be expected as the models' forecasts are less reliable at longer horizons, so the data-based information 
advising the changes in posteriors over time is limited. 
The dynamic intercept term serves as a comparison base as it moves away from zero, playing a more 
active role in BPS(4) than in the 1-step case.  
Additionally, the 4-step ahead coefficient values (indicated here by just the on-line means, of course) 
are quite different from 1-step coefficients, reasonably reflecting the differing forecasting abilities of the 
agents at differing horizons.   BPS(4) is able to adapt to the 4-step ahead forecast, differently from the 1-step BPS, 
and dominates in performance compared to all other methods as a result.

\subsection{Retrospective Analysis \label{sec:posterior}}

Based on the full MCMC analysis of all data in 1990/Q1-2014/Q4, we review aspects of retrospective 
posterior inference.

\begin{figure}[htbp!]
\centering
\includegraphics[width=0.75\textwidth]{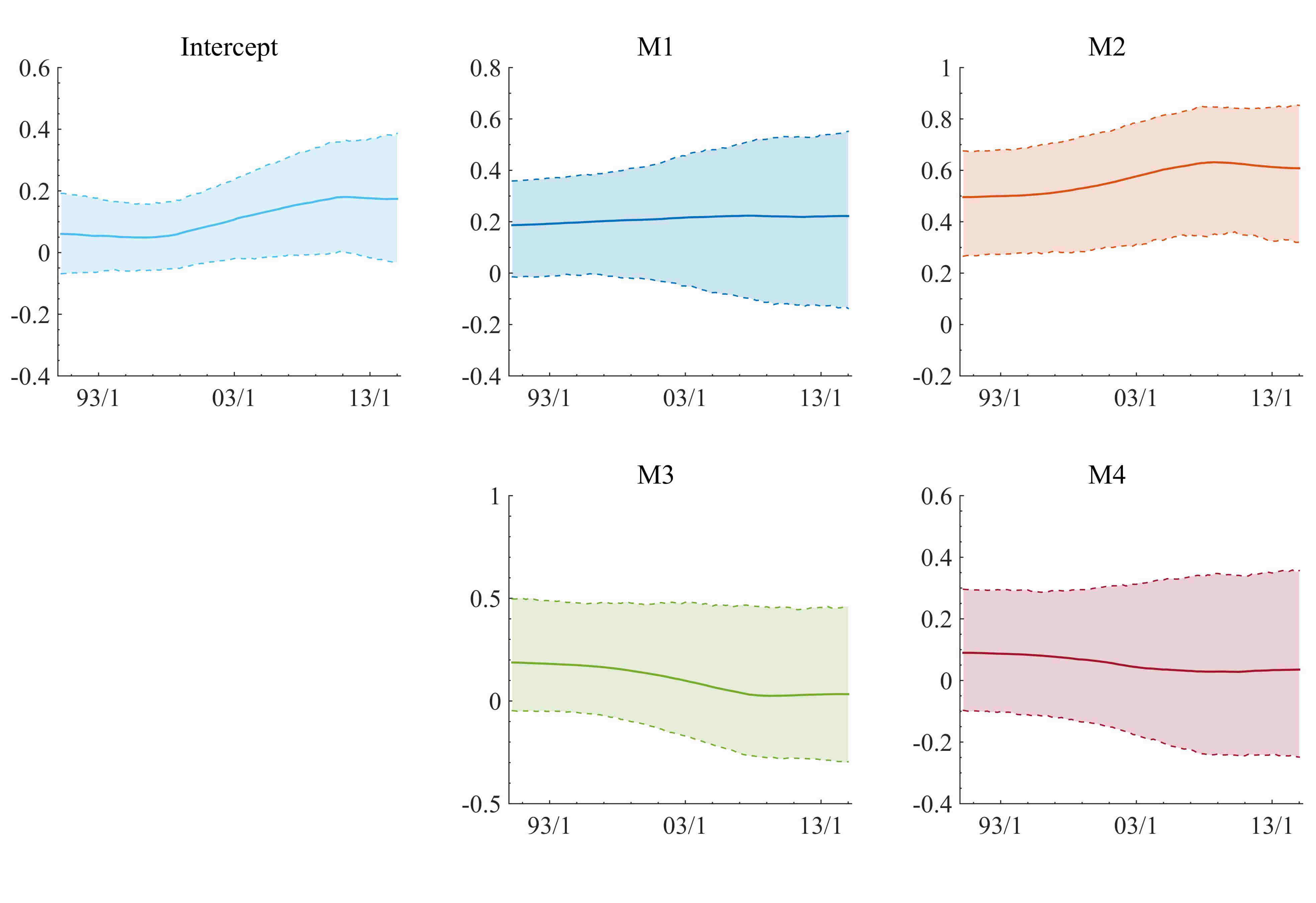} 
\caption{US inflation rate forecasting 1990/Q1-2014/Q4: Retrospective posterior trajectories of the BPS model coefficients based on data from the full $t=\seq1{100}$ quarters. Posterior means (solid) and 95$\%$ credible intervals (shaded).
\label{1posttheta}}
\end{figure}

\paragraph{BPS Coefficients.}
Figs.~\ref{1posttheta} shows trajectories with uncertainties under BPS (1-step); these can 
be compared with on-line point summaries in  Fig.~\ref{1coeff} earlier discussed. 
We see the expected smoothing of estimated trajectories of coefficients.
Similar figures are given for BPS(4) model coefficients  in supplementary Appendix B. 
To the extent that the role of the intercept terms can be regarded as reflecting (lack of) effectiveness of 
the synthesized models,  these figures confirm that the agents' predictions are much more questionable at 4-steps
ahead than at 1-step ahead. Intercepts increase up to and during the sub-prime mortgage crisis due to the increased inability of the models to forecast well during this time.
 
\paragraph{Latent Agent States and Forecast Dependencies.} 
BPS naturally allows for-- and adapts to-- dependencies among agents as they evolve over time. 
In many cases, models and data used by agents are typically unknown to the decision maker $\mD$ and therefore 
posterior inference on dependencies among agents  is of special interest; even when agents are 
chosen statistical models-- as in this example-- the questions of inter-dependence and potential redundancy in 
forecast value are hard and open questions in all approaches to aggregation.  

\begin{figure}[htbp!]
\centering
\includegraphics[width=0.75\textwidth]{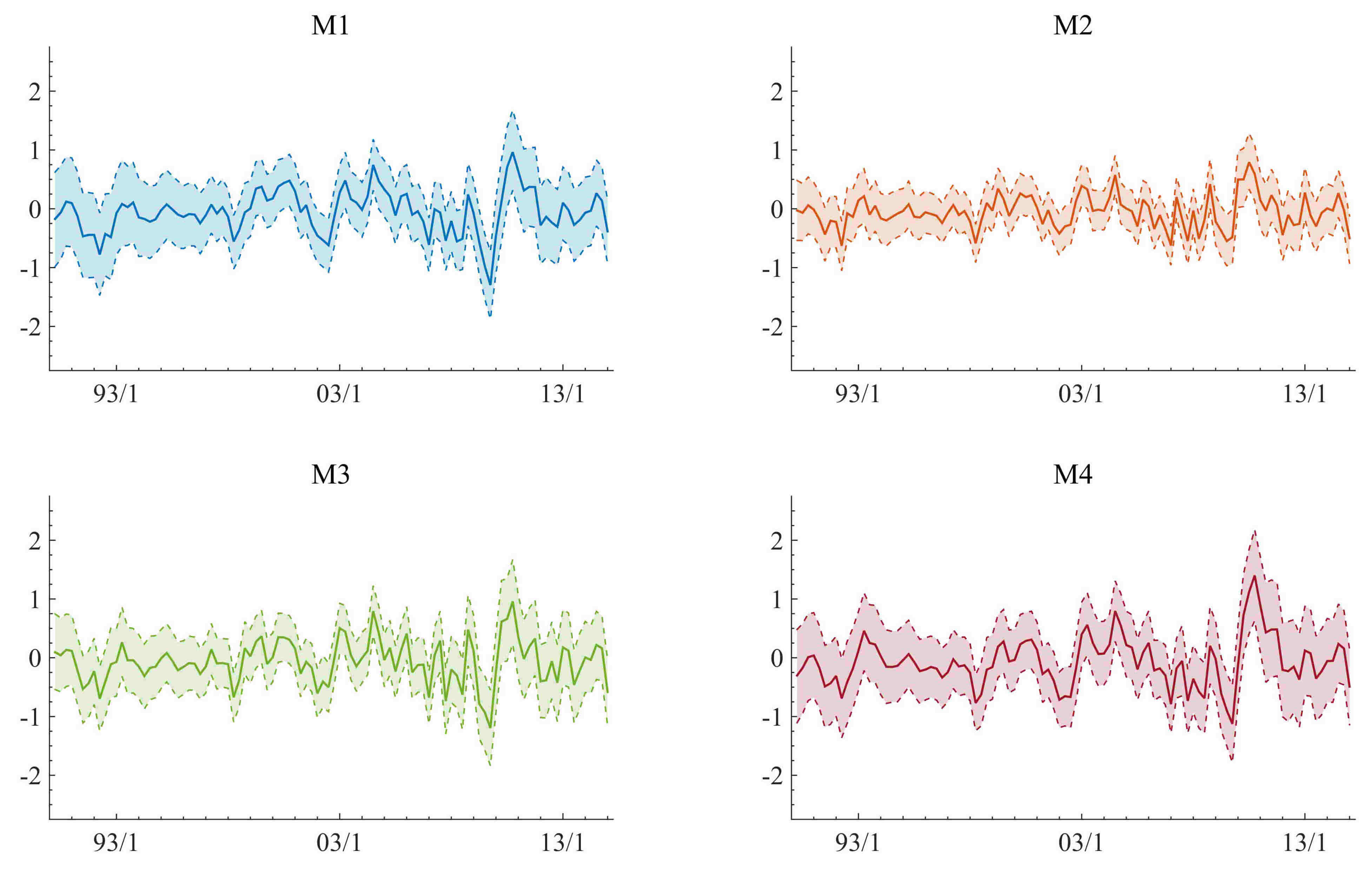} 
\caption{US inflation rate forecasting 1990/Q1-2014/Q4: BPS model-based posterior trajectories of the error in latent agents states 
 $y_t - x_{tj}$ for $j=\seq14$ over the $t=\seq1{100}$ quarters. Posterior means (solid) and
95$\%$ credible intervals (shaded) from the MCMC analysis. 
\label{1xerror}}
\end{figure}

As noted early, the
conceptual and theoretical basis of BPS allows direct investigation of agent dependencies, as the
inherent latent agent states $x_{tj}$-- when inferred based on the observed data-- carry the relevant information. 
From the full MCMC analysis to the end of the test data period, we have full posterior samples for the 
states $\x_t$-- in both the direct  BPS and customized BPS(4). 
For illustration, we focus on the 1-step BPS analysis.   
Fig.~\ref{1xerror} displays posterior trajectories for the errors in latent states, namely 
$y_t-x_{tj}$ over time $t$, in terms of posterior means and intervals (similar
trajectories of the $x_{tj}$ themselves, together with the inflation outcomes $y_t$,  
are in supplementary Appendix B).   
The patterns over time in each of these reflect the strong, positive dependencies among agents that 
are to be expected given the nature of the agent models.

\begin{figure}[htbp!]
\centering
  \includegraphics[width=3.2in]{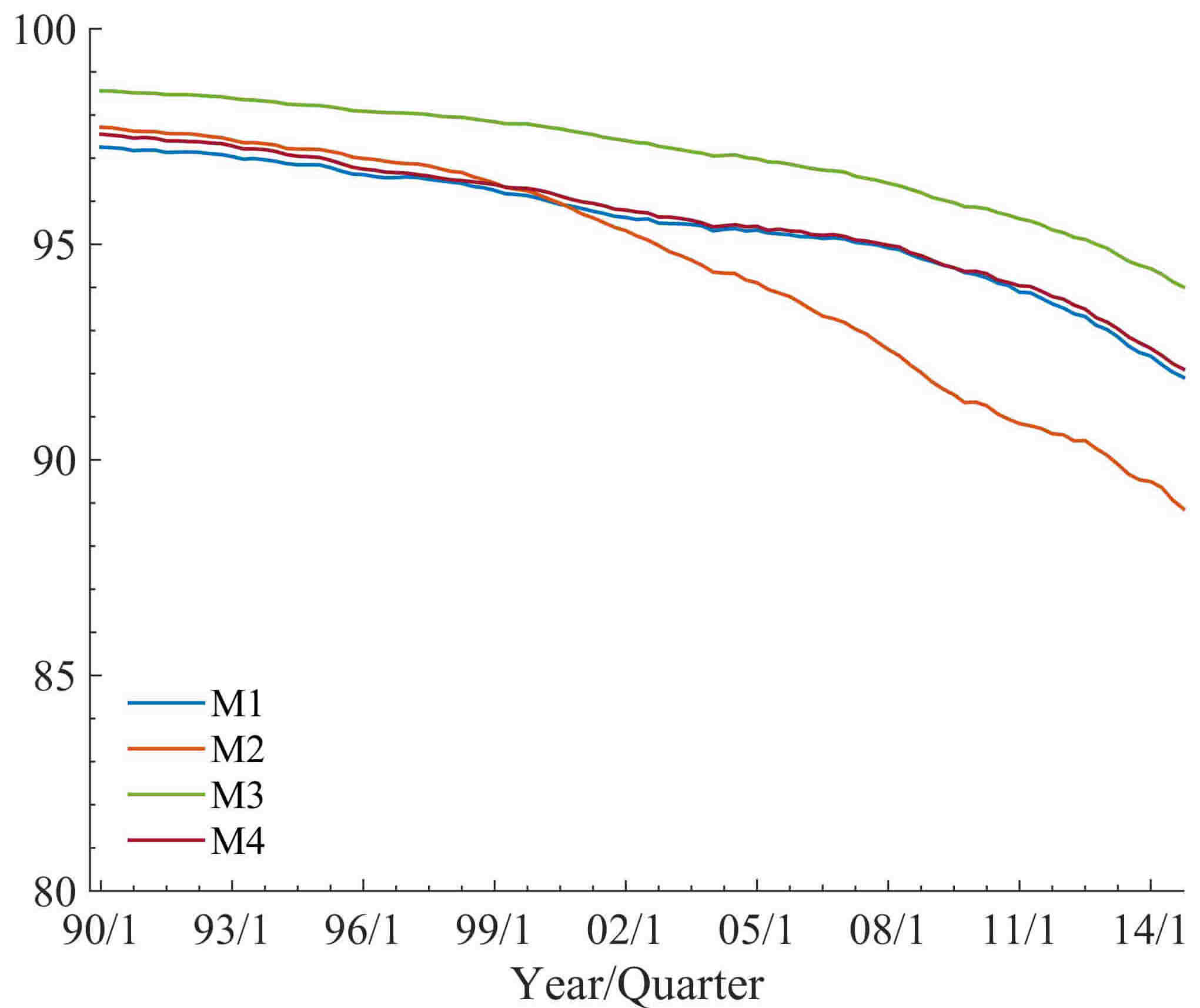} 
  \includegraphics[width=3.2in]{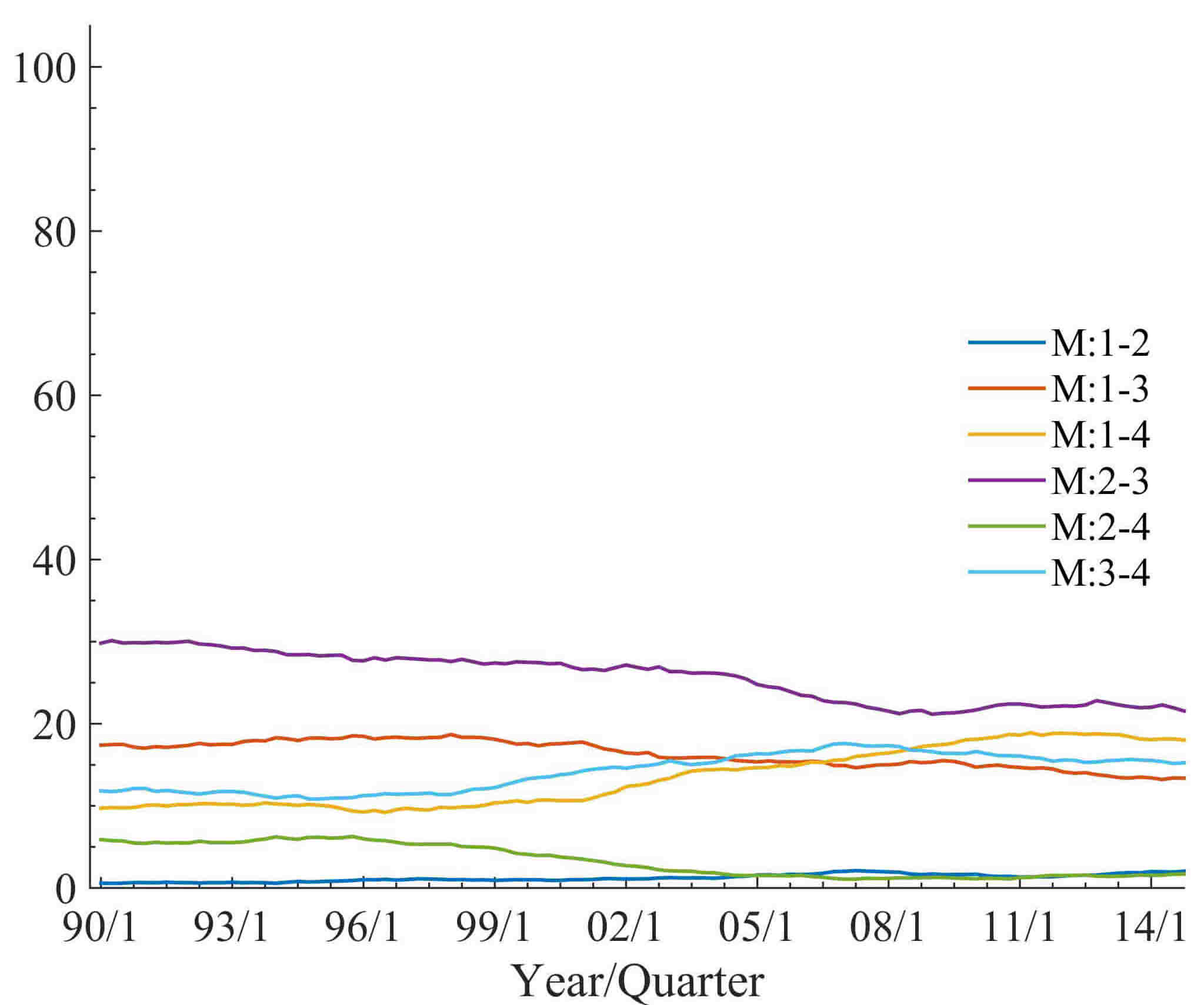}
  \caption{US inflation rate forecasting 1990/Q1-2014/Q4: BPS model-based trajectories of 1-step ahead  MC-empirical R$^2$ (left) and paired MC-empirical R$^2$ (right) in the posterior for the latent agent states $x_{jt}$  for $j=\seq14$ over the $t=\seq1{100}$ quarters.}
    \label{1r2}
\end{figure}

To explore dependencies, we simply investigate the posterior for $\x_{\seq1T}.$   This is not of standard form and
is represented in terms of the MCMC-based posterior sample.   One simple set of summaries is based on 
just computing empirical R$^2$ measures: from the MCMC sample, compute the approximate posterior 
variance matrix of $\x_t$ at each $t,$ and from that extract implied sets of conditionals variances of
any $x_{tj}$ given any subset of the other $x_{ti}, i\ne j.$    We do this for $i=\seq1J\backslash j,$ defining the 
MC-empirical R$^2$ for agent $j$ based on all other agents, i.e.,  measuring the redundancy of agent $j$ 
in the context of all $J$ agents-- the {\em complete conditional dependencies}.  We do this also using each single agent $i\ne j$, defining paired  MC-empirical R$^2$ measures of how dependent agents $i,j$ are-- the {\em bivariate dependencies}.
Fig.~\ref{1r2}  displays trajectories over time for these two measures from the  BPS 1-step analysis;
a corresponding figure from BPS(4)  is  in supplementary Appendix B. 
 
Overall, we see high complete conditional dependencies at both forecast horizons, as expected due to the
nature of the 4 models and their evaluation on the same data. Dependencies are substantial and 
much higher for 1-step forecasts than for 4-step ahead forecasts, reflecting decreasing concordance with 
increasing horizon, and all decrease over the test period. The predictability of M2 based on the others
drops at a greater rate after about the start of 2002, in part due to poorer and less reliable performance during the dot-com crisis.
The paired measures are all very low compared to the complete conditionals, and again naturally lower overall in
4-step forecasting. Concordance of M2 and M3 decreases for 1-step but increases slightly for 4-step ahead forecasts,
reflecting dynamics in relationships that differ with forecast horizon; from earlier discussion of forecast accuracy, 
this can be explained by how, in 1-step ahead forecasts, M2 improves while M3 deteriorates during the sub-prime mortgage crisis.
In contrast, for 4-step ahead forecasts 
we see forecast errors converging between the two, explaining the increase in concordance as all models performed equally poorly.

\subsection{Simulation Studies \label{sec:simulatedata}} 
We have explored a range of synthetic data sets to fully evaluate the above analysis using
the same four models but with known parameters, and with simulated data generated 
with random switching between models. The results echo and amplify those of the macroeconomic 
study. In particular, at 1-step ahead, BPS outperforms the best model and best traditional strategy by 
nearly 20$\%$ in terms of point forecasts, as well as significantly improving in terms of LPDR measures of 
density forecasts. At 4-steps ahead, BPS(4) very substantially improves on all models and on 
BPS, the latter being partly due to improved characterization of forecast uncertainties under BPS(4),
coupled with somewhat improved point forecasts. 

Detailed graphical summaries of analysis of one such synthetic data set appear in the supplementary 
Appendix~\ref{supp:sim}. 
These graphs include on-line filtering and forecast summaries
as well as aspects of the retrospective posterior analysis from the dynamic BPS model, to compare with those
from the real data analysis. 


\section{Summary Comments \label{sec:summary}}

Drawing on theory of Bayesian agent opinion analysis, 
BPS provides a theoretically and conceptually sound framework to compare and synthesize density 
forecasts that has been developed here for dynamic contexts of sequential time series forecasting.
With this new framework and extension, decision makers are able to dynamically calibrate, learn, and update weights for ranges of forecasts   from dynamic models, with multiple lags and predictors as exemplified here, as well as from 
more subjective sources such as individual forecasters or agencies.


The U.S. macroeconomic data study illustrates how effective and practical BPS is under settings that are increasingly important and topical in macroeconomics and econometrics.
By dynamically synthesizing the forecasts, BPS improves forecast performance and dominates other standard strategies, such as BMA and pooling, over short and long horizons and for both point and distribution forecasts.
Further analysis shows evidence that BPS is also robust in its forecast abilities under economic distress, which is critically important for practical applications.
Additionally, posterior inference of the full time series provides the decision maker with  information on how agents are related, 
and how that relationship dynamically evolves through time; this has potential to inform BPS modeling for continued 
forecast synthesis into the future. 

In addition to applications to U.S. macroeconomic data, BPS has the potential to be applied to other fields   where multiple forecasts, whether from forecasters or models, are available.
This includes areas such as finance (e.g., stocks, indexes, and bonds),  business (e.g., product demand and earnings),  meteorology, and risk (e.g., seismic and environmental risk). 
Methodological extensions are needed for multivariate synthesis, non-normal forecasts and discrete data, and missing or incomplete/partial forecasts.    
It will certainly be of interest and practical importance to develop studies in 
which agents are represented by sets of more elaborate macroeconomic models, such as VAR, dynamic threshold models, 
dynamic stochastic general equilibrium (DSGE) models, and to integrate forecasts coming from professional forecasters and economists.  In the latter contexts, developing BPS models that integrate partial forecast information-- such as quantiles of
agent forecast distributions rather than full density forecasts-- is of practical importance. We believe that 
further such studies will define increasing empirical support for the utility of the approach and attract applied 
researchers.  Coupled with this, some of the foundational 
theory of Bayesian agent opinion analysis in the single agent case~\citep{West1992d} explicitly addressed partial
forecast information,
but there has been no extension, to date, to multiple agents and the dynamic/time series setting.

Computational  questions are also relevant; as developed and exemplified, analysis in the sequential time series context 
relies on repeat reanalysis using MCMC, with a new simulation analysis required as each new time period arises. This
is in common with the application of Bayesian dynamic latent factor models of other forms in the sequential forecasting 
context, including, in particular, dynamic latent threshold models~\citep[e.g.][]{Nakajima2010,Nakajima2011a,Nakajima2014,Zhou2012} whose use in defining sets of 
candidate agents for BPS is of some applied interest. 
One view is that a substantial computational burden is nowadays a minor issue and, in fact, a small price to 
pay for the potential improvements in forecasting accuracy and insights  that our example illustrates.  That said, some 
methods of sequential model analysis based on sequential Monte Carlo (SMC, e.g.~\citealp{LopesTsay2011}) may provide for more efficient computations, 
at least in terms of CPU cycles,  in some stylized versions of the overall BPS model framework. 

\subsection*{Acknowledgements}

We are grateful for useful discussions with colleagues that helped with perspective in developing this work, 
including Matt Johnson (Duke University) and Jouchi Nakajima (Bank for International Settlements), and 
the organizers and participants of the 5th European Seminar on Bayesian Econometrics (ESOBE) held
at the Study Center Gerzensee, Switzerland, October 29-30th 2015. 

We also thank the editors and four anonymous referees for very detailed,  positive, and constructive comments
 on the original version of this paper.  

\bibliographystyle{elsarticle-harv}
\bibliography{McAlinnWest}

\newpage

\begin{center}
{\Large Dynamic Bayesian Predictive Synthesis\\ in Time Series Forecasting} 

\bigskip
{\large Kenichiro McAlinn \& Mike West} 

\bigskip
{\Large  Tables and Figures} 

\bigskip\bigskip
\end{center}


\begin{table}[htbp!]
\centering
\begin{tabular}{lrrrrrrr}
                              & \multicolumn{4}{c}{MSFE$_{1{:}T}$}       & \multicolumn{3}{c}{LPDR$_{1{:}T}$}                                            \\
\multicolumn{1}{l|}{}         & 1-step& $\%$  &4-step&  \multicolumn{1}{r|}{$\%$}  & 1-step & 4-step & \begin{tabular}[c]{@{}r@{}} BPS(4)\end{tabular} \\ \hline
\multicolumn{1}{l|}{M1}       & 0.0634& $-$23.83& 0.4227 &\multicolumn{1}{r|}{$-$14.68}& $-$13.84 & 71.43  & $-$94.56                                                   \\
\multicolumn{1}{l|}{M2}       & 0.0598 &$-$16.80&{0.4156}&\multicolumn{1}{r|}{$-$12.75} & $-$8.55  & 68.16  & $-$97.82                                                   \\
\multicolumn{1}{l|}{M3}       & 0.0616 &$-$20.31&{0.4208} &\multicolumn{1}{r|}{$-$14.16}& $-$9.06  & 60.08  & $-$105.90                                                   \\
\multicolumn{1}{l|}{M4}       & 0.0811 & $-$58.40&{0.4880} &\multicolumn{1}{r|}{$-$32.39}& $-$22.71 & 67.46  & $-$98.53                                                   \\
\multicolumn{1}{l|}{BMA}      & 0.0617 &$-$20.51& {0.4882} &\multicolumn{1}{r|}{$-$32.45}& $-$9.00  & 65.65  & $-$100.33                                                   \\
\multicolumn{1}{l|}{LinP}     & 0.0575 & $-$12.30&{0.4275} &\multicolumn{1}{r|}{$-$15.98} &$-$8.84  & 85.50 & $-$80.48                                                 \\
\multicolumn{1}{l|}{LogP}     & 0.0579 & $-$13.09&{0.4275} &\multicolumn{1}{r|}{$-$15.98} &$-$7.86  & 68.23  & $-$97.75                                                   \\
\multicolumn{1}{l|}{DeCo}     & 0.0571 & $-$11.52&{0.4156} &\multicolumn{1}{r|}{$-$12.75}  &-  & -  & -                                                   \\
\multicolumn{1}{l|}{BPS}      & 0.0512 & -&{0.4001} &\multicolumn{1}{r|}{$-$8.55}   & -    &     -   &       -                                                    \\
\multicolumn{1}{l|}{BPS(4)} & -       & -&{0.3686} &\multicolumn{1}{r|}{-}&  -     &    -    &          -                                                
\end{tabular}
\caption{US inflation rate forecasting 1990/Q1-2014/Q4: Forecast evaluations for quarterly U.S. inflation over the 25 years 1990/Q1-2014/Q4, 
 comparing mean squared forecast errors   and log predictive density ratios  for this $T=100$ quarters. The column $\%$ denotes improvements over BPS and BPS$(k)$ for 1- and 4-step ahead forecasts, respectively. Note: LPDR$_{\seq1T}$ is respect to BPS and BPS(4) and nonexistent for DeCo due to lack of analytic predictive distributions.} 
\label{table:inf}
\end{table}


%

\begin{figure}[htbp!]
\centering
\includegraphics[width=0.75\textwidth]{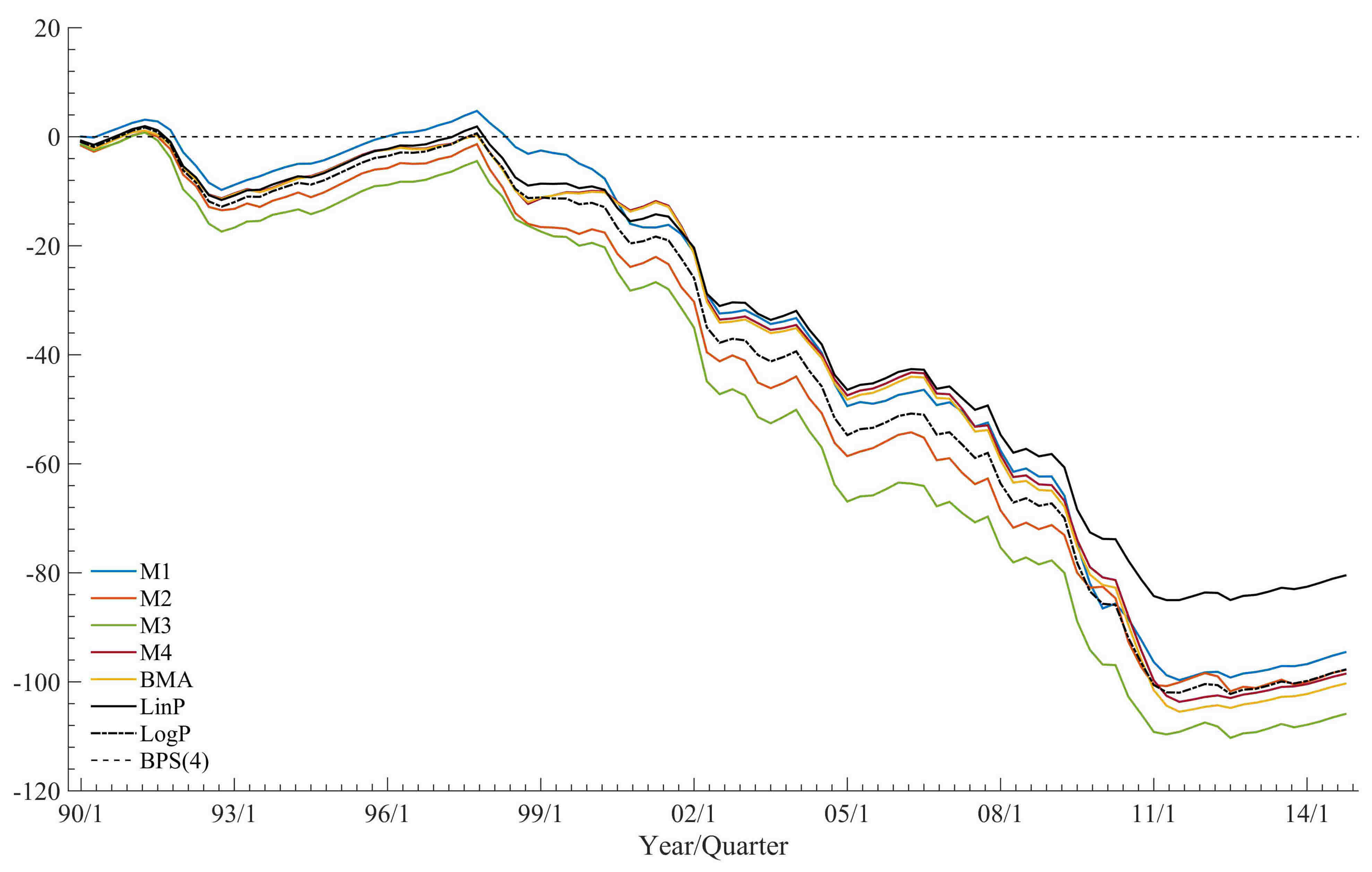} 
\caption{US inflation rate forecasting 1990/Q1-2014/Q4:  4-step ahead log predictive density ratios, LPDR$_{1{:}t}(4)$ sequentially revised at each of the $t=\seq1{100}$ quarters using the 4-step ahead
customized BPS(4) model (baseline at 0 over time). 
\label{4lpdr}}
\end{figure}

\clearpage
\setcounter{page}{1}
\begin{center}
{\Large Dynamic Bayesian Predictive Synthesis\\ in Time Series Forecasting} 

\bigskip
{\large Kenichiro McAlinn \& Mike West} 

\bigskip
{\Large  Supplementary Material} 

\bigskip\bigskip
\end{center}

\appendix

\section{Appendix: Summary of MCMC for Dynamic BPS\label{supp:comp}}

\subsection{Overview and Initialization}
This appendix summarizes algorithmic details of implementation of the MCMC computations for dynamic BPS model fitting
of Section~\ref{sec:comp}.  This involves a   standard set of steps in a customized 
two-component block Gibbs sampler: one component samples the latent agent states, 
and the second samples the dynamic BPS model states/parameters.
The latter involves the FFBS algorithm central to MCMC in all conditionally normal DLMs
(~\citealt{Schnatter1994}; \citealt[][Sect 15.2]{WestHarrison1997book2}; \citealt[][Sect 4.5]{Prado2010}).  

In our sequential learning and forecasting context, the full MCMC analysis is 
performed anew at each time point as time evolves and new data are observed.   We detail MCMC steps for a 
specific time $t$ here,   based on all data up until that time point. 

Standing at time $T$, the decision maker has historical information $\{ y_{\seq1T}, \mH_{\seq1T}\}$,   initial 
prior $\btheta_0\sim N(\m_0, \C_0 v_0/s_0 )$ and $1/v_0\sim G(n_0/2, n_0s_0/2),$ and discount factors $(\beta,\delta)$. 
The MCMC is run iteratively as follows. 

\paragraph{Initialization:} First, initialize by setting $\F_t=(1,x_{t1},...,x_{tJ})'$ for each $t=\seq1T$ at some chosen initial values of the
latent agent states.  
Initial values can be chosen arbitrarily. One obvious and appropriate choice-- our recommended default choice-- is to simply generate agent states from their priors, i.e., 
from the agent forecast distributions,  $x_{tj} \sim h_{tj}(x_{tj})$ independently for all $t=\seq1T$ and $j=\seq1J$.
This is easily implemented in cases when the agent forecasts are T or normal distributions, or can be otherwise directly
sampled; we use this in our analyses reported in the paper, and recommend it as standard.  
An obvious alternative initialization is to simply set $x_{tj}=y_t$ for each $j,t$, though we prefer to initialize with 
some inherent dispersion in starting values.  Ultimately, since the MCMC is rapidly convergent, choice of initial values is not critical.

\subsection{Two Sampling Steps in Each MCMC Iterate} 

Following initialization,  the MCMC  iterates repeatedly to resample two coupled sets of conditional posteriors to generate the 
MCMC samples from the target posterior $p(\x_{\seq1T},\bPhi_{\seq1T}|y_{\seq1T}, \mH_{\seq1T}).$   These two conditional posteriors
and algorithmic details of their simulation are as follows. 

\subsubsection{Per MCMC Iterate Step 1: Sampling BPS DLM parameters $\bPhi_{\seq1T}$ } 
Conditional on any values of
the latent agent states,  we are in the setting of a conditionally normal DLM with the agent states as
known predictors based on their specific values.  The BPS conjugate DLM form, 
\begin{align*}
		y_t&=\F_t'\btheta_t+\nu_t, \quad \nu_t\sim N(0,v_t), \label{eq:DLMa} \\
	\btheta_t&=\btheta_{t-1}+\bomega_t, \quad \bomega_t\sim N(0, v_t\W_t),
\end{align*}
has known elements $\F_t,\W_t$ and specified initial prior at $t=0.$ The implied conditional posterior 
for  $\bPhi_{\seq1T}$  then does not depend on  $\mH_{\seq1T}$, reducing to 
$p(\bPhi_{\seq1T}|\x_{\seq1T},y_{\seq1T}).$  This is simulated using the efficient and standard FFBS 
algorithm, modified to incorporate the discount stochastic volatility components for $v_t$
(e.g.~\citealt{Schnatter1994}; \citealt[][Sect 15.2]{WestHarrison1997book2}; \citealt[][Sect 4.5]{Prado2010}). 
In detail, this proceeds as follows. 

\begin{itemize} 
\item[]{\em\bf Forward filtering:} For each $t=\seq1T$ in sequence, perform the standard one-step 
filtering updates to compute and save the sequence of sufficient statistics for the on-line posteriors
$p(\btheta_t,v_t|y_{\seq1t},\x_{\seq1t})$ at each $t.$ The summary technical details are as follows: 
\begin{itemize} 
	\item[1.]{\em Time $t-1$ posterior:} 
		\begin{align*}
		\btheta_{t-1}|v_{t-1},\x_{\seq1{t-1}},y_{\seq1{t-1}}&\sim N(\m_{t-1}, \C_{t-1}v_{t-1}/s_{t-1}),\\
		v_{t-1}^{-1}|\x_{\seq1{t-1}},y_{\seq1{t-1}}&\sim G(n_{t-1}/2, n_{t-1}s_{t-1}/2),
		\end{align*}
		with point estimates $\m_{t-1}$ of $\btheta_{t-1}$ and $s_{t-1}$ of $v_{t-1}.$ 
	\item[2.]{\em Update to time $t$ prior:} 
		\begin{align*}
		\btheta_{t}|v_t,\x_{\seq1{t-1}},y_{\seq1{t-1}}&\sim N(\m_{t-1}, \R_tv_t/s_{t-1})
		\quad\textrm{with}\quad \R_{t}=\C_{t-1}/\delta, \\
		v_t^{-1}|\x_{\seq1{t-1}},y_{\seq1{t-1}}&\sim G(\beta n_{t-1}/2, \beta n_{t-1}s_{t-1}/2),
		\end{align*}
		with (unchanged) point estimates $\m_{t-1}$ of $\btheta_{t}$ and $s_{t-1}$ of $v_{t},$  but with 
		increased uncertainty relative to the time $t-1$ posteriors, the level of increased uncertainty 
			being defined by the discount factors.   
	\item[3.]{\em  1-step predictive distribution:} 
		$y_t |\x_{\seq1t},y_{\seq1{t-1}} \sim T_{\beta n_{t-1}}(f_t,q_t)$ where
		$$f_t=\F_t'\m_{t-1}\quad \textrm{and}\quad q_t=\F_t'\R_t\F_t+s_{t-1}.$$
	\item[4.]{\em  Filtering update to time $t$ posterior:}  
		\begin{align*}
		\btheta_{t}|v_{t},\x_{\seq1{t}},y_{\seq1{t}}&\sim N(\m_{t}, \C_{t}v_{t}/s_{t}),\\
		v_{t}^{-1}|\x_{\seq1{t}},y_{\seq1{t}}&\sim G(n_{t}/2, n_{t}s_{t}/2),
		\end{align*}
		 with defining parameters as follows:  
		\begin{itemize}
			\item[i.]{For $\btheta_t|v_t:$}  $\m_t=\m_{t-1}+\A_t e_t$ and $ 	\C_{t}=r_t(\R_t-q_t\A_t\A_t'),$
			\item[ii.]{For $v_t:$}  	$n_t=\beta n_{t-1}+1$ and $ s_t=r_ts_{t-1},$ 
		\end{itemize} 
		based on  1-step forecast error  $ e_t=y_t-f_t,$ the state adaptive coefficient vector (a.k.a. \lq\lq Kalman gain'') 
		$\A_t=\R_t\F_t/q_t,$  and volatility estimate ratio $r_t=(\beta n_{t-1}+e_t^2/q_t)/n_t .$ 
\end{itemize} 
	\item[]{\em\bf Backward sampling:}  Having run the forward filtering analysis up to time $T,$ the 
	 backward sampling proceeds as follows. 
	 \begin{itemize}
	 \item[a.]{\em At time $T$:} Simulate $\bPhi_T=(\btheta_T,v_T)$ from the final normal/inverse gamma posterior 
	  $p(\bPhi_T|\x_{\seq1{T}},y_{\seq1{T}})$ as follows. First, draw $v_T^{-1}$ from $G(n_{T}/2, n_{T}s_{T}/2),$ and then 
	  	draw $\btheta_T$ from $N(\m_T,\C_T v_T/s_T).$
	 \item[b.]{\em Recurse back over times $t=T-1, T-2, \ldots, 0:$}  At time $t,$ sample  
	 	$\bPhi_t=(\btheta_t,v_t)$ as follows:
	 	\begin{itemize}  
	 		\item[i.] Simulate the volatility $v_t$ via 
	 			$v_t^{-1}=\beta v_{t+1}^{-1}+\gamma_t$ where $\gamma_t$ is an independent draw from
	 				$\gamma_t  \sim G((1-\beta)n_t/2,n_ts_t/2),$
	 		\item[ii.] Simulate the state $\btheta_t$ from the conditional normal posterior 
	 			$p(\btheta_{t}|\btheta_{t+1},v_t,\x_{\seq1T},y_{\seq1T})$ with mean 
	 			vector $\m_{t}+\delta(\btheta_{t+1}-\m_{t})$ and variance matrix 
	 			$ \C_{t} (1-\delta)(v_t/s_t).$ 	 			
	 	\end{itemize} 
	 \end{itemize} 
\end{itemize} 
\subsubsection{Per MCMC Iterate Step 2:   Sampling the latent agent states $\x_{\seq1T}$}

  Conditional on most recently
sampled values of
the BPS DLM parameters $\bPhi_{\seq1T},$   the MCMC iterate completes with resampling of the 
latent agent states from their full conditional posterior
$ p( \x_{\seq1t} |  \bPhi_{\seq1t}, y_{\seq1t}, \mH_{\seq1t} ).$    It is immediate that the $\x_t$ are
conditionally independent over time $t$ in this conditional distribution, with time $t$ 
conditionals 
\begin{equation}\label{app:ccforx}
p( \x_t|  \bPhi_t, y_t, \mH_t) \propto N(y_t|\F_t'\btheta_t, v_t) \prod_{j=\seq1J} h_{tj}(x_{tj}) 
	\quad\textrm{where}\quad  \F_t=(1, x_{t1},x_{t2},...,x_{tJ})'. \end{equation}
Several comments are relevant to studies with different forms of the agent forecast densities. 
\begin{itemize} 
\item[1.] {\em Normal agent forecast densities:} 
In cases when each of the agent forecast densities is normal,  the posterior in eqn.~(\ref{app:ccforx}) 
yields a multivariate normal distribution for $\x_t.$  Computation of its defining parameters and then 
drawing a new sample vector $\x_t$  are trivial.
Specifically, suppose that  $h_{tj}(x_{tj})$ is density of the normal $ N(h_{tj},H_{tj})$, and
write $\h_t=(h_{t1},h_{t2},...,h_{tJ})'$  
and $ \H_t=\textrm{diag}(H_{t1},H_{t2},...,H_{tJ}).$   
Then the posterior distribution for each $\x_t$ is 
\begin{equation}\label{app:condx}
	p( \x_t|  \bPhi_t, y_t, \mH_t) = N(\h_{t}+\b_t c_t, \H_t-\b_t\b_t'g_t)
\end{equation}
where $c_t = y_t- \theta_{t0} - \h_t'\btheta_{t,\seq1J}$,    $g_t=v_t+  \btheta_{t,\seq1J}'\H_t\btheta_{t,\seq1J}$
and $\b_t =  \H_t\btheta_{t,\seq1J}/g_t.$ 
This is easily computed and then sampled independently for each $\seq1T$ to provide resimulated agent states
over $\seq1T.$

\item[2.] In some cases, as in our study in this paper, the agent forecast densities will be those of Student T 
distributions. In our case study the four agents represent conjugate dynamic linear models in which 
all forecast densities are T, with parameters varying over time and with step-ahead forecast horizon. 
In such cases,   standard Bayesian augmentation methods apply to enable simulation.    Each T distribution 
is expressed as a scale mixture of normals,   with the mixing scale parameters introduced as inherent
latent variables with inverse gamma distributions.   
This expansion of the parameter space makes the T distributions conditional normals, and the
mixing scales are resampled (from implied conditional posterior inverse gamma distributions) 
for each MCMC iterate along with the agent states. This is again a standard MCMC approach
and much used in Bayesian time series, as in other areas (e.g.~\citealt{Schnatter1994}; \citealt[][Chap 15]{WestHarrison1997book2}).
Then, conditional on the current values of these latent scales,  sampling the $\x_t$ reduces technically to that 
conditional normals above. 

Specifically, suppose that  $h_{tj}(x_{tj})$ is density of the normal $ T_{n_{tj}}(h_{tj},H_{tj})$; 
the notation means that $ (x_{tj}-h_{tj})/\sqrt{H_{tj}}$ has a standard 
Student T distribution with $n_{tj}$ degrees of freedom.   Then latent scale factors $\phi_{tj}$ exist such that:
{\em (i)} conditional on $\phi_{tj},$  latent agent factor $x_{tj}$ has a conditional normal density
$x_{tj}|\phi_{tj} \sim N(h_{tj},H_{tj}/\phi_{tj})$ independently over $t,j;$ {\em (ii)}  the $\phi_{tj}$ are 
independent over $t,j$ with gamma distributions, $\phi_{tj} \sim G(n_{tj}/2,n_{tj}/2).$ 
Then, at each MCMC step, the above normal update for latent agent states is replaced by the following two steps:
\begin{itemize}
\item[i. ] Based on current values of all $\phi_{tj}$ simulated at the last MCMC iterate, 
the normal update above applies to resample each $\x_t$ vector; the only modification in the multivariate normal
conditional of eqn.~(\ref{app:condx}) is that the diagonal matrix $\H_t$ is now given by
 $ \H_t=\textrm{diag}(H_{t1}/\phi_{t1},H_{t2}/\phi_{t2},...,H_{tJ}/\phi_{tJ}).$ 
\item[ii.] Conditional on these new samples of the $\x_t,$  updated samples of the latent scales are drawn--
 independently for all $t,j$-- 
from the implied set of conditional gamma posteriors $\phi_{tj}|x_{tj} \sim G((n_{tj}+1)/2,(n_{tj}+d_{tj})/2)$
where $d_{tj}= (x_{tj}-h_{tj})^2/H_{tj}.$ 
\end{itemize} 
 
\item[3.] In some cases, agent densities may be more elaborate mixtures of normals, such as (discrete or
continuous) location and/or scale mixtures that represent asymmetric distributions.  The same augmentation
strategy can be applied in such cases, with augmented parameters including location shifts in place of, or 
in addition to, scale shifts. 

\item[4.]  In other cases,  we may be able to directly simulate the agent forecast distributions and evaluate 
forecast density functions at any point,   but do not have access to analytic forms.  One class of examples
is when the agents are simulation models, e.g., DSGE models. Another involves forecasts in terms of 
histograms.  In such cases,   MCMC will proceed using some form of 	
  Metropolis-Hastings algorithm, or accept/reject methods, or importance sampling for the 
  latent agent states. 
  
  For example,  suppose we only have access to simulations from the agent forecast distributions, 
   in terms of $I$ independent draws from each collated in the simulated vectors $\x_t^{(i)}$ for 
    $i=\seq1I.$  We can apply importance sampling as follows: 
  {\em (a)} compute the marginal likelihood
  values  $p( y_t|  \bPhi_t, \x_t^{(i)}, \mH_t)$ for each $i=\seq1I$; {\em (b)} compute and normalize the
   implied importance sampling weights $w_{ti} \propto N( y_t|  \bPhi_t, \x_t^{(i)}, \mH_t), $ 
  and then {\em (c)} resample latent agent states for this MCMC stage according to the probabilities these
   weights define.     
\end{itemize}


 \newpage

\section{Appendix: Additional  Graphical Summaries from Inflation Forecasting Analysis  \label{supp:moreoncasestudy}}

\renewcommand{\thefigure}{B\arabic{figure}}\setcounter{figure}{0}  

This appendix lays out additional graphical summaries of results from the inflation forecasting analysis in the paper, 
providing material supplementary to that discussed in Section~\ref{sec:Inf}.

\begin{figure}[htbp!]
\centering
\includegraphics[width=0.75\textwidth]{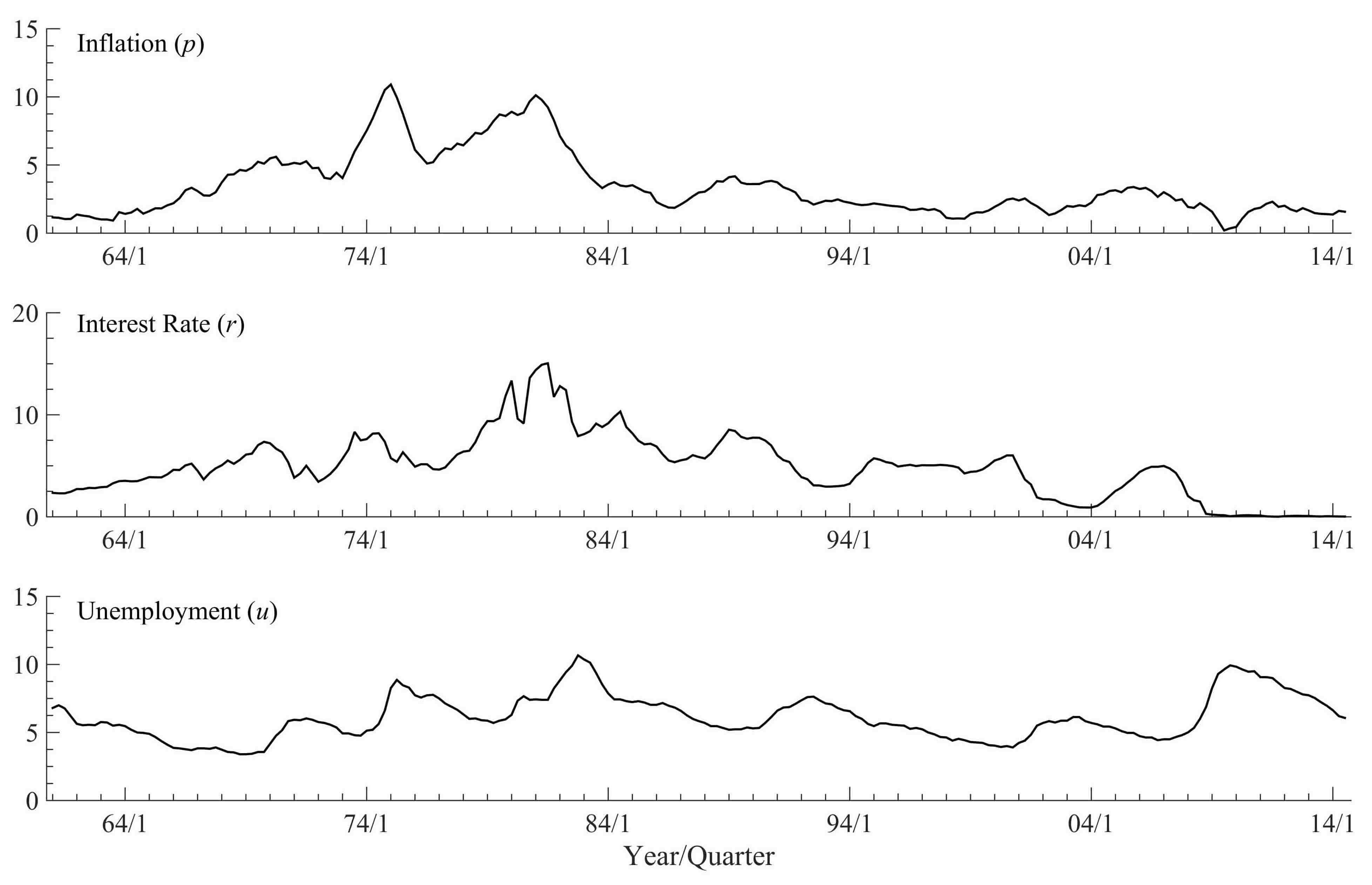} 
\caption{US inflation rate forecasting 1990/Q1-2014/Q4: U.S. macroeconomic time series (indices $\times$100 for $\%$ basis):  
annual inflation rate (p), short-term nominal interest rate (r), 
and unemployment rate $(u)$.
\label{iru}}
\end{figure}

\begin{figure}[htbp!]
\centering
\includegraphics[width=0.75\textwidth]{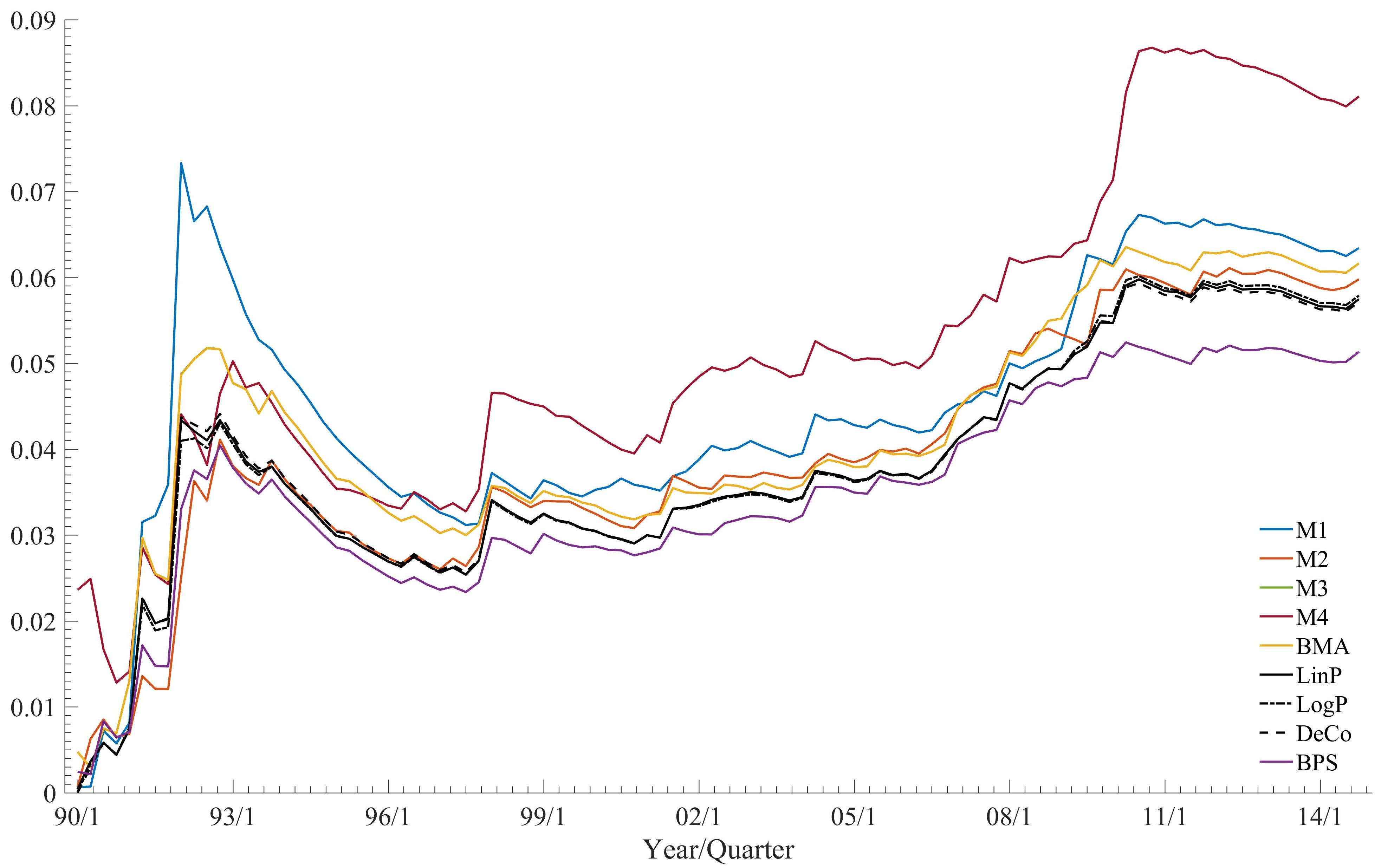} 
\caption{US inflation rate forecasting 1990/Q1-2014/Q4: Mean squared 1-step ahead forecast errors MSFE$_{1{:}t}(1)$ sequentially revised at each of the $t=\seq1{100}$ quarters. 
As with forecast uncertainties, point forecast accuracy is particularly improved under BPS at crisis periods, with 
MSFE staying relatively level    while  significantly increasing for other models and methods. 
\label{1mse}}
\end{figure}

%
%

\begin{figure}[htbp!]
\centering
\includegraphics[width=0.75\textwidth]{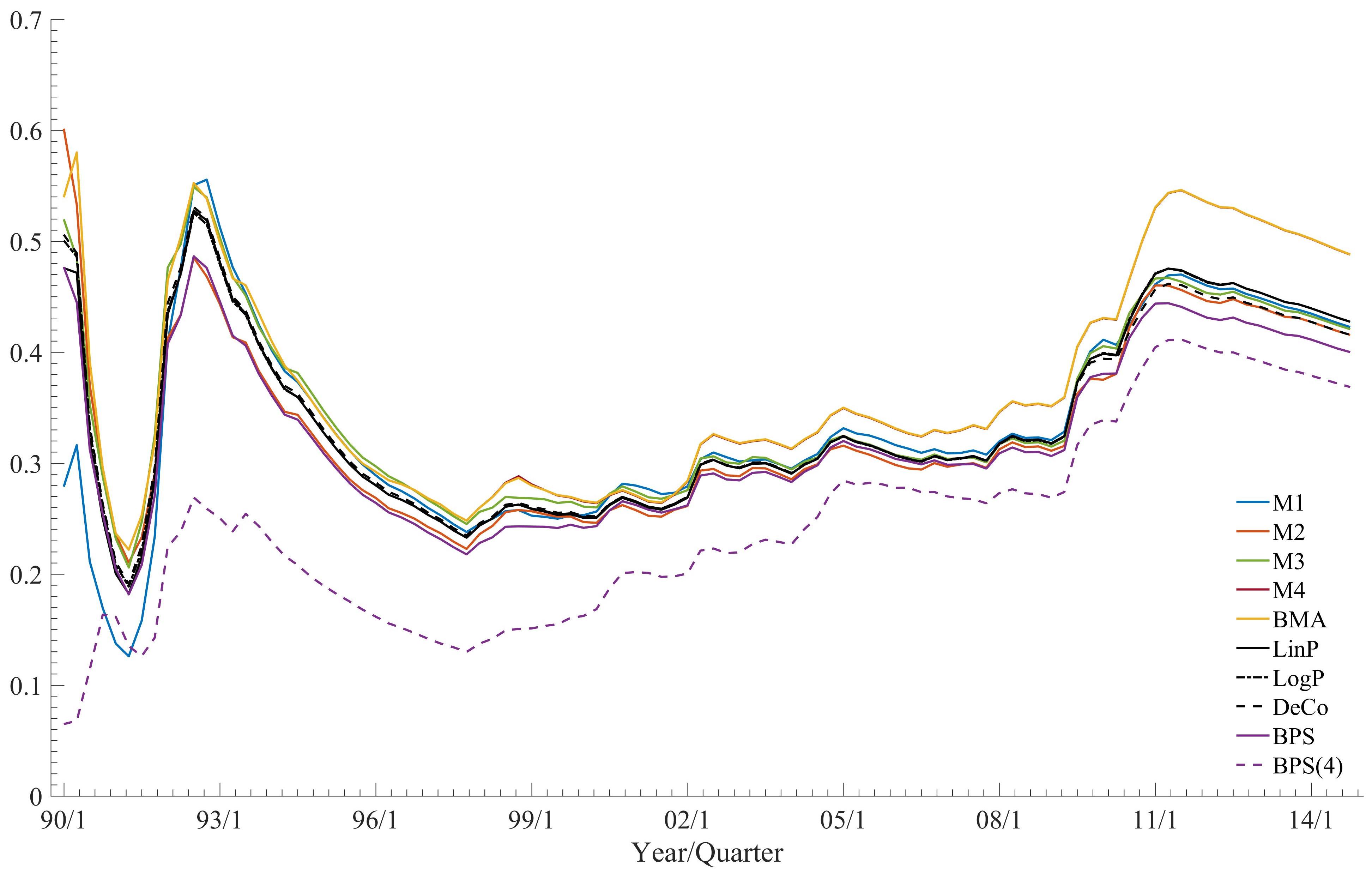} 
\caption{US inflation rate forecasting 1990/Q1-2014/Q4: Mean squared 4-step ahead forecast errors MSFE$_{1{:}t}(4)$  sequentially revised at each of the $t=\seq1{100}$ quarters.  Customized to the 4-step ahead horizon, 
BPS(4) dominates in point forecast accuracy over all models and methods,  including the direct BPS extrapolation. 
\label{14mse}}
\end{figure}

\begin{figure}[htbp!]
\centering
\includegraphics[width=0.75\textwidth]{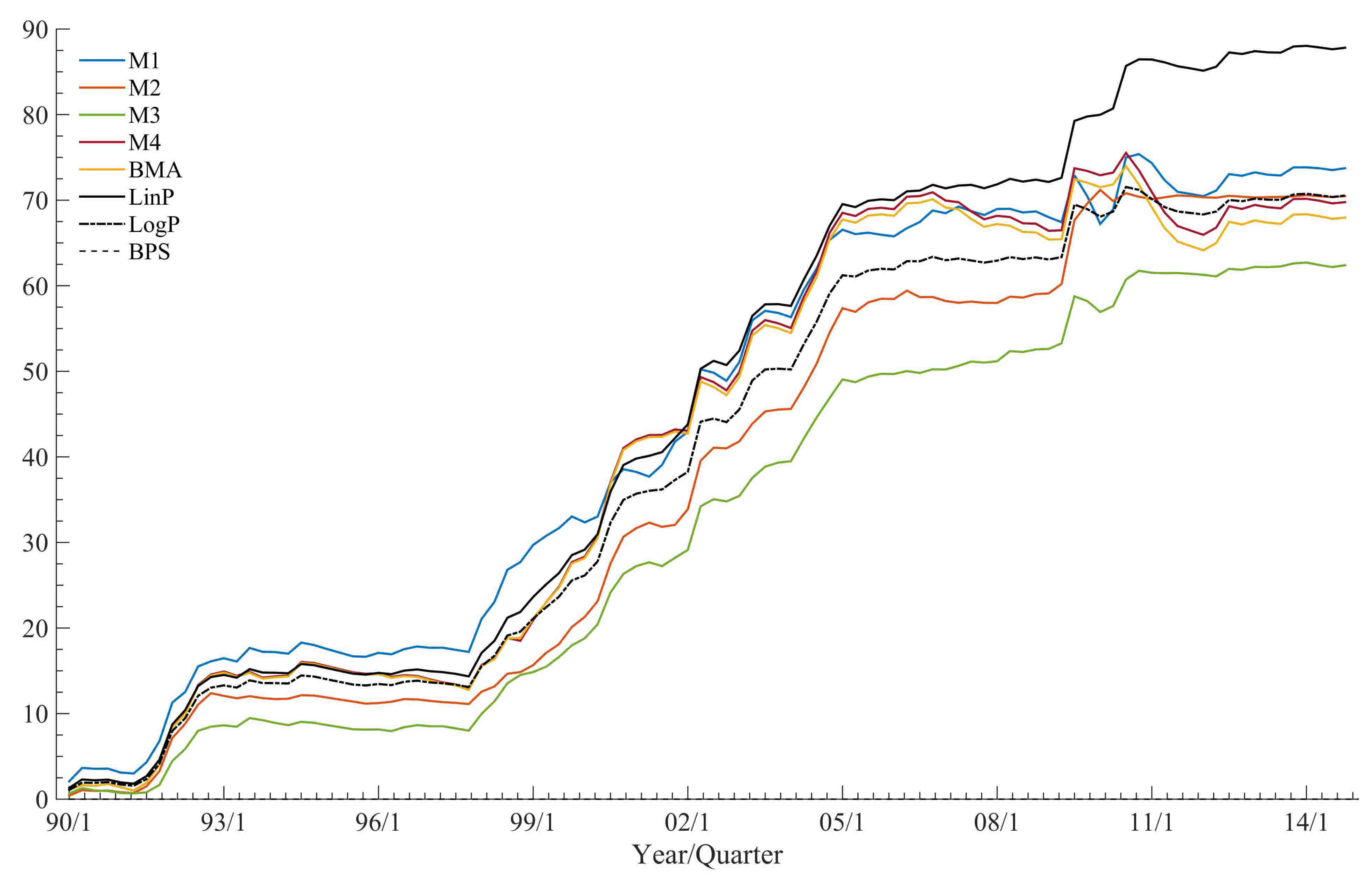} 
\caption{US inflation rate forecasting 1990/Q1-2014/Q4:  4-step ahead log predictive density ratios LPDR$_{1{:}t}(4)$  sequentially revised at each of the $t=\seq1{100}$ quarters using direct projection from the 1-step ahead BPS model (baseline at 0 over time). 
 Direct BPS extrapolation performs relatively poorly  as-- being inherently calibrated to 1-step model fit-- it
fails to adequately represent the increased uncertainty associated with long term forecasts in addition to replying on 
less accurate model forecasts. In contrast,  BPS(4) is able to improve 
by adjusting to the increased forecast uncertainties.  
\label{14lpdr}}
\end{figure}

\begin{figure}[htbp!]
\centering
\includegraphics[width=0.75\textwidth]{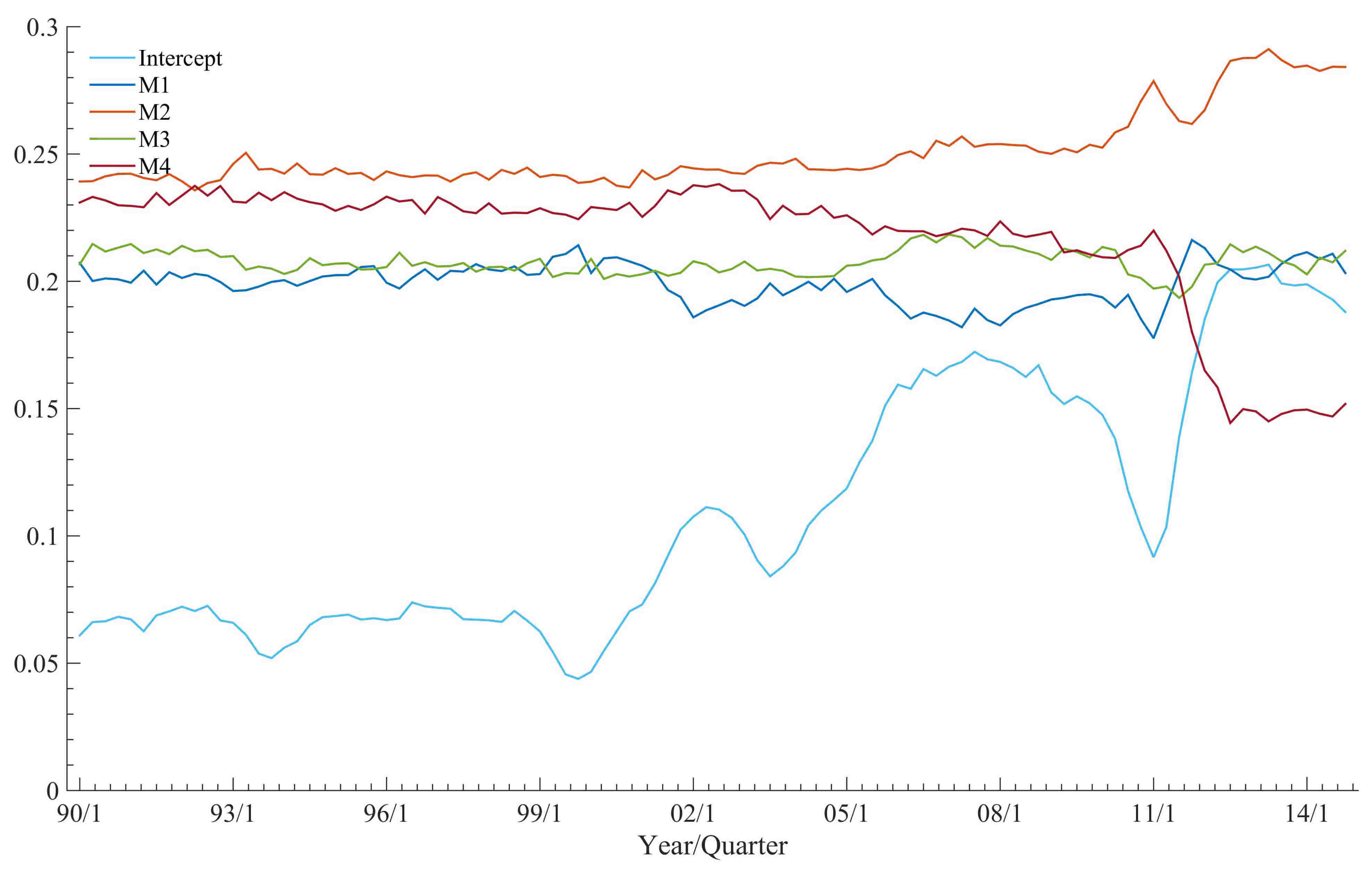} 
\caption{US inflation rate forecasting 1990/Q1-2014/Q4: On-line posterior means of BPS(4) model
 coefficients $\btheta_t$ sequentially computed at each of the $t=\seq1{100}$ quarters. 
There is a notable reduction in adaptability over time relative to the 1-step BPS coefficients; 
this is expected as the agents' forecasts are less reliable at longer horizons, so the data-based information 
advising the changes in posteriors over time is relatively limited. 
Further,  the 4-step ahead trajectories  
are quite different from the 1-step trajectories, reasonably reflecting the differing forecasting abilities of the 
models at differing horizons. 
\label{4coeff}}
\end{figure}


\begin{figure}[htbp!]
\centering
\includegraphics[width=0.75\textwidth]{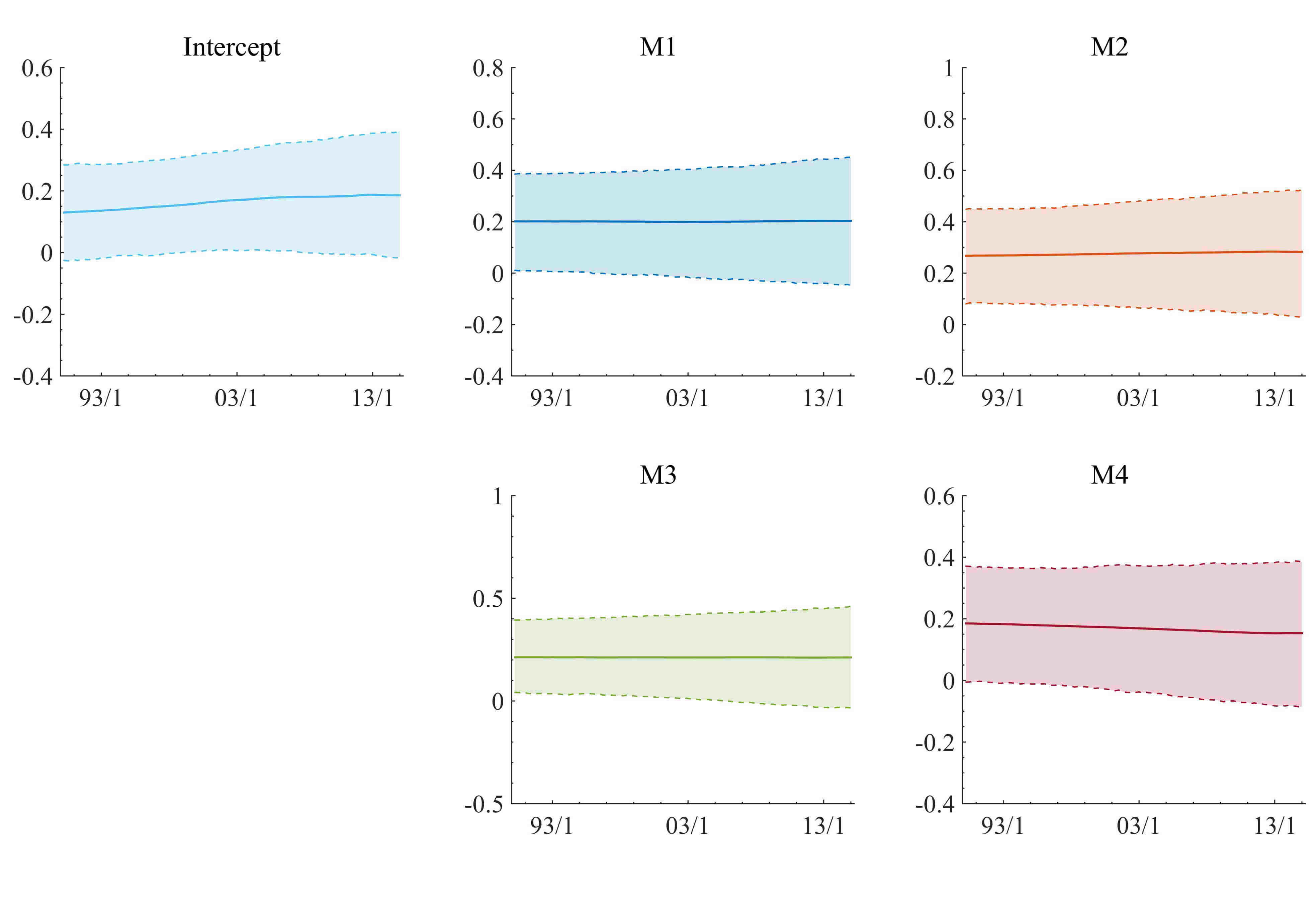} 
\caption{US inflation rate forecasting 1990/Q1-2014/Q4: Retrospective posterior trajectories of the BPS(4) model coefficients based on data from the full $t=\seq1{100}$ quarters. Posterior means (solid) and 95$\%$ credible intervals (shaded).
\label{4posttheta}}
\end{figure}

\begin{figure}[htbp!]
\centering
\includegraphics[width=0.75\textwidth]{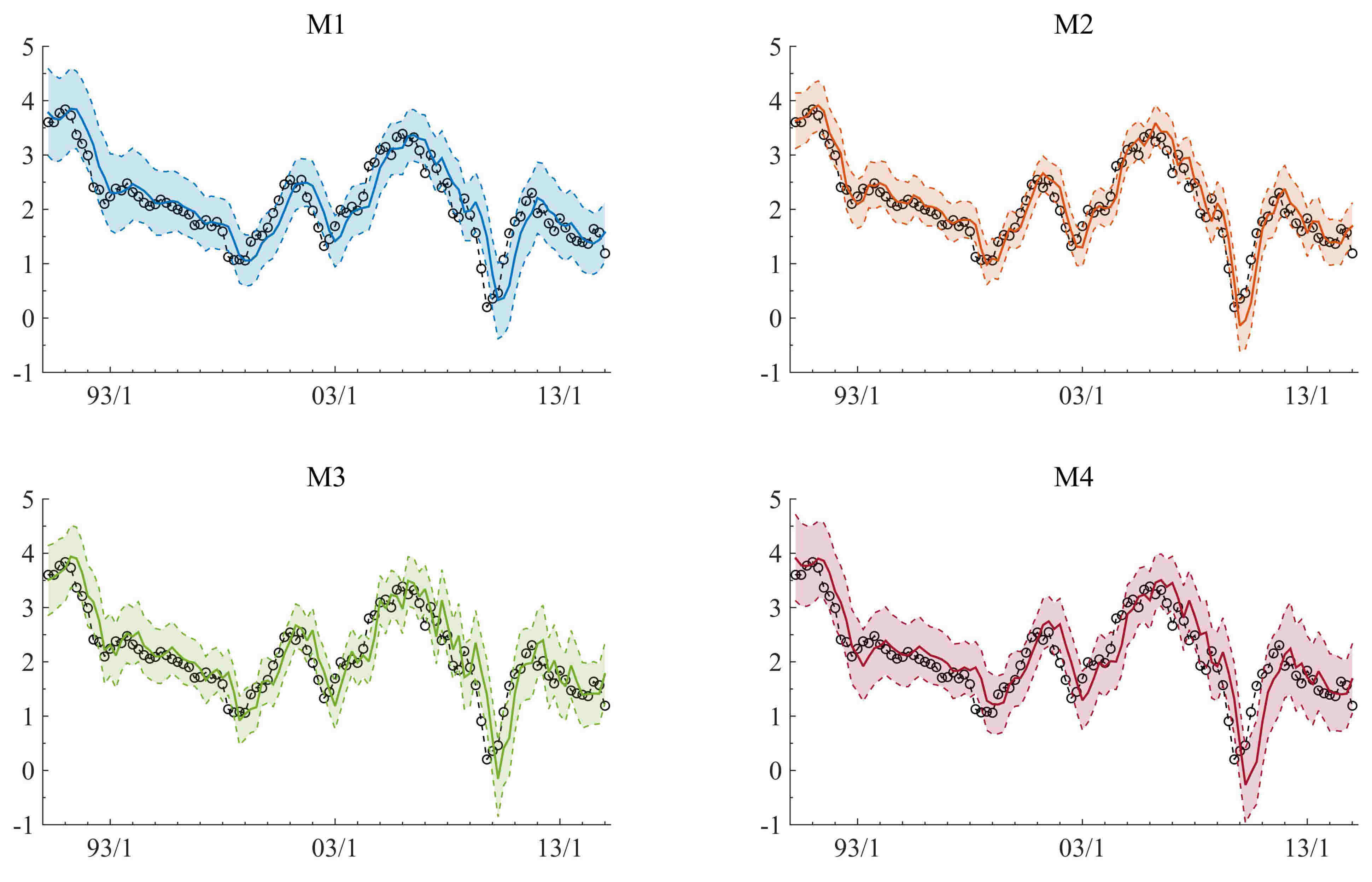} 
\caption{US inflation rate forecasting 1990/Q1-2014/Q4: BPS model-based posterior trajectories of the latent agent states $x_{tj}$ for $j=\seq14$ over the $t=\seq1{100}$ quarters. Posterior means (solid) and
95$\%$ credible intervals (shaded) from the MCMC analysis, with data $y_t\equiv p_t$ (circles). 
\label{1postx}}
\end{figure}

\begin{figure}[htbp!]
\centering
  \includegraphics[width=3.2in]{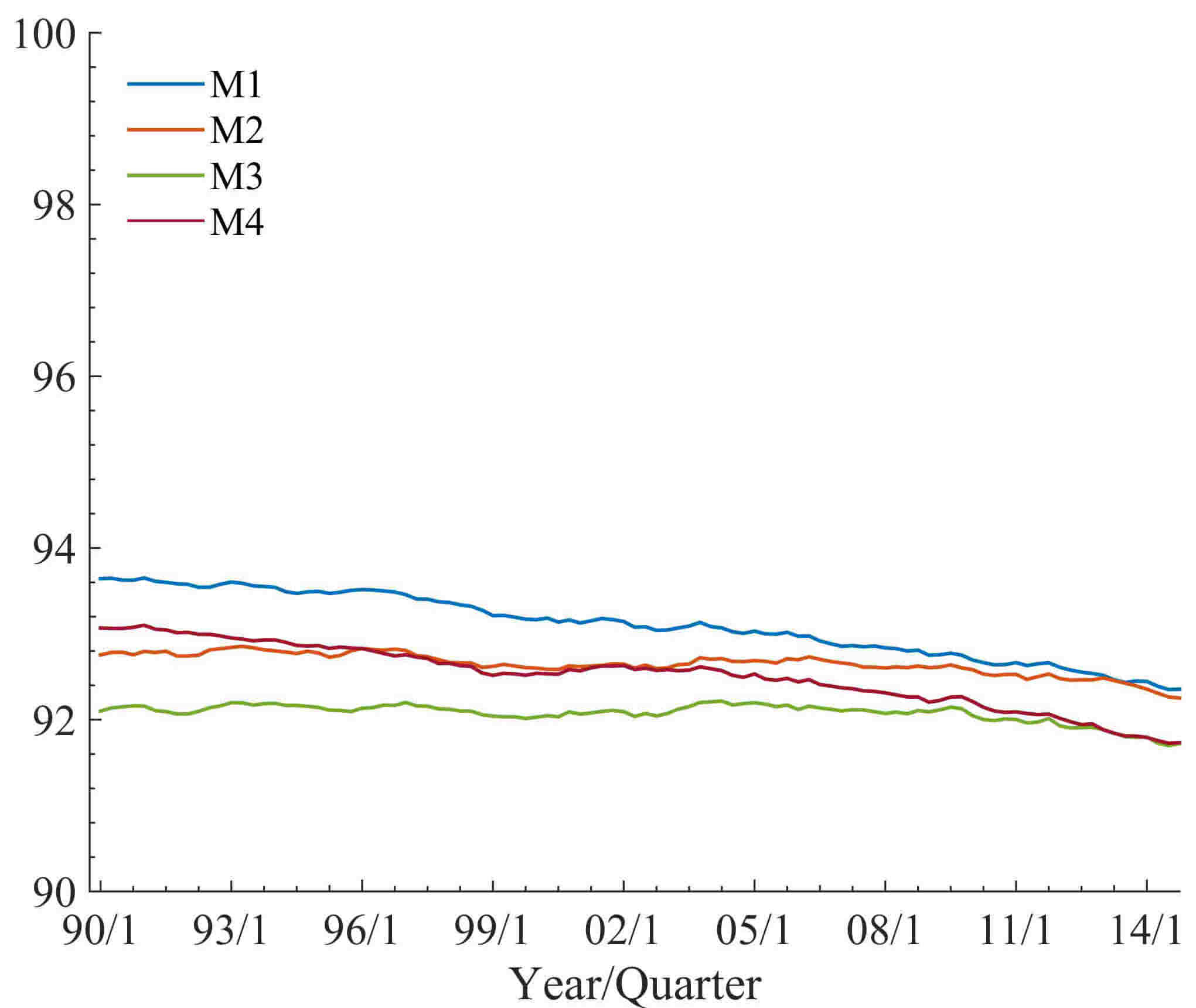} 
  \includegraphics[width=3.2in]{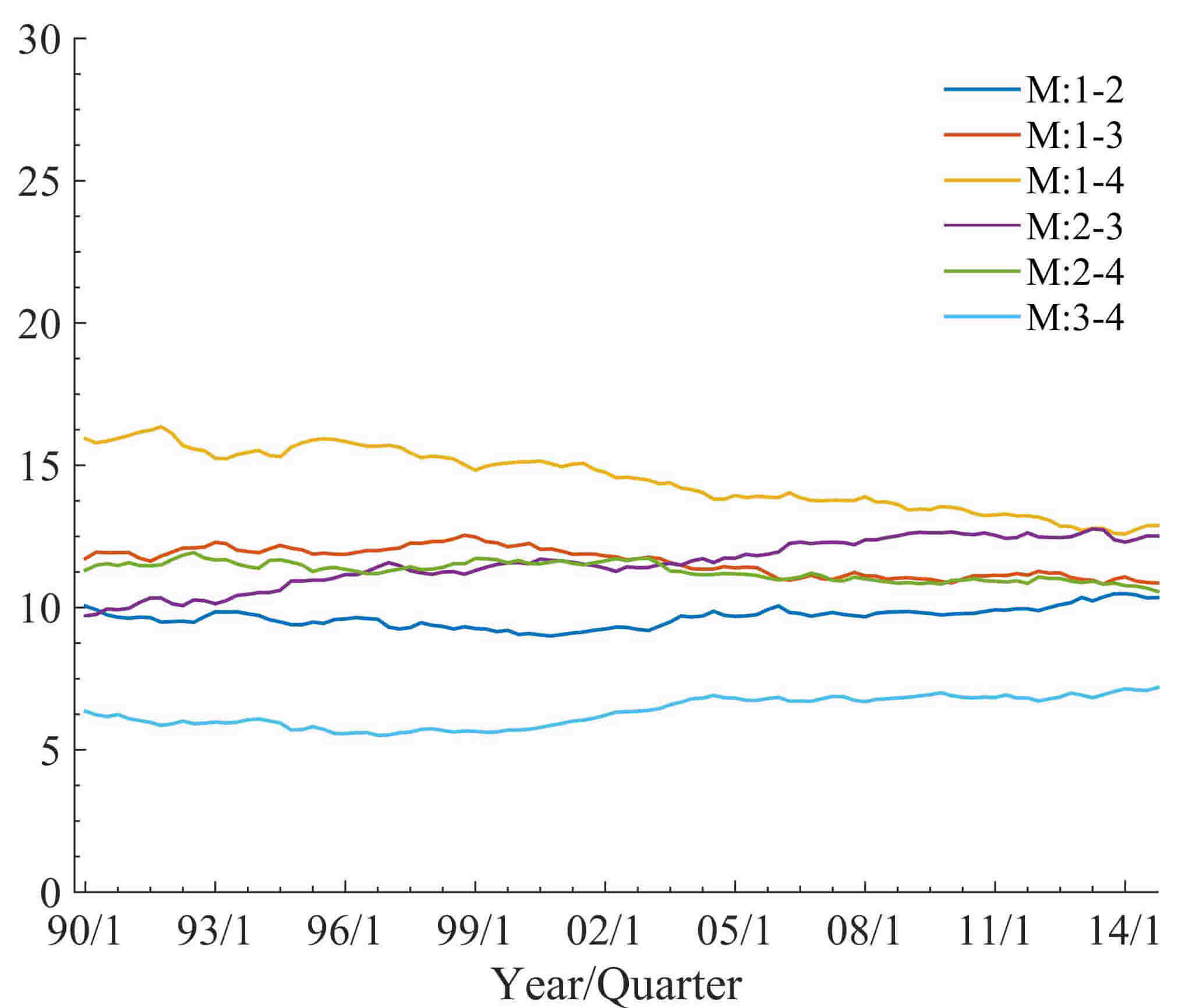}
  \caption{US inflation rate forecasting 1990/Q1-2014/Q4: BPS(4) model-based trajectories of 4-step ahead  MC-empirical R$^2$ (left) and paired MC-empirical R$^2$ (right) in the posterior for the latent agent states $x_{jt}$  for $j=\seq14$ over the $t=\seq1{100}$ quarters.}
    \label{4r2}
\end{figure}

\FloatBarrier
 
\vskip3in\phantom{.}


\newpage

\section{Appendix: Additional Assessment Summaries for Simulation Study \label{supp:sim}}

\renewcommand{\thefigure}{C\arabic{figure}}\setcounter{figure}{0}  

This appendix lays out graphical summaries of results of analysis on one simulated data as noted 
in Section~\ref{sec:simulatedata}.   

\subsection{Sequential Forecast Performance and On-line Inference}

\begin{figure}[htbp!]
\centering
\includegraphics[width=5in]{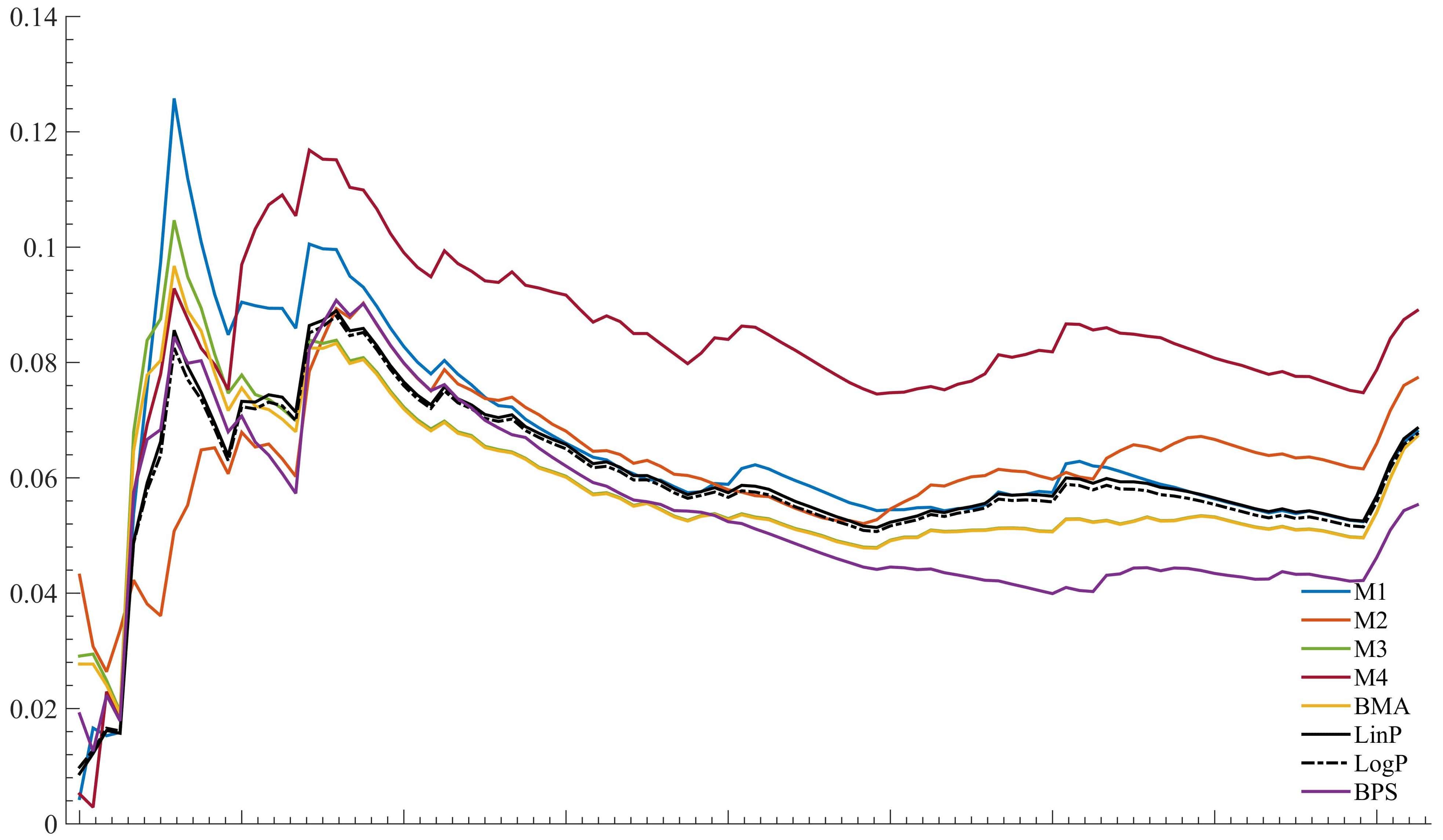} 
\caption{Simulated data forecasting: Mean squared 1-step ahead forecast errors MSFE$_{1{:}t}$ in the analysis of the synthetic data. Note that BPS outperforms for the latter half of the analysis.
\label{S1mse}}
\end{figure}

\begin{figure}[htbp!]
\centering
\includegraphics[width=5in]{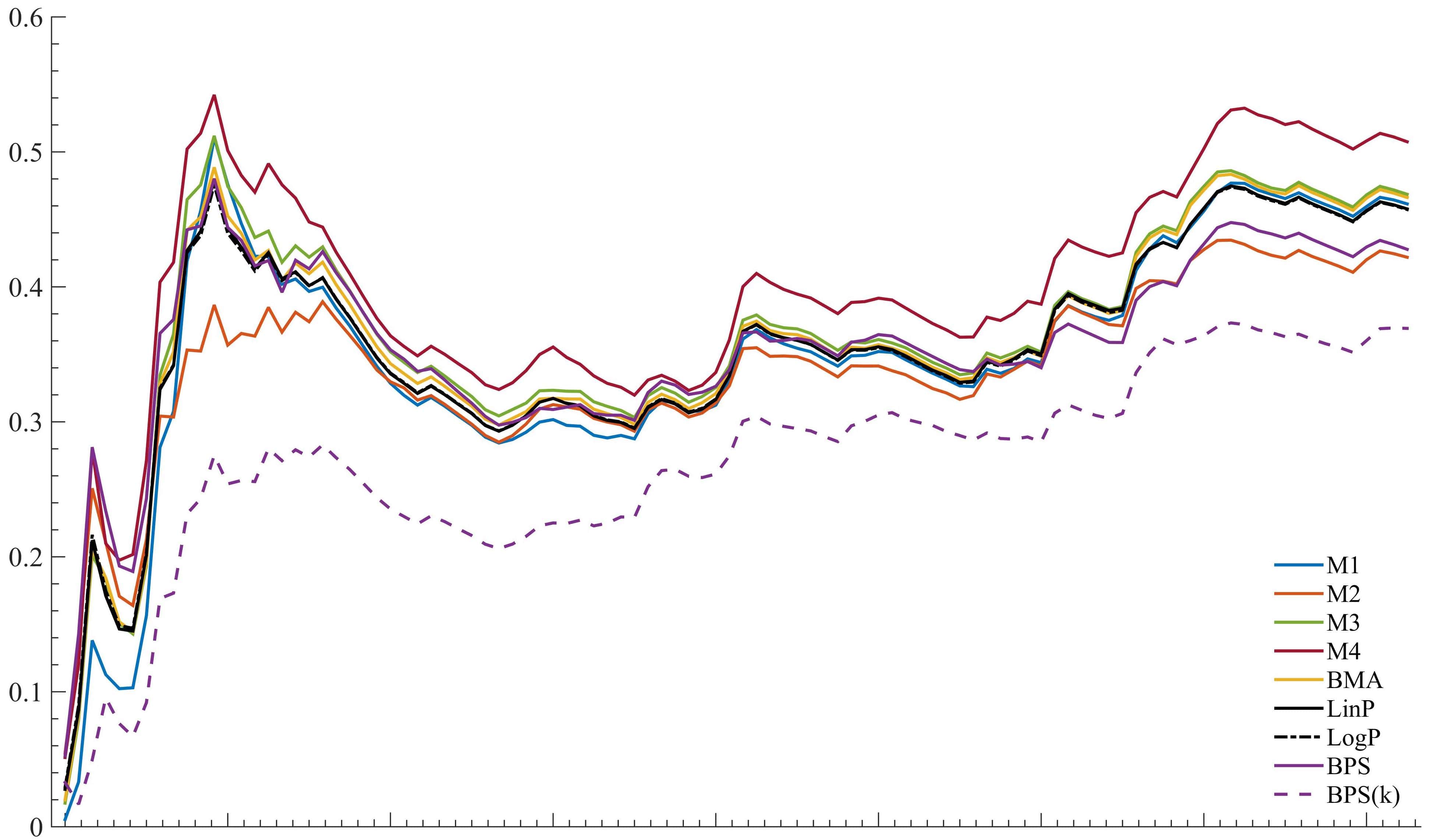} 
\caption{Simulated data forecasting: Mean squared 4-step ahead forecast errors MSFE$_{1{:}t}$ in the analysis of the synthetic data. Note that, while BPS underperforms compared to the best model and BMA, BPS($k$) dominates all models and strategies through $1{:}T$.
\label{S14mse}}
\end{figure}

\begin{figure}[htbp!]
\centering
\includegraphics[width=5in]{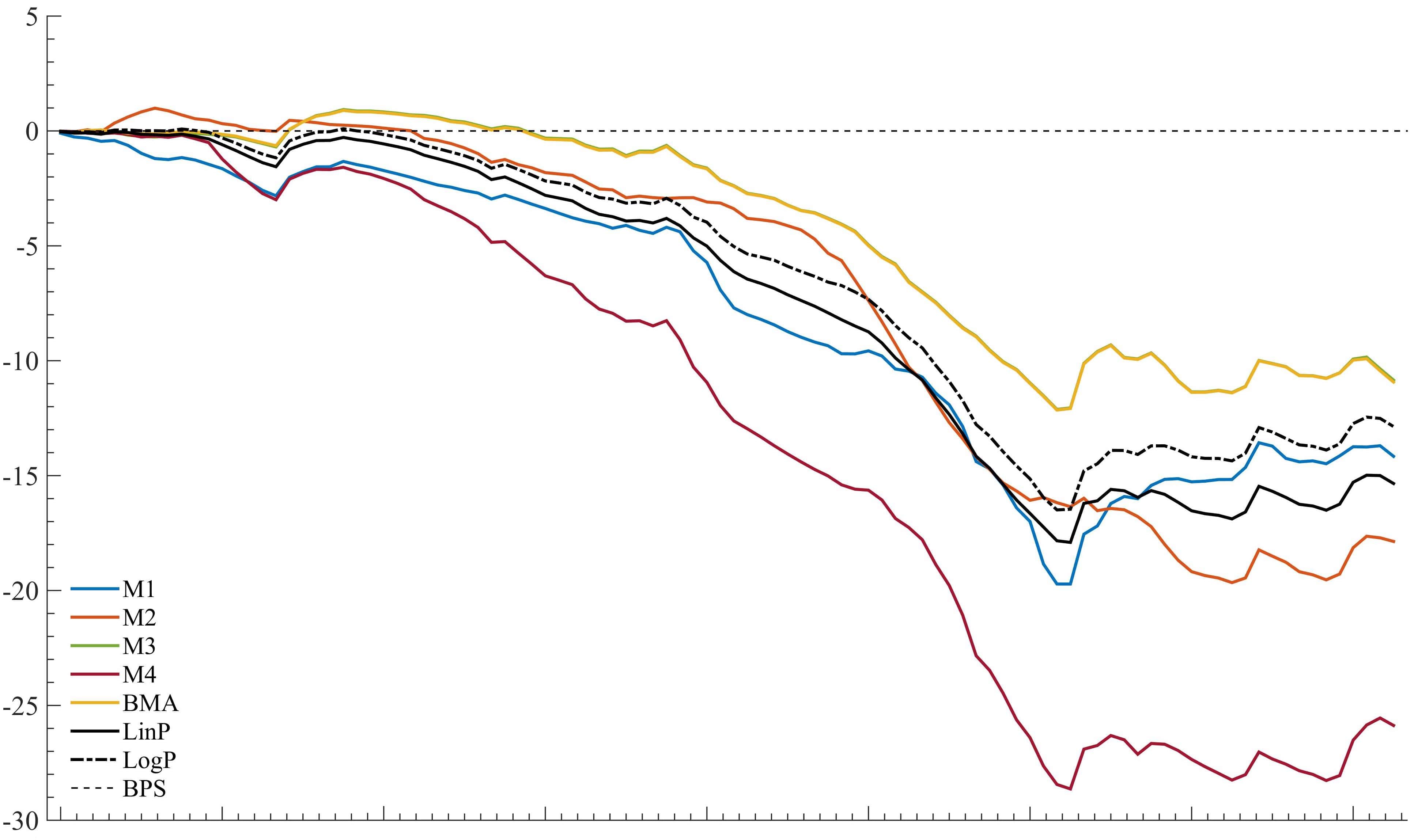} 
\caption{Simulated data forecasting: Cumulative sum of 1-step ahead log predictive density ratios LPDR$_{1{:}t}$ in the analysis of the synthetic data. Though BPS underperforms slightly in the beginning, it dominates the others in the later half of the analysis.
\label{S1lpdr}}
\end{figure}

\begin{figure}[htbp!]
\centering
\includegraphics[width=5in]{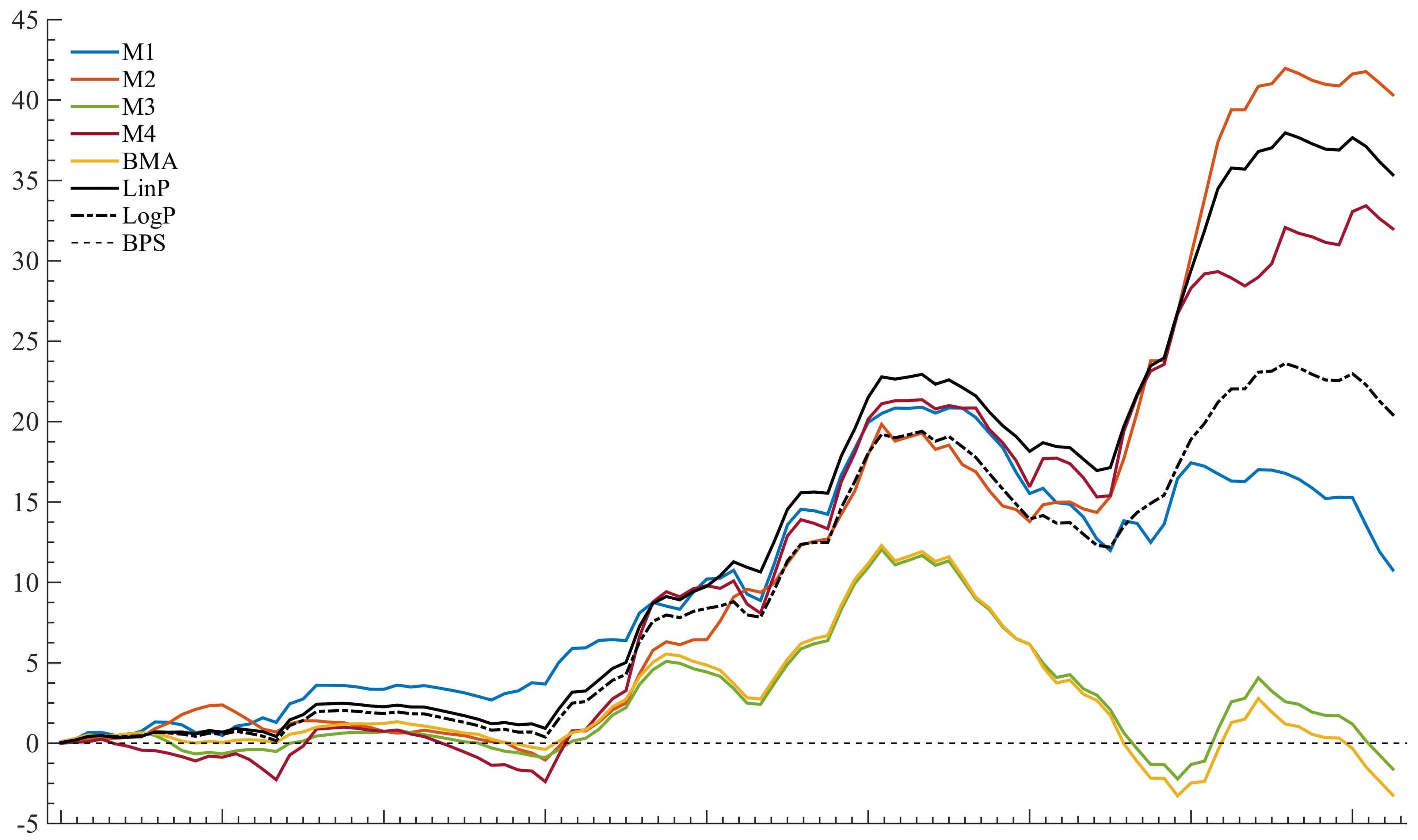} 
\caption{Simulated data forecasting: Cumulative sum of 4-step ahead log predictive density ratios LPDR$_{1{:}t}$ in the analysis of the synthetic data. BPS overfits to the forecasts and fails to correctly estimate the forecast standard deviation, leading to poor density forecasts.
\label{S14lpdr}}
\end{figure}

\begin{figure}[htbp!]
\centering
\includegraphics[width=5in]{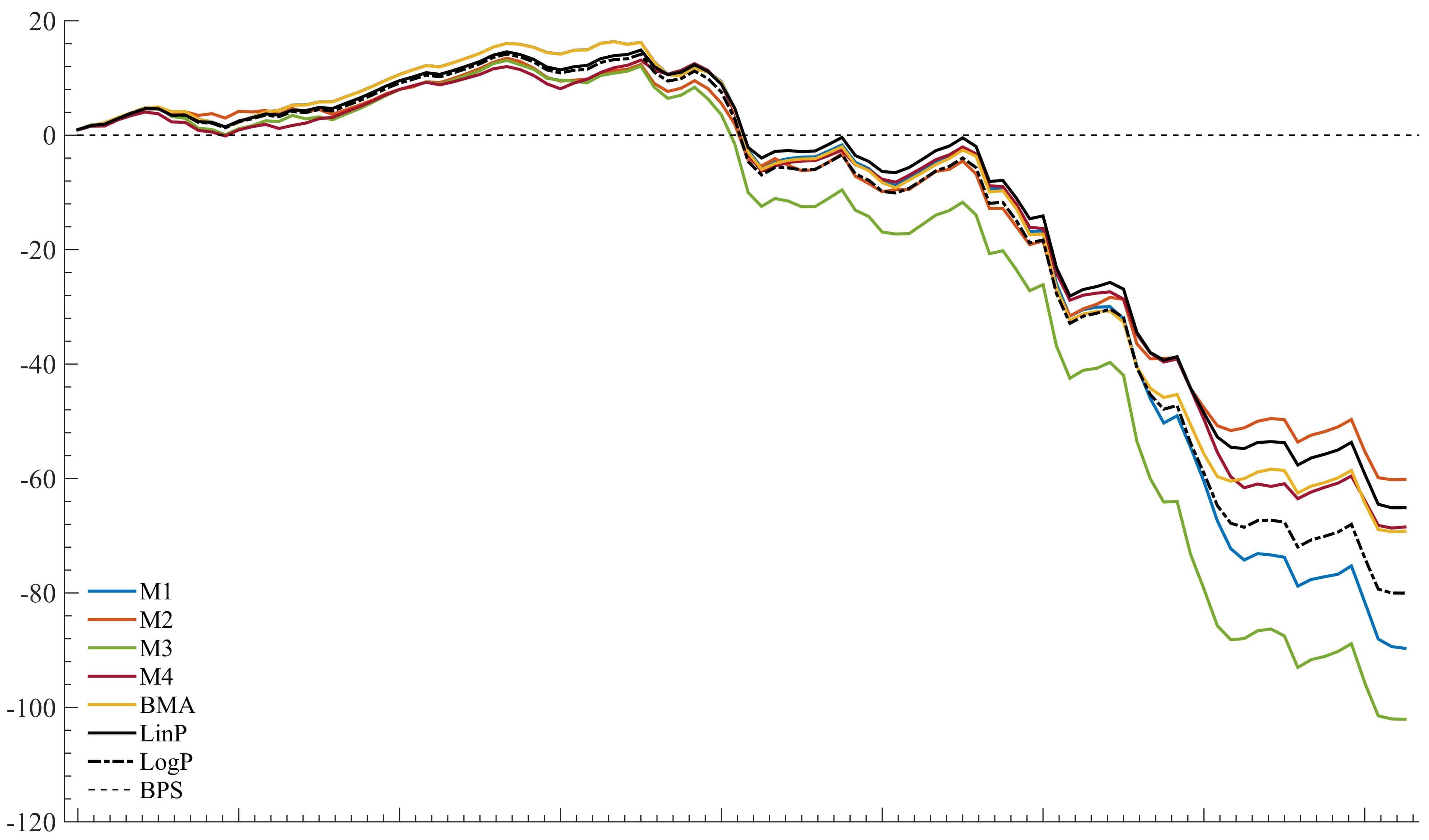} 
\caption{Simulated data forecasting: Cumulative sum of 4-step ahead log predictive density ratios LPDR$_{1{:}t}$ in the analysis of the synthetic data. Contrary to Fig.~\ref{S14lpdr}, BPS($k$) succeeds in forecasting uncertainty and is able to achieve superior density forecasts. Note that, BPS($k$) does better when the model switching is less frequent (Fig.~\ref{regimes}).
\label{S4lpdr}}
\end{figure}

\begin{figure}[htbp!]
\centering
\includegraphics[width=5in]{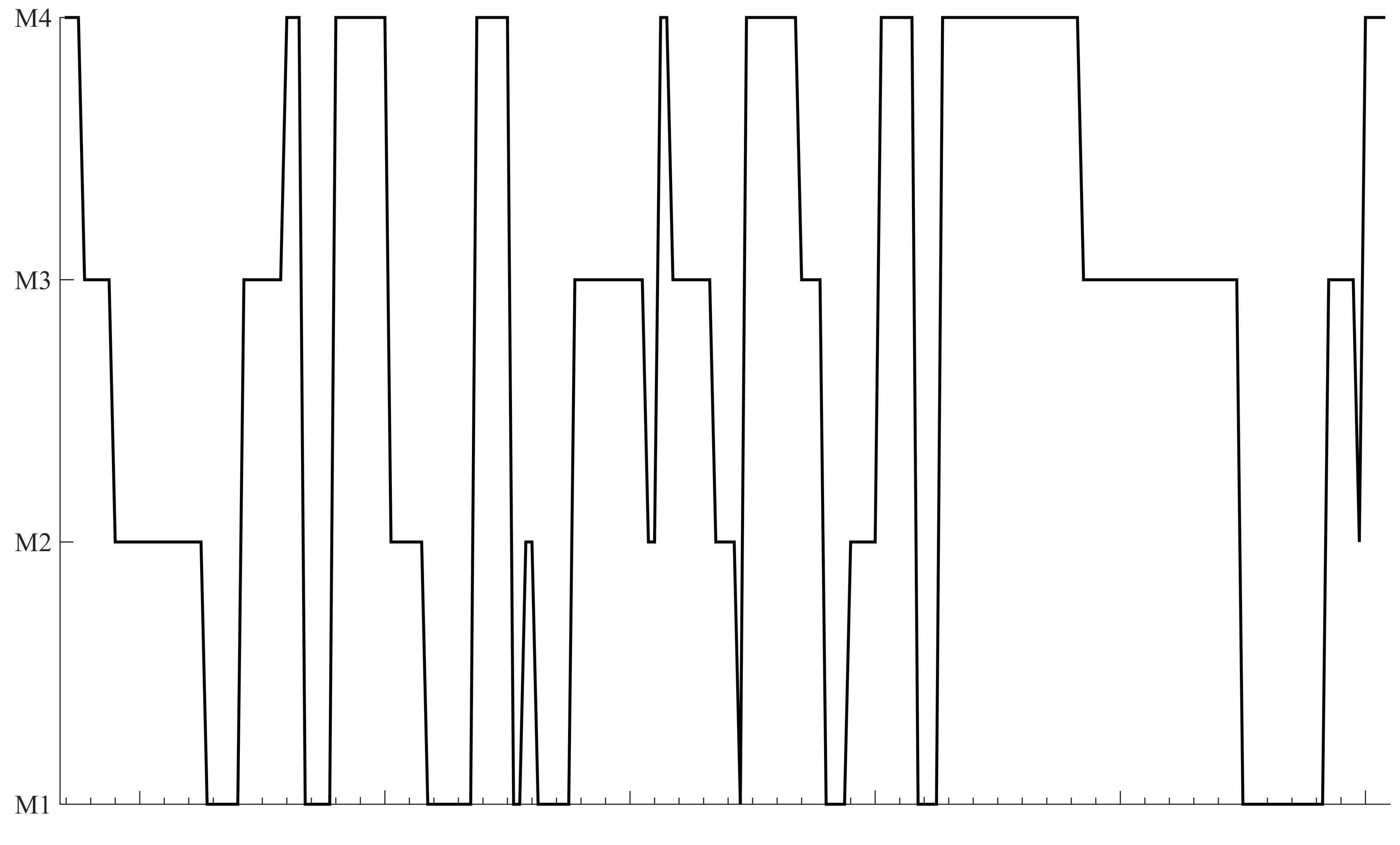} 
\caption{Simulated data forecasting: Models chosen in simulating the synthetic data.
\label{regimes}}
\end{figure}

\begin{figure}[htbp!]
\centering
\includegraphics[width=5in]{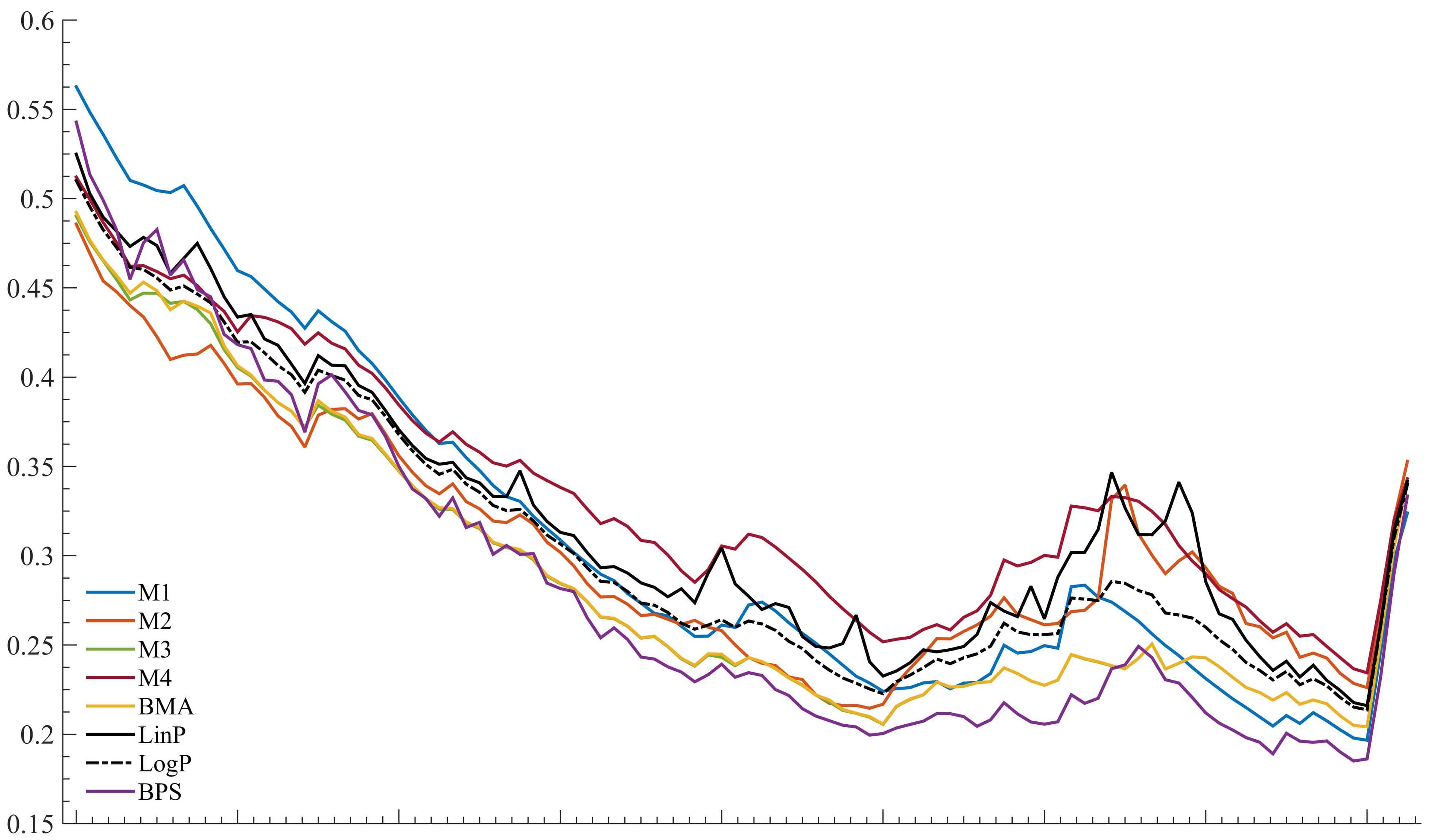} 
\caption{Simulated data forecasting: 1-step ahead forecast standard deviations in the analysis of the synthetic data.
\label{S1sd}}
\end{figure}

\begin{figure}[htbp!]
\centering
\includegraphics[width=5in]{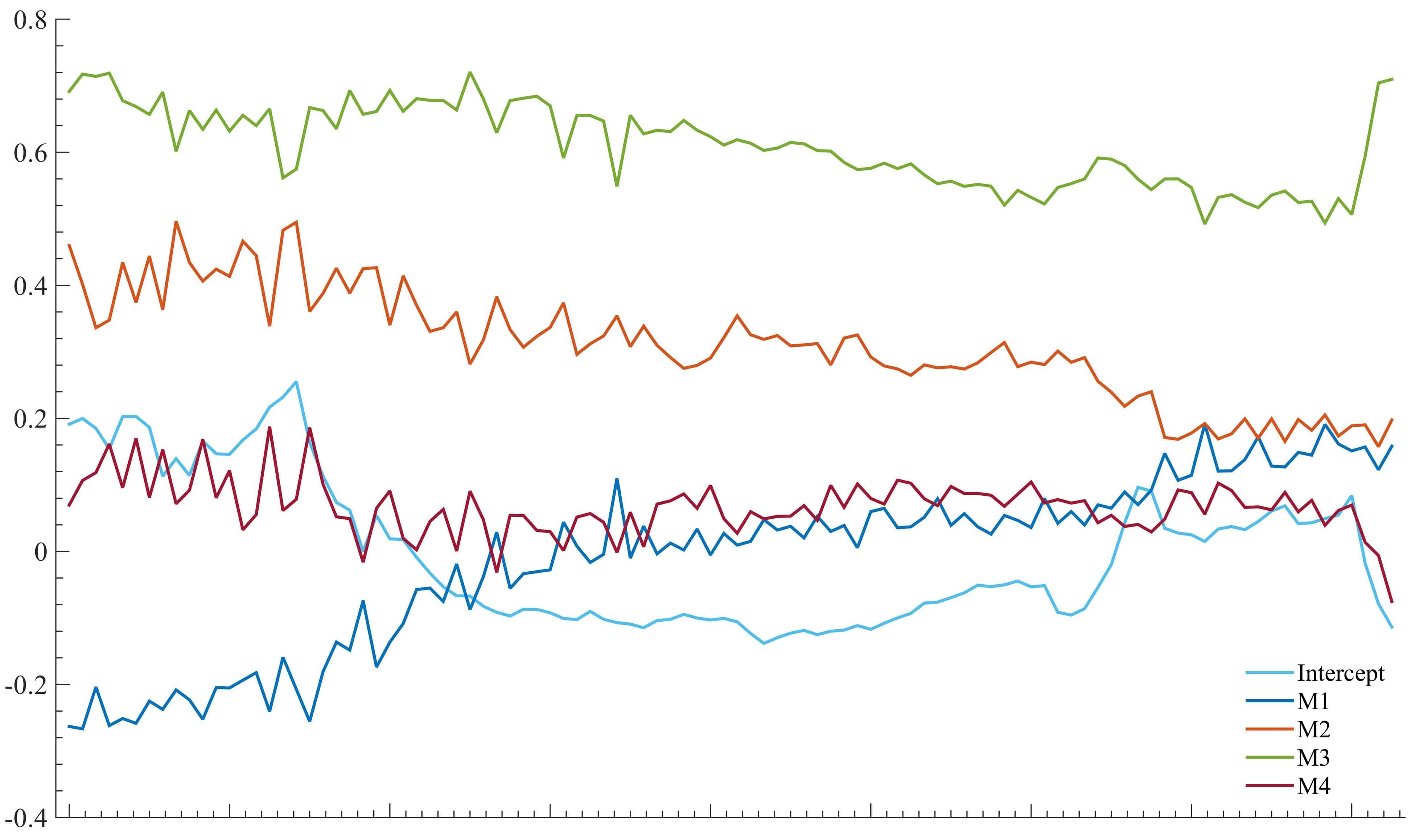} 
\caption{Simulated data forecasting: 1-step ahead forecast coefficients of BPS in the analysis of the synthetic data.
\label{S1coeff}}
\end{figure}

\begin{figure}[htbp!]
\centering
\includegraphics[width=5in]{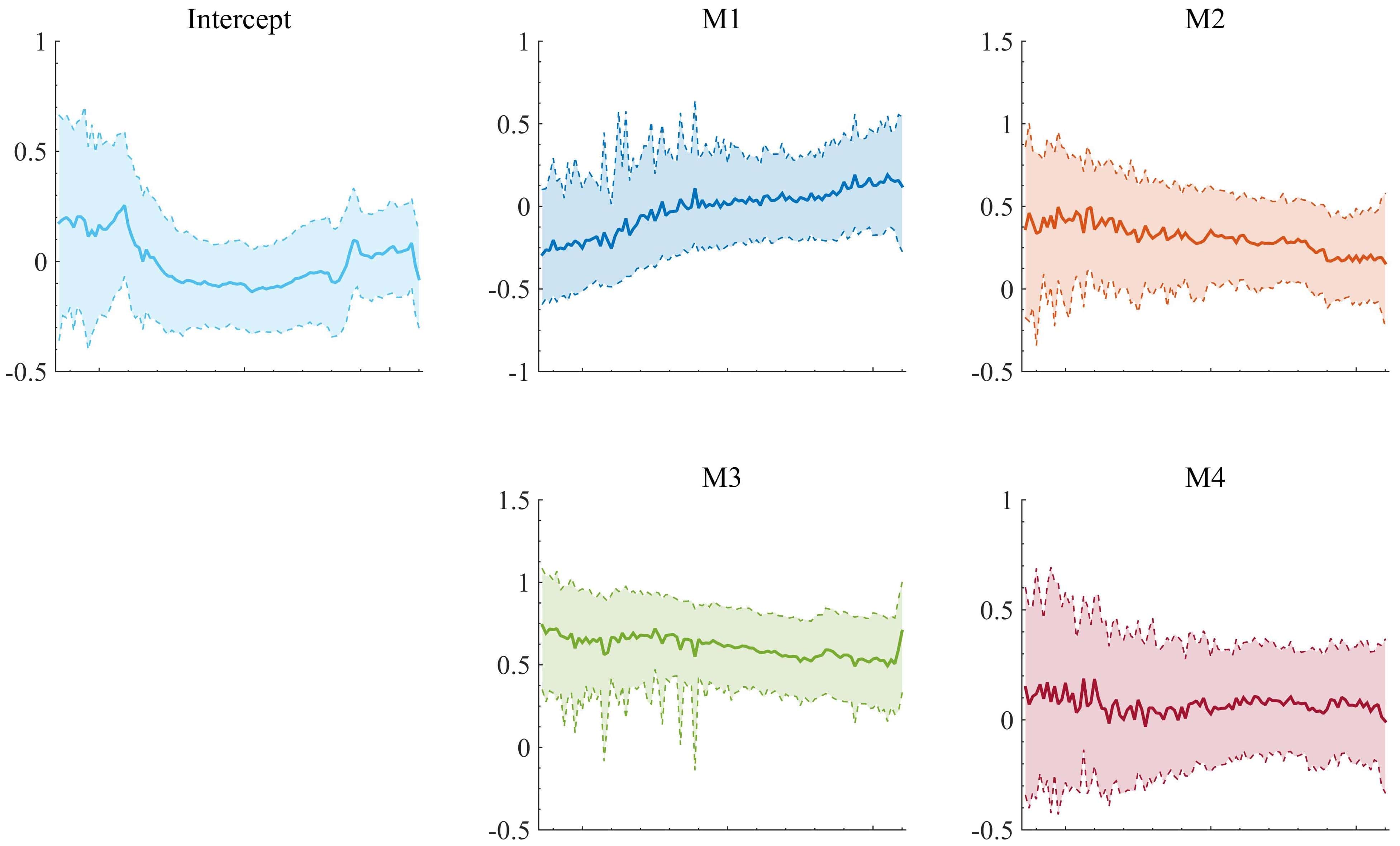} 
\caption{Simulated data forecasting: 1-step ahead forecast coefficients of BPS in the analysis of the synthetic data with credible intervals: posterior means (solid) and 95$\%$ credible intervals (dotted).
\label{S1coefCI}}
\end{figure}

\begin{figure}[htbp!]
\centering
\includegraphics[width=5in]{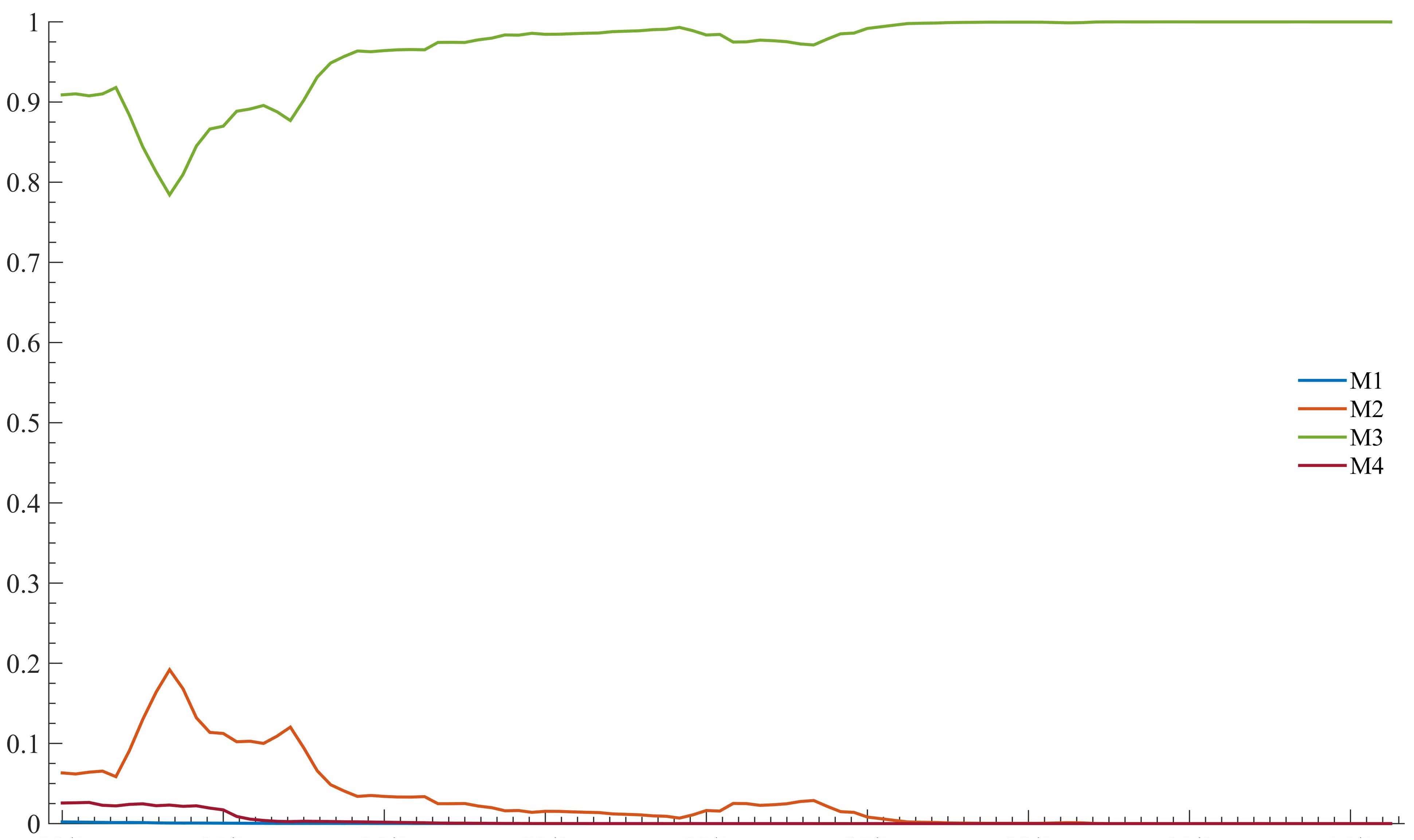} 
\caption{Simulated data forecasting: 1-step ahead forecast weights of BMA in the analysis of the synthetic data.
\label{S1BMAcoeff}}
\end{figure}

\begin{figure}[htbp!]
\centering
\includegraphics[width=5in]{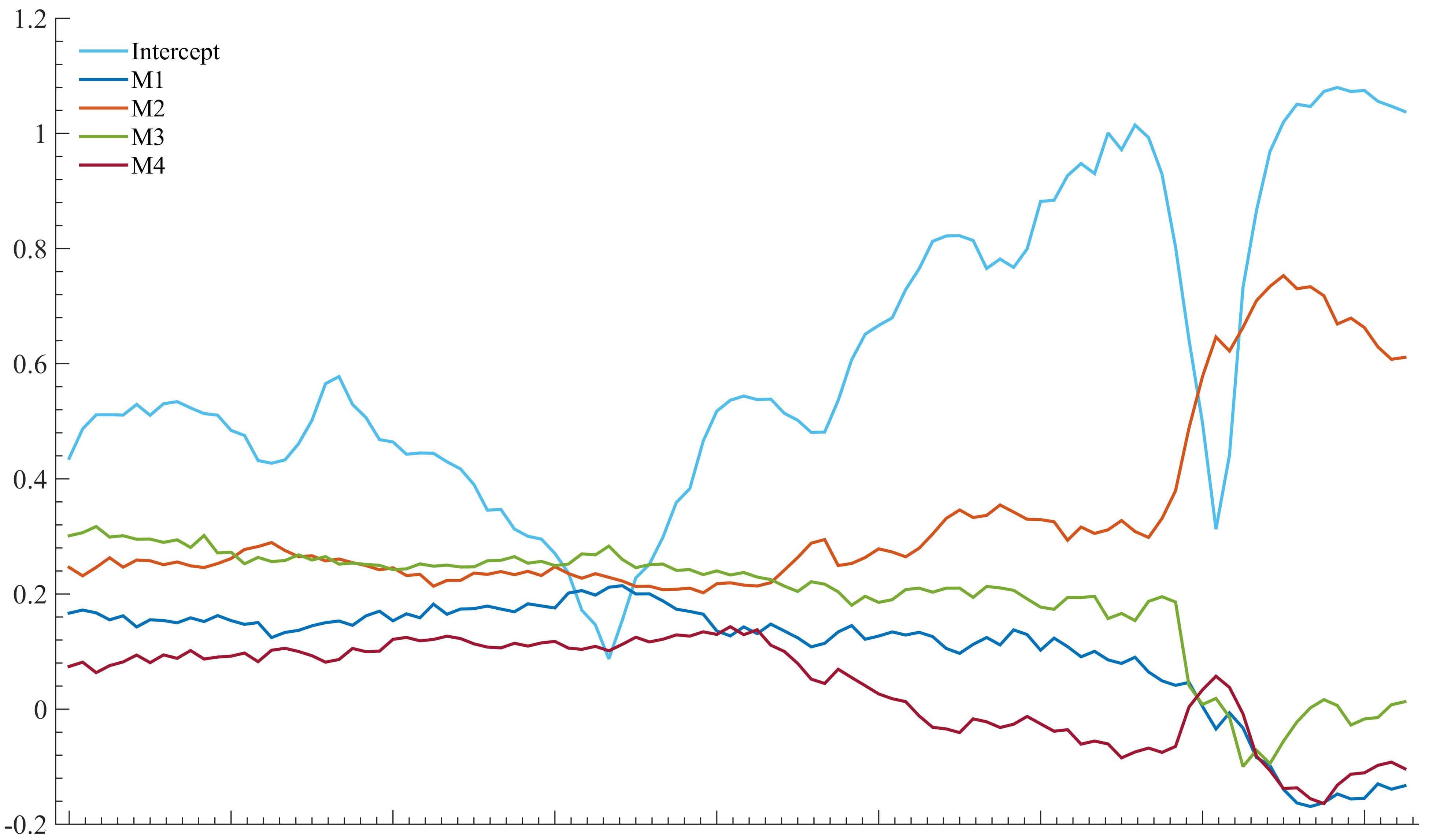} 
\caption{Simulated data forecasting: 4-step ahead forecast coefficients of BPS in the analysis of the synthetic data.
\label{S4coeff}}
\end{figure}

\begin{figure}[htbp!]
\centering
\includegraphics[width=5in]{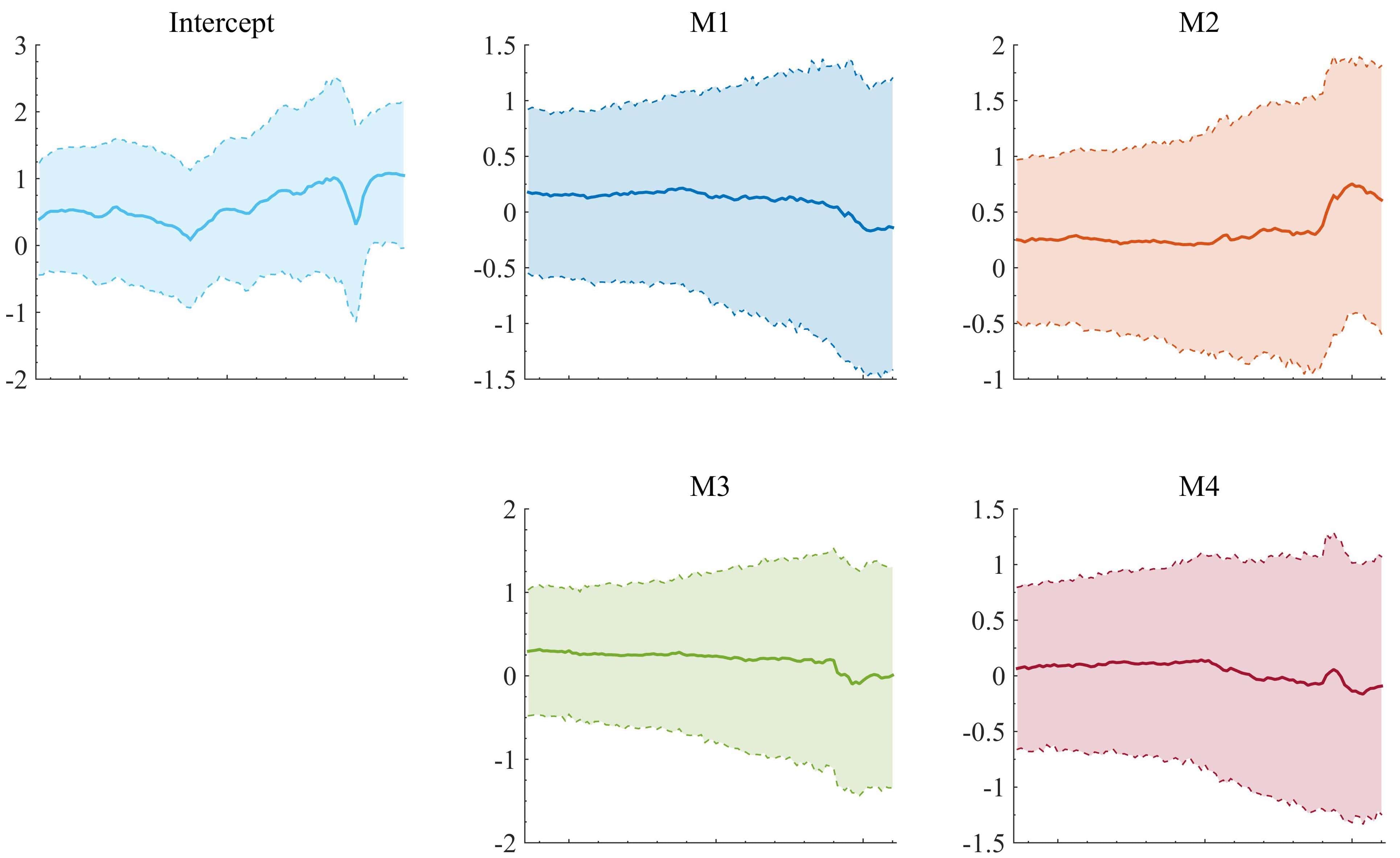} 
\caption{Simulated data forecasting: 4-step ahead forecast coefficients of BPS in the analysis of the synthetic data with credible intervals: posterior means (solid) and 95$\%$ credible intervals (dotted)..
\label{S4coefCI}}
\end{figure}

\FloatBarrier

\subsection{Retrospective Posterior Analysis}

\begin{figure}[htbp!]
\centering
\includegraphics[width=5in]{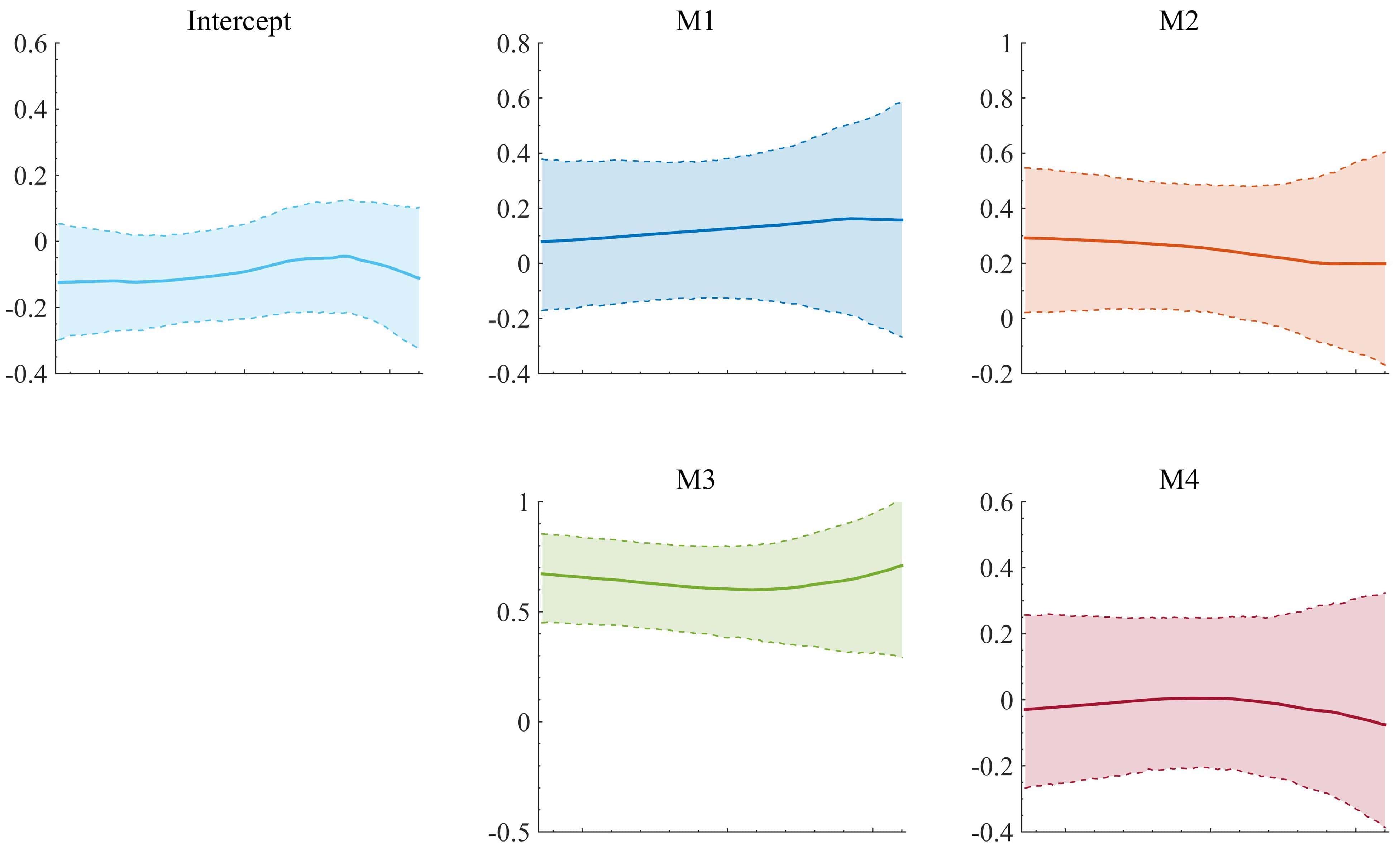} 
\caption{Simulated data forecasting: Posterior trajectories of the BPS coefficients in the analysis of the synthetic data for 1-step ahead forecasts.
\label{S1posttheta}}
\end{figure}

\begin{figure}[htbp!]
\centering
\includegraphics[width=5in]{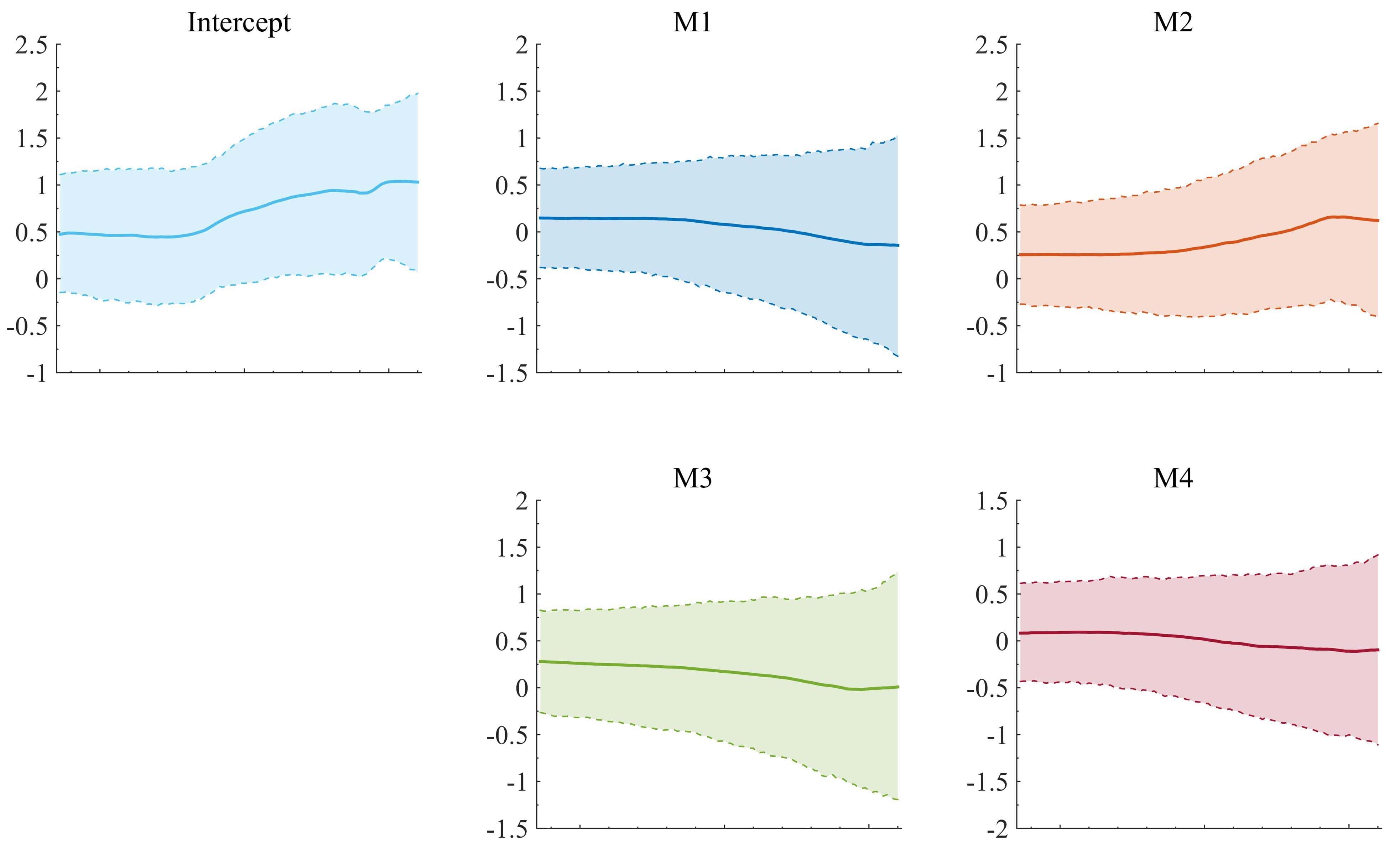} 
\caption{Simulated data forecasting: Posterior trajectories of the BPS coefficients in the analysis of the synthetic data for 4-step ahead forecasts.
\label{S4posttheta}}
\end{figure}

\begin{figure}[htbp!]
\centering
\begin{minipage}{.4\linewidth}
  \includegraphics[width=2in]{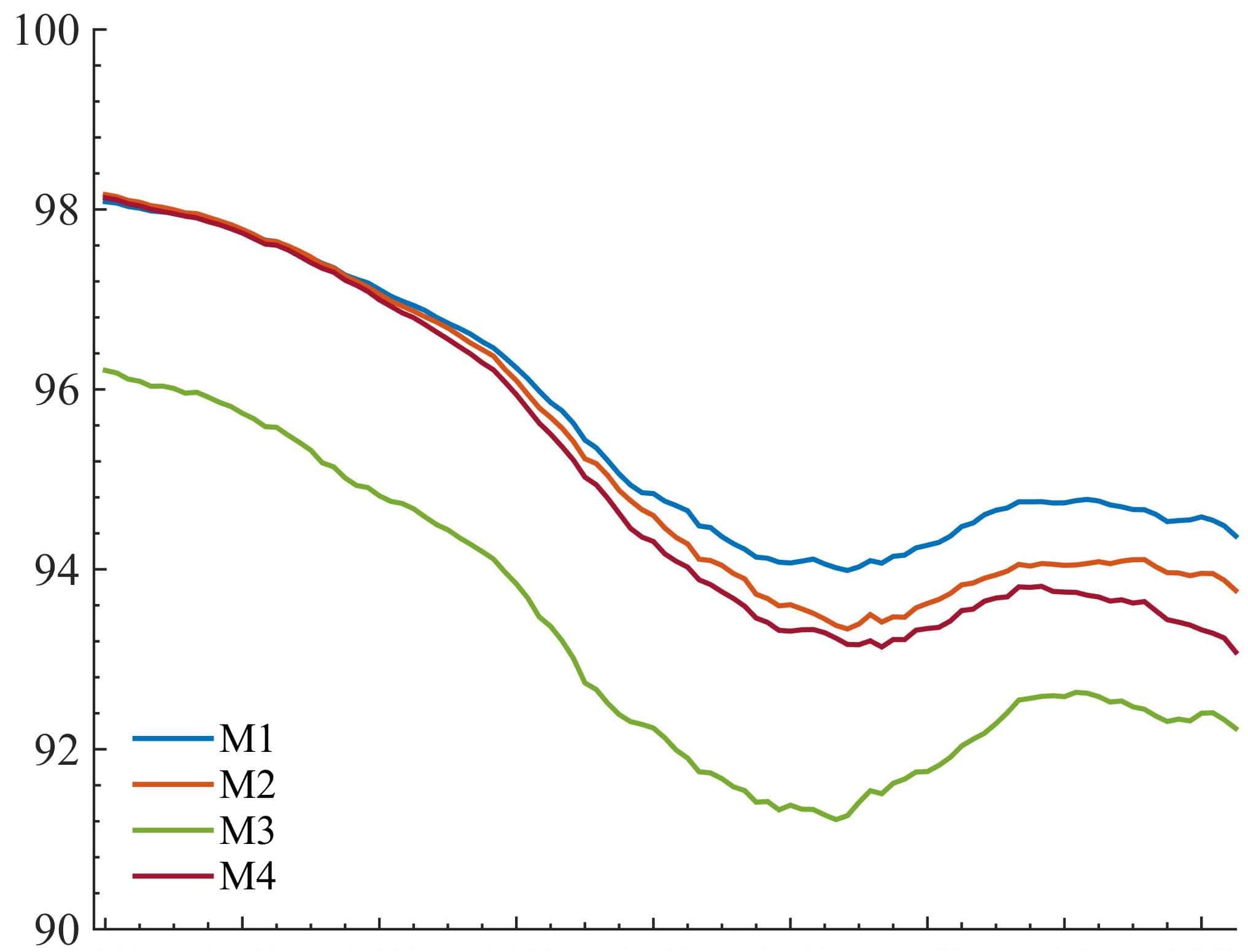} 
\end{minipage}
\hspace{.01\linewidth}
\begin{minipage}{.4\linewidth}
  \includegraphics[width=2in]{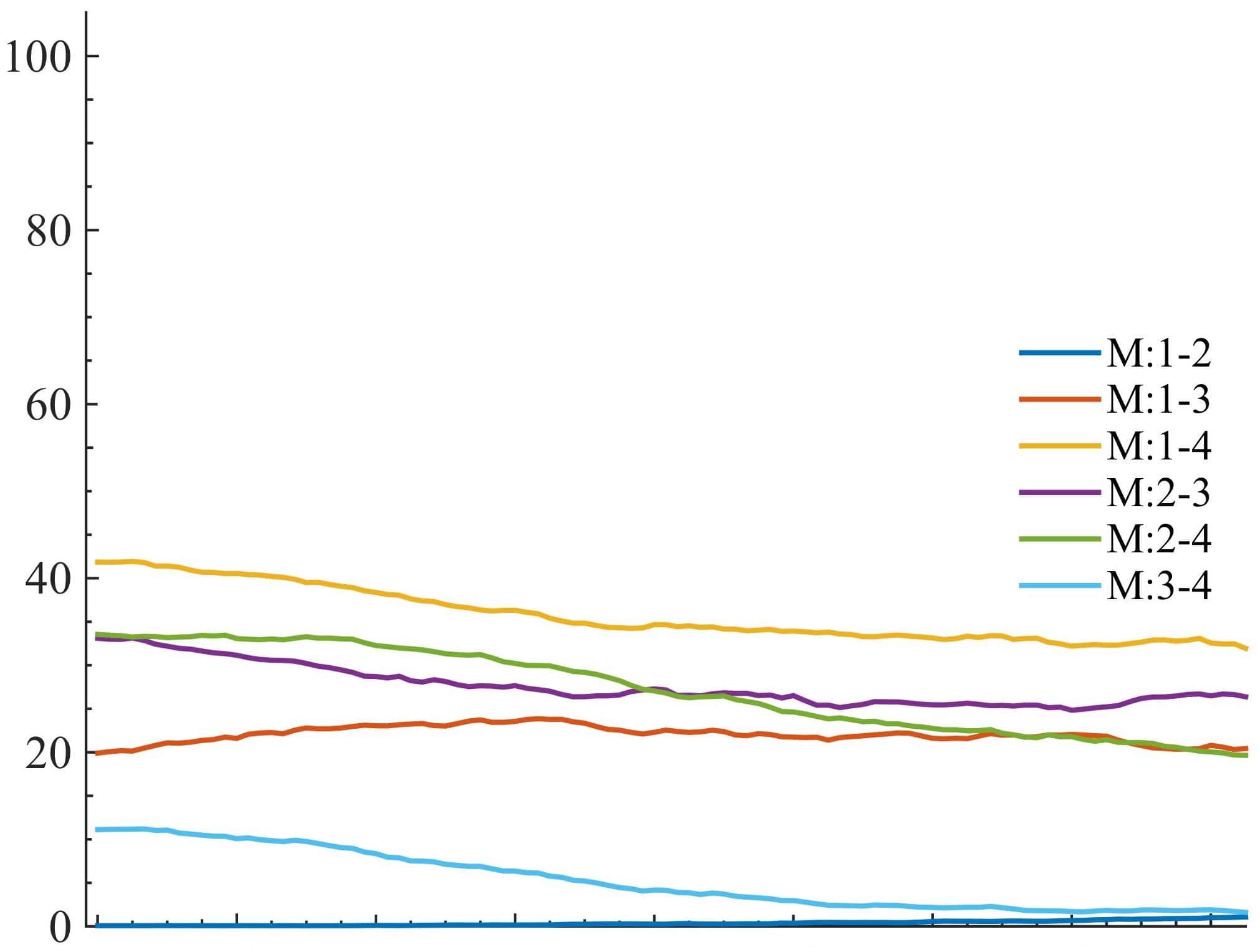}
  \end{minipage}
  \caption{Simulated data forecasting: MC-empirical R$^2$ (Left) and Paired MC-empirical R$^2$ (Right) of the forecasts in the analysis of synthetic data using 1-step ahead forecasts.}
    \label{S1r2}
\end{figure}

\begin{figure}[htbp!]
\centering
\begin{minipage}{.4\linewidth}
  \includegraphics[width=2in]{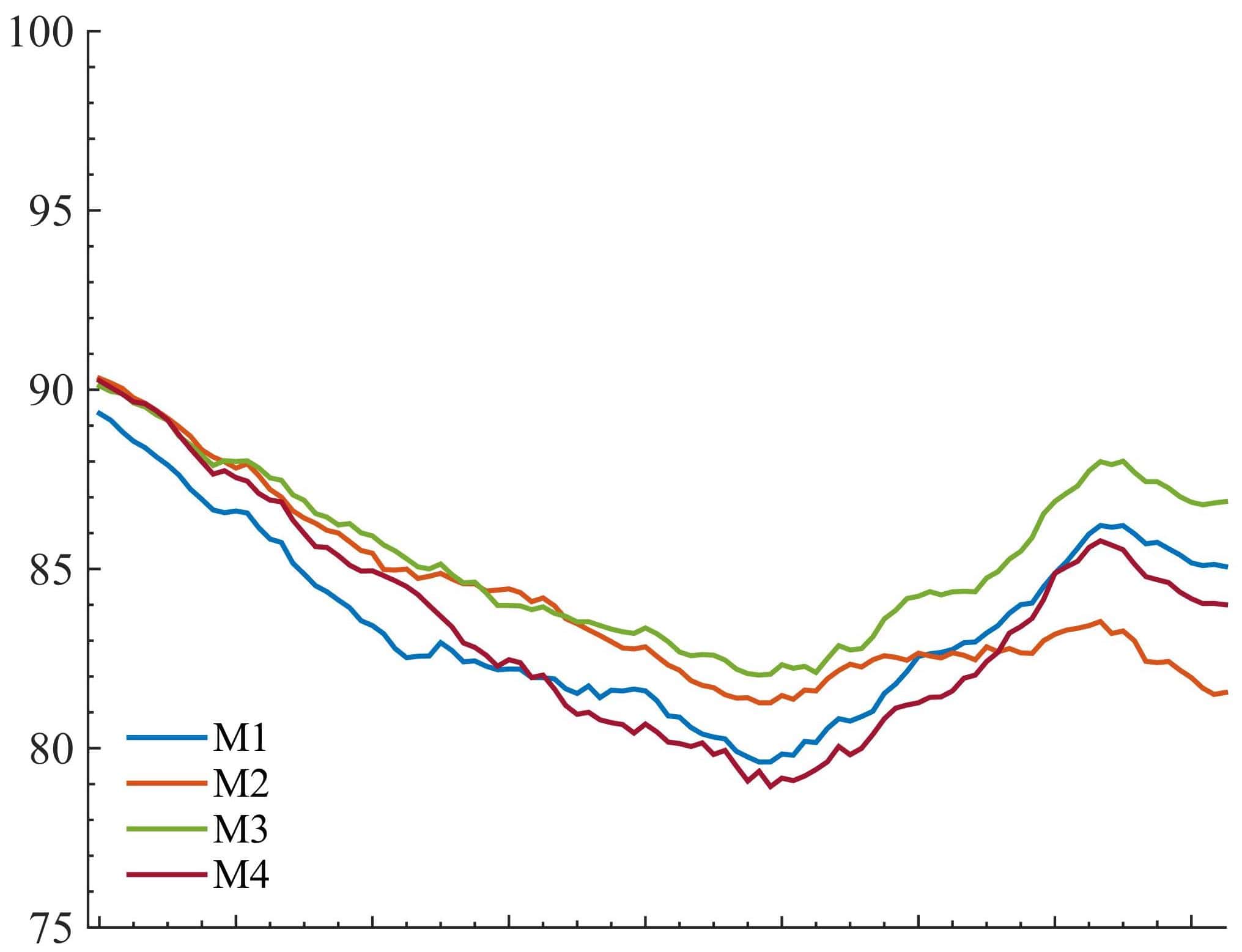} 
\end{minipage}
\hspace{.01\linewidth}
\begin{minipage}{.4\linewidth}
  \includegraphics[width=2in]{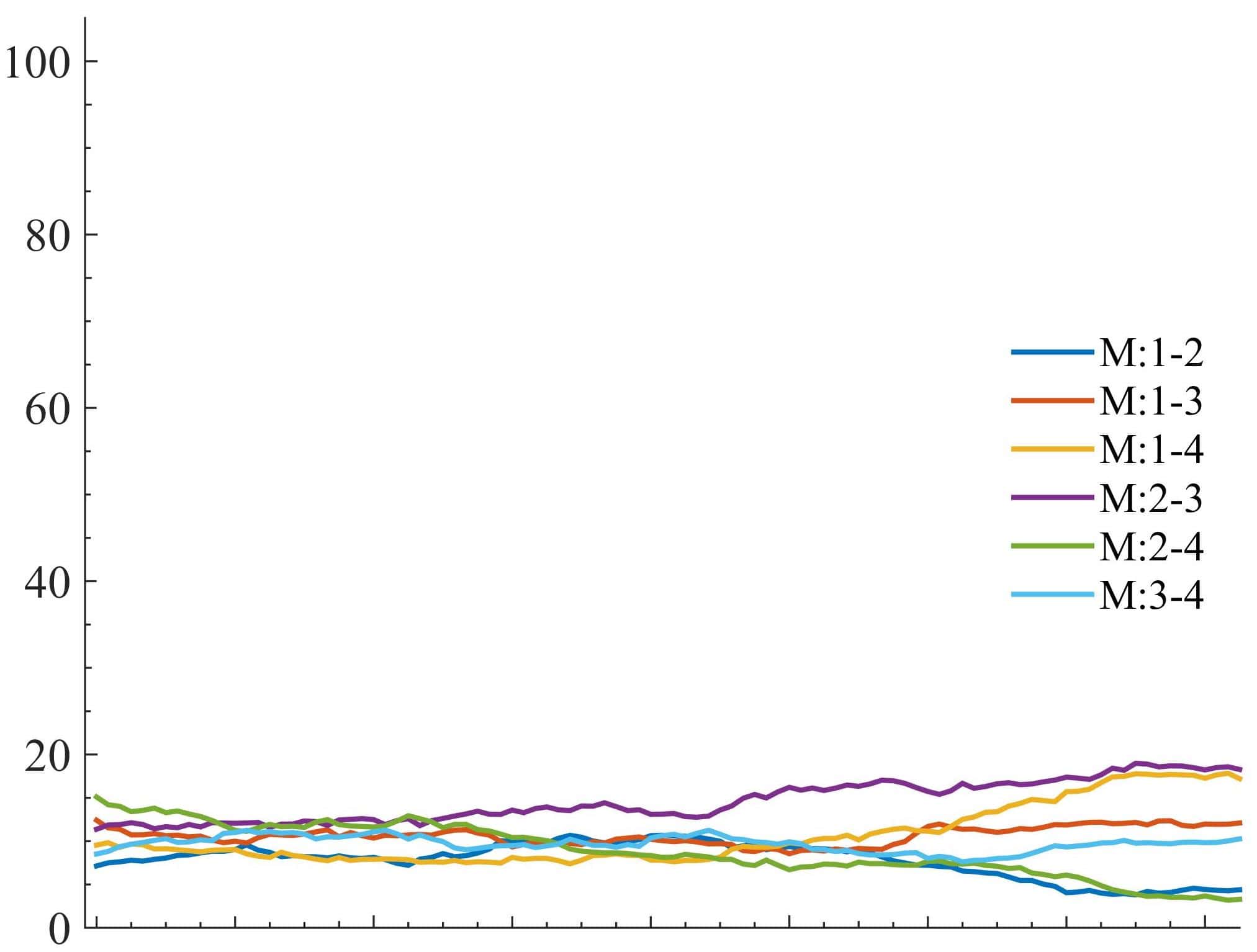}
  \end{minipage}
  \caption{Simulated data forecasting: MC-empirical R$^2$ (Left) and Paired MC-empirical R$^2$ (Right) of the forecasts in the analysis of synthetic data using 4-step ahead forecasts.}
    \label{S4r2}
\end{figure}

\end{document}